\documentclass[twoside,10pt]{article}

\usepackage{amsmath,amsfonts}
\usepackage{amssymb,latexsym}
\usepackage{amsxtra}
\usepackage{upref}
\usepackage{exscale}
\usepackage[mathscr]{eucal}

\newcommand{\be}{\begin{equation}}
\newcommand{\ee}{\end{equation}}
\newcommand{\beq}{\begin{eqnarray}}
\newcommand{\eeq}{\end{eqnarray}}
\newcommand{\no}{\nonumber}
\newcommand{\bea}{\begin{array}}
\newcommand{\eea}{\end{array}}
\newcommand{\lb}{\label}
\newcommand{\mcal}{\mathcal}
\newcommand{\mscr}{\mathscr}
\newcommand{\mfrak}{\mathfrak}
\newcommand{\ve}{\varepsilon}

\newcommand{\ts}{\textstyle}
\newcommand{\pp}{\partial}
\newcommand{\im}{\imath}
\newcommand{\ppr}{^{\boldsymbol{\prime}}}
\newcommand{\pppr}{^{\boldsymbol{\prime\prime}}}

\newcommand{\pdag}{^{\dagger}}
\newcommand{\wt}{\widetilde}
\newcommand{\ovv}{\overline}
\newcommand{\uvv}{\underline}

\newcommand{\ph}{\phantom}
\newcommand{\scr}{\scriptstyle}
\newcommand{\scrscr}{\scriptscriptstyle}
\newcommand{\scz}{\scriptsize}
\newcommand{\trab}{\raisebox{-4pt}{$\mbox{tr}\atop {\scriptstyle a,b}$}}
\newcommand{\trpq}{\raisebox{-4pt}{$\mbox{tr}\atop {\scriptstyle p,q}$}}
\newcommand{\TRAB}{\raisebox{-4pt}{$\mbox{Tr}\atop {\scriptstyle a,b;p,q}$}}
\newcommand{\sdelta}{{\scr\Delta}}
\numberwithin{equation}{section}
\allowdisplaybreaks[4]

\begin{document}

\begin{center}
{\large\bf Ensemble averaged coherent state path integral for disordered bosons} \vspace*{0.1cm}\\
{\large\bf with a repulsive interaction and derivation of a nonlinear sigma model} \vspace*{0.3cm}\\
{\bf Bernhard Mieck}
\footnote{e-mail: "bjmeppstein@arcor.de"; freelance activity 2010; current location :
Zum Kohlwaldfeld 16, D-65817 Eppstein, Germany.} \vspace*{0.2cm}\\ 
\end{center}
\begin{abstract}
A coherent state path integral is considered for bosons with an ensemble average of a random potential and with an additional,
repulsive interaction in the context of BEC under inclusion of specially prepared disorder. The essential normalization of the coherent
state path integral, as a generating function of observables, is obtained from the non-equilibrium time contour for 'forward' and 'backward'
propagation so that a time contour metric has to be taken into account in the ensemble average with the random potential. Therefore,
the respective symmetries for the derivation of a nonlinear sigma model follow from the involved time contour metric which leads to a
coset decomposition $\mbox{Sp}(4)\,/\,\mbox{U}(2)\otimes\mbox{U}(2)$ of the symplectic group $\mbox{Sp}(4)$ with the 
subgroup $\mbox{U}(2)$ for the unitary invariance of the density-related vacuum or ground state; the corresponding spontaneous
symmetry breaking gives rise to anomalous- or 'Nambu'-doubled field degrees of freedom within self-energy matrices which are finally
regarded by remaining coset matrices. The notion of a 'return probability', according to the original 'Anderson-localization', is thus
naturally contained within coherent state path integrals of a non-equilibrium contour time for equivalent 'forward' and 'backward'
propagation.\newline
\vspace*{0.3cm}

\noindent {\bf Keywords:} Bose-Einstein condensation, ensemble averages for random potentials,
coherent state path integral, many-particle physics, non-equilibrium or Keldysh time contour.\newline
\noindent {\bf PACS 03.75.Nt , 03.75.Kk , 03.75.Hh}
\end{abstract}

\tableofcontents

\newpage

\section{Coherent state path integral and averaging method for disorder} \lb{s1}

\subsection{Ensemble averages for static and dynamic disorder} \lb{s11}

The original concept of 'Anderson localization' is combined to a 'return probability' of a wave packet, which starts out to
propagate from an initial space point and which is scattered back by the impurities and the disordered potential, for a
measure of its 'localization' \cite{Gonis}. This 'return probability' can therefore be specified by the propagation of ensemble averaged,
retarded and advanced Green functions \cite{Efetov}. Our presented approach in terms of coherent state path integrals, adapted to a system 
of BEC in an external trap potential, naturally contains this concept by a time contour metric with 'forward' and 'backward'
propagation\cite{Neg}-\cite{Mosk}. 
The ensemble average of a random potential involves the time contour metric which thus determines the
symmetries for the derivation of a nonlinear sigma model\cite{pop2,pop1}. 
The coset decomposition \(\mbox{Sp}(4)\,/\,\mbox{U}(2)\otimes\mbox{U}(2)\)
of an anomalous-doubled self-energy gives rise to a spontaneous symmetry breaking (SSB) with the subgroup \(\mbox{U}(2)\)
for an invariant vacuum or ground state of density-related field degrees of freedom. The remaining coset matrices 
\(\mbox{Sp}(4)\,/\,\mbox{U}(2)\) finally comprise the anomalous- or 'Nambu'-doubled field degrees of freedom from the
off-diagonal block parts of the total self-energy which is accomplished by a Hubbard-Stratonovich transformation (HST)
of anomalous-doubled, dyadic products of boson fields with incorporation of 'hinge' fields from the SSB \cite{Nambu,Gold,Strato}.

We consider two models (\ref{s1_1}) with Hamiltonians \(\hat{H}_{I}(\hat{\psi}\pdag,\hat{\psi},V_{I})\) and
\(\hat{H}_{II}(\hat{\psi}\pdag,\hat{\psi},V_{II})\) in normal ordering of bosonic creation and annihilation 
operators and with static and dynamic disorder potentials \(V_{I}(\vec{x})\), \(V_{II}(\vec{x},t)\) in parallel. 
Both Hamiltonians (\ref{s1_1}) contain the same one-particle part \(\hat{h}(\vec{x})\) (\ref{s1_2}) with the
kinetic energy and with an external trap-potential \(u(\vec{x})\), which is shifted by a reference energy \(\mu_{0}\)
as a chemical potential. Apart from the same one-particle part \(\hat{h}(\vec{x})\) (\ref{s1_2}), we take the identical
repulsive interaction with parameter \(V_{0}>0\) for a quartic contact interaction of bosonic operators 
'\(\hat{\psi}_{\vec{x}}\pdag\:\hat{\psi}_{\vec{x}}\pdag\:\hat{\psi}_{\vec{x}}\:\hat{\psi}_{\vec{x}}\)' (\ref{s1_1}).
Furthermore, even- and complex-valued, spatially local source fields \(j_{\psi;\vec{x}}(t)\),
\(j_{\psi\psi;\vec{x}}(t)\) are included in common in order to allow for a SSB with a coherent, macroscopic
wavefunction and with anomalous (or 'Nambu') paired bosons, respectively. These source fields can also be used for the
determination of observables from differentiating various manners of the generating function. According to a
presupposed, spatially spherical symmetry of \(u(\vec{x})\), we normalize the spatial summations
\(\sum_{\vec{x}}\ldots\) (\ref{s1_3}) by the spherical system volume \(V^{(D)}\) of '$D$' dimensions
and by the spatial unit cell \((\sdelta x)^{D}\) which yields a parameter 
\(\mcal{N}_{x}=V^{(D)}/(\sdelta x)^{D}\) to be applied for various approximations, as e.g. in a saddle point
computation. The time parameter \(t_{p}\) is restricted to the range \(0\ldots +T_{0}\) with the discrete
intervals \(\sdelta t>0\) so that one is limited by the order of maximum energy 
\(\hbar\Omega_{max}=\hbar/\sdelta t\) within the propagation of the two Hamiltonians
\beq \lb{s1_1}
\hat{H}_{I,II}(\hat{\psi}\pdag,\hat{\psi},V_{I,II}) &=&
\sum_{\vec{x}}\hat{\psi}_{\vec{x}}\pdag\;\Big(
\hat{h}(\vec{x})+\left(V_{I}(\vec{x})\;\mbox{or}\;V_{II}(\vec{x},t)\right)+
V_{0}\;\hat{\psi}_{\vec{x}}\pdag\;\hat{\psi}_{\vec{x}}\bigg)\;\hat{\psi}_{\vec{x}}+
\\ \no &+&\sum_{\vec{x}}\Big(j_{\psi;\vec{x}}^{*}(t)\;\hat{\psi}_{\vec{x}}+
\hat{\psi}_{\vec{x}}\pdag\;j_{\psi;\vec{x}}(t)\Big)+\frac{1}{2}
\sum_{\vec{x}}\bigg(j_{\psi\psi;\vec{x}}^{*}(t)\;\;\hat{\psi}_{\vec{x}}\hat{\psi}_{\vec{x}}+
\hat{\psi}_{\vec{x}}\pdag\hat{\psi}_{\vec{x}}\pdag\;\;j_{\psi\psi;\vec{x}}(t)\bigg) \\  \lb{s1_2}
\hat{h}(\vec{x})&=&-\frac{\hbar^{2}}{2m}\bigg(\frac{\pp}{\pp\vec{x}}\cdot\frac{\pp}{\pp\vec{x}}\bigg)+
u(\vec{x})-\mu_{0}\;; \\     \lb{s1_3}
\sum_{\vec{x}}\ldots&=&\sum_{\vec{x}_{i}}^{|\vec{x}_{i}|<L}
\frac{(\sdelta x)^{D}}{V^{(D)}}\ldots =\int_{|\vec{x}|<L}\frac{d^{D}\!x}{V^{(D)}}\ldots\;;\;\;\;\;
V^{(D)}=\frac{S^{(D)}}{D}\;L^{D}\;;  \\ \no
S^{(D=1)} &=&1\;;\;\;\;\;S^{(D=2)}=2\pi\;;\;\;\;\;S^{(D=3)}=4\pi\;;\;\;\;
\mcal{N}_{x}=\frac{V^{(D)}}{(\sdelta x)^{D}}\;;  \\ \no
\Omega_{max} &=& \frac{1}{\sdelta t}\;;\;\;\;0<t_{p}<+T_{0}\;;\;\;\;T_{0}=T_{max}\;.
\eeq
The two Gaussian distributions for random potentials \(V_{I}(\vec{x})\), \(V_{II}(\vec{x},t)\) are
determined by the second moments (\ref{s1_4},\ref{s1_5}) for static and dynamic disorder, respectively,
and vanishing mean values. Both distributions are delta-function like, concerning the spatially 'contact'
Kronecker-delta and concerning the white-noise delta-function of time. Moreover, we emphasize the two
different normalizations of second moments in (\ref{s1_4},\ref{s1_5}) which are important in subsequent
transformations and derivations for a proper, finite scaling of energy ranges within the nonlinear sigma
models (cf. the second moments and their normalization in random matrix theories). Therefore, there occur
two different disorder parameters \(R_{I}\), \(R_{II}\) of physical dimensions [energy$\times$time] and
[energy$\times$$\mbox{(time)}^{1/2}$] in the two moments (\ref{s1_4},\ref{s1_5}). 
We apply the normalized generating functions \(Z[\hat{\mscr{J}},V_{I}]\), \(Z_{II}[\hat{\mscr{J}},V_{II}]\)
(\ref{s1_6}) with the time development operators (\ref{s1_7}) which are composed of the prevailing
Hamiltonian 'I' or 'II', each having an additional source matrix \(\hat{\mscr{J}}\) for calculating
observables by differentiating the coherent state path integral. The latter path integral (\ref{s1_6},\ref{s1_7})
results into proper normalization of unity (\ref{s1_8})
for vanishing source matrix \(\hat{\mscr{J}}\equiv0\) and identical
values (\ref{s1_9}) of symmetry breaking fields 
\(j_{\psi;\vec{x}}(t)\), \(j_{\psi\psi;\vec{x}}(t)\) on the two branches
of propagating time development operators in forward and backward direction
\beq \lb{s1_4}
\ovv{V_{I}(\vec{x}_{1})\;V_{I}(\vec{x}_{2})}&=&
\frac{R_{I}^{2}\;\Omega_{max}^{2}}{\mcal{N}_{x}}\;\; \delta_{\vec{x}_{1},\vec{x}_{2}} \;;
\hspace*{0.5cm}\mbox{static disorder}\\ \lb{s1_5}
\ovv{V_{II}(\vec{x}_{1},t_{1})\;V_{II}(\vec{x}_{2},t_{2})}&=&
R_{II}^{2}\;\;\delta_{\vec{x}_{1},\vec{x}_{2}}\;\;
\delta(t_{1}-t_{2}) \;;\hspace*{0.5cm}\mbox{dynamic disorder}\;\;\;;  \\ \lb{s1_6}
Z[\hat{\mscr{J}},V_{I,II}]&=&\big\langle 0\big|\hat{U}_{I,II}\big(t=0,+T_{0};V_{I,II};\hat{\mscr{J}}\big)
\;\;\hat{U}_{I,II}\big(+T_{0},t=0;V_{I,II};\hat{\mscr{J}}\big)\big|0\big\rangle \;\;;   \\ \lb{s1_7}
\hat{U}_{I,II}\big(t,0;V_{I,II};\hat{\mscr{J}}\big) &=&
\overleftarrow{\exp}\bigg\{-\frac{\im}{\hbar}\int_{0}^{t}d\tau\;\;
\hat{H}_{I,II}\big(\hat{\psi}\pdag,\hat{\psi},V_{I,II};\hat{\mscr{J}}\big)\bigg\}_{\mbox{;}}  \\ \lb{s1_8}
Z[\hat{\mscr{J}}\equiv 0,V_{I,II}]\Big|_{\{j_{\psi},j_{\psi\psi}\}} &\equiv & 1 \;\;; \\  \lb{s1_9}
\ldots\Big|_{\{j_{\psi},j_{\psi\psi}\}} &:=& \Big\{j_{\psi;\vec{x}}(t_{+})=j_{\psi;\vec{x}}(t_{-})\;;\;
j_{\psi\psi;\vec{x}}(t_{+})=j_{\psi\psi;\vec{x}}(t_{-})\Big\} \;\;; \\ \lb{s1_10}
\ovv{Z_{I,II}[\hat{\mscr{J}}]}&=&\ovv{\big\langle 0\big|\hat{U}_{I,II}\big(t=0,+T_{0};V_{I,II};\hat{\mscr{J}}\big)
\;\;\hat{U}_{I,II}\big(+T_{0},t=0;V_{I,II};\hat{\mscr{J}}\big)\big|0\big\rangle}\Big|_{\{j_{\psi},j_{\psi\psi}\}\;\;.} 
\eeq
In accordance to the forward '+' and backward '-' propagation of time development operators, we introduce a contour
time '\(t_{p=\pm}\)' and contour integral \(\int_{C}dt_{p}\ldots\) with a contour time metric
'\(\eta_{p=\pm}=\pm1\)' in order to regard the changing sign of phases in the exponent of the time
evolution operators (\ref{s1_11}-\ref{s1_14}) \cite{nonequi}-\cite{Bm6}. In the remainder we will also briefly outline the various steps
of transformations to a nonlinear sigma model with incorporation of the precise time steps and intervals within
the coherent state path integral which is specified by the normal ordering of creation and annihilation operators.
These precise time steps, with additional time shift '\(\sdelta t_{p}\)' in the complex conjugated coherent
state fields, are usually omitted for brevity in literature, but are ubiquitous in many-particle physics; this
problem of proper time steps does not prevent transformations and derivations to sigma models with SSB
and a coset decomposition and has necessarily to be regarded as soon as quantum mechanical field integration
variables at neighbouring, but still different times '\(t_{p}\)' and '\(t_{p}+\sdelta t_{p}\)', are
considered within normal ordered coherent state path integrals. We briefly hint at the problem of the 
appropriate, precise time steps by defining a slightly 
modified time contour (\ref{s1_13},\ref{s1_14}) for simplified representation of the exact time steps  
(cf. also additional boundary conditions of coherent state fields in  (\ref{s2_22}-\ref{s2_25}))
\beq \lb{s1_11}
\int_{C}d t_{p}\ldots &=&\int_{0}^{+T_{0}}dt_{+}\;\ldots+\int_{+T_{0}}^{0}dt_{-}\;\ldots=
\int_{0}^{+T_{0}}dt_{+}\;\ldots-\int_{0}^{+T_{0}}dt_{-}\;\ldots \;\;; \\ \lb{s1_12}
\int_{C}d t_{p}\ldots &=&\sum_{p=\pm}\int_{0}^{+T_{0}}dt_{p}\;\eta_{p}\;\ldots\;\;;\hspace{0.6cm}
\eta_{p}=\Big\{\underbrace{\eta_{+}=+1}_{p=+}\;;\;\underbrace{\eta_{-}=-1}_{p=-}\Big\}\;;   \\ \lb{s1_13}
\int_{\breve{C}}d t_{p}\ldots &=&\int_{-\sdelta t}^{+T_{0}}
dt_{+}\ldots+\int_{+(T_{0}+\sdelta t)}^{0}dt_{-}\ldots=
\int_{-\sdelta t}^{+T_{0}}dt_{+}\ldots-
\int_{0}^{+(T_{0}+\sdelta t)}d t_{-}\ldots \;\;; \\ \lb{s1_14}
\int_{\breve{C}}d t_{p}\ldots &=&\sum_{p=\pm}
\int_{\breve{0}_{p}}^{+\breve{T}_{p}}dt_{p}\;\;\eta_{p}\ldots \;; \hspace*{0.3cm}\breve{0}_{+}=-\sdelta t\;;\;
\breve{T}_{+}=T_{0}\;;\;\breve{0}_{-}=0\;;\;\breve{T}_{-}=T_{0}+\sdelta t\;\;.
\eeq
Since coherent state path integrals allow for the exact time sequence of coherent state fields with proper,
additional time shifts '\(\sdelta t_{p}\)' of the corresponding complex-conjugated fields, 
one can also investigate other kinds of coherent state path integrals as those of Eqs. (\ref{s1_4}-\ref{s1_10}) \cite{bmprecise,bmdelta,bmultra}.
Apart from the presented problem, we have also applied coherent state path integrals to a trace representation
of delta functions with maximal commuting sets of symmetry operators (\ref{s1_15},\ref{s1_16}). As these
symmetry operators (\ref{s1_17}-\ref{s1_20}) are given in terms of normal ordered creation and
annihilation operators, one achieves a similar coherent state path integral on time contours as
(\ref{s1_4}-\ref{s1_10},\ref{s1_11}-\ref{s1_14})
\beq\lb{s1_15}
\varrho_{j,m_{z}}(E,n_{0}) &=&\mbox{Coherent state path integral representation of the trace} \\ \no &:=&
\mbox{Tr}\Big[\delta(E-\hat{H})\;\delta(n_{0}-\hat{N})\;
\delta(\hbar^{2}\,j(j+1)-\hat{\vec{J}}{\boldsymbol\cdot}\hat{\vec{J}})\;
\delta(\hbar\,m_{z}-\hat{J}_{z})\Big] \;;   \\ \lb{s1_16}
\hat{H},\,\hat{N},\,\hat{\vec{J}}{\boldsymbol\cdot}\hat{\vec{J}},\,\hat{J}_{z} &:=&
\mbox{maximal commuting set of second quantized operators
according to symmetries};  \\ \lb{s1_17}
\hat{H}(\hat{\psi}\pdag,\hat{\psi}) &=& \sum_{\vec{x},s}\hat{\psi}_{\vec{x},s}\pdag
\bigg(-\frac{\hbar^{2}}{2\,m}\;\frac{\pp}{\pp\vec{x}}\cdot\frac{\pp}{\pp\vec{x}}-\frac{n_{0}\;e^{2}}{|\vec{x}|}+
\sum_{\vec{x}\ppr,s\ppr}\frac{e^{2}}{|\vec{x}-\vec{x}\ppr|}\;\hat{\psi}_{\vec{x}\ppr,s\ppr}\pdag\;
\hat{\psi}_{\vec{x}\ppr,s\ppr}\bigg)\;\hat{\psi}_{\vec{x},s} \;;  \\ \lb{s1_18}
\hat{N}(\hat{\psi}\pdag,\hat{\psi}) &=&\sum_{\vec{x},s}\hat{\psi}_{\vec{x},s}\pdag\;
\hat{\psi}_{\vec{x},s} \;;  \\  \lb{s1_19}
\hat{\vec{J}}(\hat{\psi}\pdag,\hat{\psi}) &=& \sum_{\vec{x},s_{1,2}}\hat{\psi}_{\vec{x},s_{2}}\pdag\;
\Big(\vec{x}\times\hat{\vec{p}}+{\ts\frac{\hbar}{2}}\vec{\sigma}\Big)_{\!\!s_{2}s_{1}}
\hat{\psi}_{\vec{x},s_{1}} \;;   \\   \lb{s1_20}
\hat{J}_{z}(\hat{\psi}\pdag,\hat{\psi}) &=&\sum_{\vec{x},s_{1,2}}
\hat{\psi}_{\vec{x},s_{2}}\pdag\;\Big(\big(\vec{x}\times\hat{\vec{p}}\big)_{z}+
{\ts\frac{\hbar}{2}}\;\hat{\sigma}_{z}\Big)_{\!\!s_{2}s_{1}}\hat{\psi}_{\vec{x},s_{1}} \;.
\eeq
This is accomplished by the application of the 
Dirac identity (\ref{s1_21}) to the principal value '\(\mfrak{P}\)' and
delta function of the symmetry operators (\ref{s1_22}-\ref{s1_25}) so that propagation with exponentials
is also implied on two branches of a '{\it disconnected}' time contour for the integral representation
of delta functions. Therefore, one can also perform ensemble averages (\ref{s1_26}) of trace relations
with delta-functions of symmetry operators in their representation with coherent state path integrals,
very similar to (\ref{s1_4}-\ref{s1_14}). However, it is important to distinguish between 
one-particle (\ref{s1_19},\ref{s1_20}) and two-particle operators (\ref{s1_17},\ref{s1_27}) 
in the various transformations to a nonlinear sigma model with inclusion of a
coset decomposition for a SSB with a HST
\beq \lb{s1_21}
\lim_{y\rightarrow0_{+}}\frac{1}{x+\im\;y} &=&\mfrak{P}\frac{1}{x}-\im\;\pi\;\delta(x)\;; \\  \lb{s1_22}
\frac{1}{E-\hat{H}(\hat{\psi}\pdag,\hat{\psi})-\im\;\ve_{+}} &=&
\mfrak{P}\frac{1}{E-\hat{H}(\hat{\psi}\pdag,\hat{\psi})}+\im\;\pi\;\delta(E-\hat{H}(\hat{\psi}\pdag,\hat{\psi})\,)  \;; \\  \lb{s1_23}
\frac{1}{\mfrak{o}-\hat{\mfrak{O}(\hat{\psi}\pdag,\hat{\psi})}-\im\;\ve_{+}} &=&
\mfrak{P}\frac{1}{\mfrak{o}-\hat{\mfrak{O}}(\hat{\psi}\pdag,\hat{\psi})}
+\im\;\pi\;\delta(\mfrak{o}-\hat{\mfrak{O}}(\hat{\psi}\pdag,\hat{\psi})\,)\;;  \\  \lb{s1_24}
\delta(\mfrak{o}-\hat{\mfrak{O}}(\hat{\psi}\pdag,\hat{\psi})\,) &=& \sum_{p=\pm}\int_{0}^{+\infty}
\frac{dt_{p}}{2\pi\,\hbar}\;\exp\big\{-\im\,\eta_{p}\:(t_{p}/\hbar)\:
\big(\mfrak{o}-\hat{\mfrak{O}}(\hat{\psi}\pdag,\hat{\psi})-\im\;\ve_{p}\big)\big\} \;;  \\   \lb{s1_25}
\hat{\mfrak{O}}(\hat{\psi}\pdag,\hat{\psi}) &=& 
\hat{H}(\hat{\psi}\pdag,\hat{\psi}),\,\hat{N}(\hat{\psi}\pdag,\hat{\psi}),\,
\hat{\vec{J}}(\hat{\psi}\pdag,\hat{\psi}){\boldsymbol\cdot}\hat{\vec{J}}(\hat{\psi}\pdag,\hat{\psi}),\,
\hat{J}_{z}(\hat{\psi}\pdag,\hat{\psi})  \;\;\;;  \\ \lb{s1_26}
\ovv{\varrho_{j,m_{z}}(E_{1},E_{2},n_{0})} &=&\ovv{\mbox{Tr}\Big[\delta(E_{1}-\hat{H}(\hat{\psi}\pdag,\hat{\psi})\,)\;
\delta(E_{2}-\hat{H}(\hat{\psi}\pdag,\hat{\psi})\,)\;
\delta(n_{0}-\hat{N}(\hat{\psi}\pdag,\hat{\psi})\,)\;\times}  \\ \no &&\ovv{\times\delta(\hbar^{2}\,j(j+1)-
\hat{\vec{J}}(\hat{\psi}\pdag,\hat{\psi}){\boldsymbol\cdot}\hat{\vec{J}}(\hat{\psi}\pdag,\hat{\psi})\,)\;
\delta(\hbar\,m_{z}-\hat{J}_{z}(\hat{\psi}\pdag,\hat{\psi})\,)\Big]} \;;   \\  \lb{s1_27}
\hat{\vec{J}}(\hat{\psi}\pdag,\hat{\psi}) \boldsymbol{\cdot}\hat{\vec{J}}(\hat{\psi}\pdag,\hat{\psi}) &=& 
\bigg(\sum_{\vec{x},s_{1,2}}\hat{\psi}_{\vec{x},s_{2}}\pdag\;
\Big(\vec{x}\times\hat{\vec{p}}+{\ts\frac{\hbar}{2}}\vec{\sigma}\Big)_{\!\!s_{2}s_{1}}
\hat{\psi}_{\vec{x},s_{1}} \bigg) \boldsymbol{\cdot} \bigg(     \sum_{\vec{x}\ppr,s\ppr_{1,2}}
\hat{\psi}_{\vec{x}\ppr,s\ppr_{2}}\pdag\;
\Big(\vec{x}\ppr\times\hat{\vec{p}}\,\ppr+{\ts\frac{\hbar}{2}}\vec{\sigma}\Big)_{\!\!s_{2}\ppr s_{1}\ppr}
\hat{\psi}_{\vec{x}\ppr,s_{1}\ppr}\bigg)  \;.
\eeq

\section{Ensemble averages in model I and II for a normal-ordered Hamiltonian}\lb{s2}

\subsection{Precise time steps with shifts '$\sdelta t_{p}$' of
the complex conjugated fields '$\psi_{\vec{x}}^{*}(t_{p}\!\!+\!\!\sdelta t_{p})$'} \lb{s21}

The ensemble averages of random potentials involve an additional, mathematical aspect which concerns
the combination of fields of the two different branches (\(p=\pm\)) of the time contour. This
formal aspect is also implied by the original formulation of Anderson localization where one
examines a 'return' probability of a test wave-packet (as an initial delta-spike at a particular
space point). This 'return' probability measures the 'forward' propagation of the wave-packet, away
from an initial space point, and the corresponding 'backward' propagation of remaining wave-packet
parts, back again to the same initial space point. Localization of wave-packets within a disordered
potential is specified by a 'finite' 'return' probability which approaches vanishing values towards
delocalization. According to this physical picture, we have to combine the doubling of bosonic
coherent state fields \(\psi_{\vec{x}}(t_{+})=\psi_{\vec{x},+}(t)\), 
\(\psi_{\vec{x}}(t_{-})=\psi_{\vec{x},-}(t)\), apart from their usual anomalous or 'Nambu' doubling
\((\psi_{\vec{x}}(t_{p})\,{\boldsymbol,}\,\psi_{\vec{x}}^{*}(t_{p})\,)\) of ordered systems with
solely hermitian operators, with the two distinct branches of contour time. Consequently, there occur
two different kinds of 'Nambu' doubling which we term 'anomalous-doubled ordering' (\ref{s2_1}-\ref{s2_4})
and 'contour time ordering' (\ref{s2_5}-\ref{s2_8}), the latter being marked by a bar under the
doubled fields \(\uvv{\Psi}_{\vec{x}}^{a}(t_{p})\) (\ref{s2_5}),
\(\uvv{\Psi}_{\vec{x}}^{\dag,a}(t_{p})\) (\ref{s2_6}), 
\(\uvv{\breve{\Psi}}_{\vec{x}}^{a}(t_{p})\) (\ref{s2_7}), 
\(\uvv{\breve{\Psi}}_{\vec{x}}^{\sharp,a}(t_{p})\) (\ref{s2_8}).
The 'anomalous-doubled' ordering (\ref{s2_1}-\ref{s2_4}) groups the fields
\(\Psi_{\vec{x}}^{a}(t_{p})=(\psi_{\vec{x},p=\pm}(t)\,(a=1)\,{\boldsymbol;}\,\psi_{\vec{x},p=\pm}^{*}(t)\,
(a=2)\,)^{T}\) according to complex conjugation regardless of the branches of contour time \(t_{p=\pm}\)
whereas the 'contour time ordering' (\ref{s2_5}-\ref{s2_8}) comprises fields of identical contour metric
sign \(\eta_{p}\), regardless of complex conjugation. Aside from the 'equal time' and the hermitian
conjugation of 'equal time' 'Nambu' doubled fields \(\Psi_{\vec{x}}^{a}(t_{p})\),
\(\Psi_{\vec{x}}^{\dag,a}(t_{p})\) (\ref{s2_1},\ref{s2_2}) or
\(\uvv{\Psi}_{\vec{x}}^{a}(t_{p})\), \(\uvv{\Psi}_{\vec{x}}^{\dag,a}(t_{p})\) (\ref{s2_5},\ref{s2_6}),
one has also to introduce time shifted versions of anomalous doubled fields
\(\breve{\Psi}_{\vec{x}}^{a}(t_{p})\), \(\breve{\Psi}_{\vec{x}}^{\sharp,a}(t_{p})\) (\ref{s2_3},\ref{s2_4}),
\(\uvv{\breve{\Psi}}_{\vec{x}}^{a}(t_{p})\), \(\uvv{\breve{\Psi}}_{\vec{x}}^{\sharp,a}(t_{p})\)
(\ref{s2_7},\ref{s2_8}) for the exact proper sequence of time steps within the considered 'quantum' problem
which is represented by the coherent state path integrals of normal ordered Hamiltonians. The time
shifted, doubled fields (\ref{s2_3},\ref{s2_4}), (\ref{s2_7},\ref{s2_8}) are marked 
by the symbol '\(\breve{\ph{\psi}}\)' to be distinguished from the equal-time, doubled fields
(\ref{s2_1},\ref{s2_2}), (\ref{s2_5},\ref{s2_6}). Note that the underbar '\(\uvv{\ph{\psi}}\)' always
hints to 'contour time ordering' (\ref{s2_5}-\ref{s2_8}) as in doubled fields, self-energy or
source matrices and fields, independent of additional 'equal time' or 'time shifted' doubling
\beq
\no\mbox{{\bf{\sf 'anomalous-doubled ordering'}}}   &:& t_{p}=\sdelta t\tfrac{1-\eta_{p}}{2},\ldots\ldots,
T_{0}-\sdelta t\tfrac{1+\eta_{p}}{2}\;; \\
\no \mbox{(1)}&:& \mbox{'{\it equal time}', anomalous-doubled field :}  \\  \lb{s2_1}
\Psi_{\vec{x}}^{a(=1/2)}(t_{p(=\pm)})&=&
\Big(\underbrace{\psi_{\vec{x},+}(t)\:,\:\psi_{\vec{x},-}(t)}_{a=1}\;;\;
\underbrace{\psi_{\vec{x},+}^{*}(t)\:,\:\psi_{\vec{x},-}^{*}(t)}_{a=2}\Big)^{T} \;;   \\ 
\no \mbox{(2)} &:& \mbox{'hermitian-conjugation' '$\pdag$' of '{\it equal time}', anomalous-doubled field :} \\  \lb{s2_2}
\Psi_{\vec{x}}^{\dag a(=1/2)}(t_{p(=\pm)})&=&\Big(\big(\hat{\tau}_{1}\big)^{ab}\;
\Psi_{\vec{x}}^{b}(t_{p(=\pm)})\Big)^{T,a} =
\Big(\underbrace{\psi_{\vec{x},+}^{*}(t)\:,\:\psi_{\vec{x},-}^{*}(t)}_{a=1}\;;\;
\underbrace{\psi_{\vec{x},+}(t)\:,\:\psi_{\vec{x},-}(t)}_{a=2}\Big)   \;; \\
\no \mbox{(1)}&:& \mbox{'{\it time shifted}' $\sdelta\!t_{p}$, anomalous-doubled field denoted by
'$\boldsymbol{\breve{\ph{\Psi}}}$' :}  \\   \lb{s2_3}
\breve{\Psi}_{\vec{x}}^{a(=1/2)}(t_{p(=\pm)})&=&
\Big(\underbrace{\psi_{\vec{x},+}(t)\:,\:\psi_{\vec{x},-}(t)}_{a=1}\;;\;
\underbrace{\psi_{\vec{x},+}^{*}(t+\sdelta t)\:,\:\psi_{\vec{x},-}^{*}(t-\sdelta t)}_{a=2}\Big)^{T} \;;  \\      \no \mbox{(2)} &:& 
\mbox{'hermitian-conjugation' '$^{\sharp}$' with '{\it time shift correction}' \(\sdelta\!t_{p}\) in the complex part :}   \\     \lb{s2_4}
\breve{\Psi}_{\vec{x}}^{a(=1/2)}(t_{p(=\pm)}) &\stackrel{'\sharp'}{\Longrightarrow}&
\breve{\Psi}_{\vec{x}}^{\sharp a(=1/2)}(t_{p(=\pm)})= \Big(\big(\hat{\tau}_{1}\big)^{ab}\;
\breve{\Psi}_{\vec{x}}^{b}(t_{p(=\pm)})\Big)^{T,a} = \\ \no &=&
\Big(\underbrace{\psi_{\vec{x},+}^{*}(t+\sdelta t)\:,\:\psi_{\vec{x},-}^{*}(t-\sdelta t)}_{a=1}\;;\;
\underbrace{\psi_{\vec{x},+}(t)\:,\:\psi_{\vec{x},-}(t)}_{a=2}\Big)   \;;  \\ 
\no\mbox{{\bf{\sf 'contour time ordering'}}}   &:& t_{p}=\sdelta t\tfrac{1-\eta_{p}}{2},\ldots\ldots,
T_{0}-\sdelta t\tfrac{1+\eta_{p}}{2}\;; \\
\no \mbox{(1)}&:& \mbox{'{\it equal time}', contour time doubled field :}  \\  \lb{s2_5}
\uvv{\Psi}_{\vec{x}}^{a(=1/2)}(t_{p(=\pm)})&=&
\Big(\underbrace{\overbrace{\psi_{\vec{x},+}(t)\:,\:\psi_{\vec{x},+}^{*}(t)}^{p=+}\;;\;
\overbrace{\psi_{\vec{x},-}(t)\:,\:\psi_{\vec{x},-}^{*}(t)}^{p=-}\Big)^{T}}_{\mbox{\scz contour time ordering}} \;; \\
\no \mbox{(2)} &:& \mbox{'hermitian-conjugation' '$\pdag$' of '{\it equal time}', 
contour time doubled field :} \\  \lb{s2_6}
\uvv{\Psi}_{\vec{x}}^{\dag a(=1/2)}(t_{p(=\pm)})&=& 
\Big(\underbrace{\overbrace{\psi_{\vec{x},+}^{*}(t)\:,\:\psi_{\vec{x},+}(t)}^{p=+}\;;\;
\overbrace{\psi_{\vec{x},-}^{*}(t)\:,\:\psi_{\vec{x},-}(t)}^{p=-}\Big)}_{\mbox{\scz contour time ordering}}
\;;\\ \no \mbox{(1)}&:& \mbox{'{\it time shifted}' $\sdelta\!t_{p}$, contour time doubled field 
'$\boldsymbol{\breve{\ph{\Psi}}}$' :}  \\  \lb{s2_7}
\uvv{\breve{\Psi}}_{\vec{x}}^{a(=1/2)}(t_{p(=\pm)})&=& 
\Big(\underbrace{\overbrace{\psi_{\vec{x},+}(t)\:,\:\psi_{\vec{x},+}^{*}(t+\sdelta t)}^{p=+}\;;\;
\overbrace{\psi_{\vec{x},-}(t)\:,\:\psi_{\vec{x},-}^{*}(t-\sdelta t)}^{p=-}\Big)^{T}}_{\mbox{\scz contour time ordering}} \;;
\\ \no \mbox{(2)} &:& \mbox{'hermitian-conjugation' '$^{\sharp}$' 
with '{\it time shift correction}' \(\sdelta\!t_{p}\) in the complex part :}   \\   \lb{s2_8}
\uvv{\breve{\Psi}}_{\vec{x}}^{a(=1/2)}(t_{p(=\pm)}) &\stackrel{'\sharp'}{\Longrightarrow}&
\uvv{\breve{\Psi}}_{\vec{x}}^{\sharp a(=1/2)}(t_{p(=\pm)})=  
\Big(\underbrace{\overbrace{\psi_{\vec{x},+}^{*}(t+\sdelta t)\:,\:\psi_{\vec{x},+}(t)}^{p=+}\;;\;
\overbrace{\psi_{\vec{x},-}^{*}(t-\sdelta t)\:,\:\psi_{\vec{x},-}(t)}^{p=-}\Big)}_{\mbox{\scz contour time ordering}}\;.
\eeq
The Gaussian distributions (\ref{s1_4},\ref{s1_5}) with random potentials \(V_{I}(\vec{x})\), \(V_{II}(\vec{x},t)\)
lead to the averaged coherent state path integrals \(\ovv{Z_{I}[\hat{\mscr{J}}]}\),
\(\ovv{Z_{II}[\hat{\mscr{J}}]}\) which are composed of the same functional part (\ref{s2_11}) with the
doubled source fields \(J_{\psi;\vec{x}}^{a}(t_{p})\) (\ref{s2_12})
and source matrices \(\hat{J}_{\psi\psi;\vec{x}}^{ba}(t_{p})\) (\ref{s2_13},\ref{s2_14})
(\(\sdelta t_{p}=\eta_{p}\:\sdelta t\), \(\sdelta t_{q}=\eta_{q}\:\sdelta t\), \(\sdelta t>0\))
\beq \lb{s2_9}
\ovv{Z_{I}[\hat{\mscr{J}}]}\hspace*{-0.3cm}&=&\hspace*{-0.3cm}\boldsymbol{\bigg\langle}
F[\psi^{*},\psi;J_{\psi},\hat{J}_{\psi\psi};\hat{\mscr{J}}] \;
\exp\bigg\{\hspace*{-0.2cm}-\frac{R_{I}^{2}\Omega_{max}^{2}}{2\hbar^{2}\mcal{N}_{x}}\sum_{\vec{x}}
\bigg(\int_{\breve{C}}d t_{p}\;\;\psi_{\vec{x}}^{*}(t_{p}\!\!+\!\!\sdelta t_{p})\;\psi_{\vec{x}}(t_{p})\bigg)
\bigg(\int_{\breve{C}}d t_{q}\ppr\;\;\psi_{\vec{x}}^{*}(t_{q}\ppr\!\!+\!\!\sdelta t_{q}\ppr)\;
\psi_{\vec{x}}(t_{q}\ppr)\bigg)\bigg\} \boldsymbol{\bigg\rangle_{\hspace*{-0.2cm}F[\psi^{*},\psi]\mbox{;}}}  \\  \lb{s2_10}
\ovv{Z_{II}[\hat{\mscr{J}}]} \hspace*{-0.3cm}&=&\hspace*{-0.3cm}\boldsymbol{\bigg\langle}
F[\psi^{*},\psi;J_{\psi},\hat{J}_{\psi\psi};\hat{\mscr{J}}] \;
\exp\bigg\{\hspace*{-0.2cm}-\frac{R_{II}^{2}}{2\hbar^{2}}
\int_{\breve{0}}^{+\breve{T}}
dt\sum_{\vec{x}}\sum_{p,q=\pm} \Big(\psi_{\vec{x}}^{*}(t_{p}\!\!+\!\!\sdelta t_{p})\;\eta_{p}\;\psi_{\vec{x}}(t_{p})\Big)\;\;
\Big(\psi_{\vec{x}}^{*}(t_{q}\!\!+\!\!\sdelta t_{q})\;\eta_{q}\;\psi_{\vec{x}}(t_{q})\Big)\bigg\}
\boldsymbol{\bigg\rangle_{\hspace*{-0.2cm}F[\psi^{*},\psi]\mbox{;}}} \\   \lb{s2_11}
\lefteqn{\hspace*{-0.4cm}F[\psi^{*},\psi;J_{\psi},\hat{J}_{\psi\psi};\hat{\mscr{J}}]  =
\int d[\psi_{\vec{x}}(t_{p})]\;\exp\bigg\{-\frac{i}{\hbar}\int_{\breve{C}}
d t_{p}\sum_{\vec{x}}
\psi_{\vec{x}}^{*}(t_{p}\!\!+\!\!\sdelta t_{p})\;\;
\hat{H}_{p}(\vec{x},t_{p})\;\;\psi_{\vec{x}}(t_{p})\bigg\}  } \\ \no &\times&
\exp\bigg\{-\frac{\im}{2\hbar}\int_{\breve{C}}d t_{p}\sum_{\vec{x}}\bigg(\Big[
J_{\psi;\vec{x}}^{+a}(t_{p})\;\breve{\Psi}_{\vec{x}}^{a}(t_{p})+
\breve{\Psi}_{\vec{x}}^{\sharp a}(t_{p})\;J_{\psi;\vec{x}}^{a}(t_{p})\Big]+
\Big[\breve{\Psi}_{\vec{x}}^{\sharp b}(t_{p})\;\;\hat{J}_{\psi\psi;\vec{x}}^{ba}(t_{p})\;\;
\breve{\Psi}_{\vec{x}}^{a}(t_{p})\Big]\bigg)\bigg\}    \\ \no &\times&
\exp\bigg\{-\frac{\im}{2\hbar}\int_{\breve{C}}d t_{p}\;d t_{q}\ppr\sum_{\vec{x},\vec{x}\ppr}
\breve{\Psi}_{\vec{x}\ppr}^{\sharp b}(t_{q}\ppr)\;\;
\hat{\mscr{J}}_{\vec{x}\ppr,\vec{x}}^{ba}(t_{q}\ppr,t_{p})\;\;
\breve{\Psi}_{\vec{x}}^{a}(t_{p})\bigg\}\; \exp\bigg\{-\frac{\im}{\hbar}
\int_{\breve{C}}d t_{p}\sum_{\vec{x}}V_{0}\;\big(\psi_{\vec{x}}^{*}(t_{p}\!\!+\!\!\sdelta t_{p})\big)^{2}\;\;
\big(\psi_{\vec{x}}(t_{p})\big)^{2}\bigg\}  \;; 
\eeq
\beq        \lb{s2_12} &  & \mbox{'ordering for anomalous-doubled source field' :}\\ \no
J_{\psi;\vec{x}}^{a(=1/2)}(t_{p(=\pm)})&=&
\Big(\underbrace{j_{\psi;\vec{x}}(t_{+})\:,\: j_{\psi;\vec{x}}(t_{-})}_{(a=1)}\;;\;
\underbrace{j_{\psi;\vec{x}}^{*}(t_{+})\:,\:j_{\psi;\vec{x}}^{*}(t_{-})}_{(a=2)}
\Big)^{T}\;;  \\ \no 
J_{\psi;\vec{x},p(=\pm)}^{a(=1/2)}(t)&=&
\Big(\underbrace{j_{\psi;\vec{x},+}(t)\:,\:j_{\psi;\vec{x},-}(t)}_{(a=1)}\;;\;
\underbrace{j_{\psi;\vec{x},+}^{*}(t)\:,\:j_{\psi;\vec{x},-}^{*}(t)}_{(a=2)}
\Big)^{T}_{\mbox{;}}     \\  \lb{s2_13}
 &  & \mbox{'ordering for anomalous-doubled source matrix of pair condensates' :}\\ \no
\hat{J}_{\psi\psi;\vec{x}}^{ab}(t_{p})&=&\left(
\bea{cccc}
0 & 0 & j_{\psi\psi;\vec{x}}(t_{+}) & 0 \\
0 & 0 & 0 & j_{\psi\psi;\vec{x}}(t_{-}) \\
j_{\psi\psi;\vec{x}}^{*}(t_{+}) & 0 & 0 &  0 \\
0 & j_{\psi\psi;\vec{x}}^{*}(t_{-}) &  0 & 0
\eea\right)_{\mbox{;}}  \\ \no 
\hat{J}_{\psi\psi;\vec{x},p}^{ab}(t)&=&\left(
\bea{cccc}
0 & 0 & j_{\psi\psi;\vec{x},+}(t) & 0 \\
0 & 0 & 0 & j_{\psi\psi;\vec{x},-}(t) \\
j_{\psi\psi;\vec{x},+}^{*}(t) & 0 & 0 &  0 \\
0 & j_{\psi\psi;\vec{x},-}^{*}(t) &  0 & 0
\eea\right)_{\mbox{;}}    \\   \lb{s2_14}
\hat{J}_{\psi\psi;\vec{x}\ppr\!,\vec{x}}^{ba}(t_{q}\ppr,t_{p})&=&
\hat{J}_{\psi\psi;\vec{x}\ppr\!,q;\vec{x},p}^{ba}(t\ppr,t)
= \delta_{pq}\;\eta_{p}\;\delta(t_{p}-t_{q}\ppr)\;\delta_{\vec{x},\vec{x}\ppr}\;\;
\hat{J}_{\psi\psi;\vec{x}}^{ab}(t_{p}) \\ \no &=& \delta_{pq}\;\eta_{p}\;\delta(t_{p}-t_{q}\ppr)\;
\delta_{\vec{x},\vec{x}\ppr}\;\;\hat{J}_{\psi\psi;\vec{x},p}^{ab}(t)\;\;\;.
\eeq
The one-particle part of (\ref{s2_11}) is listed in relations (\ref{s2_15}-\ref{s2_17}) with Kronecker-deltas
of time for precise time steps following from properly normal ordered Hamiltonians 'I' and 'II'.
We also give a 'laxed' kind of the one-particle part in relations (\ref{s2_16},\ref{s2_17}) which is
appropriate for a classical approximation, as a first order variation of fields, within the exponentials of the
path integrals, but fails to result into the correct quantum expressions if integrations of coherent state fields
and their complex conjugates have to be performed
\beq \lb{s2_15}
\lefteqn{\hspace*{-1.9cm}\int_{\breve{C}}d t_{p}\sum_{\vec{x}}
\psi_{\vec{x}}^{*}(t_{p}+\sdelta\!t_{p})\;
\hat{H}_{p}(\vec{x},t_{p})\;\psi_{\vec{x}}(t_{p}) = } \\ \no
&\hspace*{-3.3cm}=&\hspace*{-1.8cm} \int_{\breve{C}}d t_{p}\sum_{\vec{x}}\bigg[
-\im\hbar\;\psi_{\vec{x}}^{*}(t_{p}+\sdelta\!t_{p})\;
\frac{\psi_{\vec{x}}(t_{p}+\sdelta\!t_{p})-\psi_{\vec{x}}(t_{p})}{\sdelta\!t_{p}}+ 
\psi_{\vec{x}}^{*}(t_{p}+\sdelta\!t_{p})\;\big(\hat{h}(\vec{x})-\im\;\hat{\ve}_{p}\big)\;
\psi_{\vec{x}}(t_{p})\bigg]  \\ 
\no &\hspace*{-3.3cm}=&\hspace*{-1.8cm} \int_{\breve{C}}d t_{p}\;dt_{q}\ppr 
\sum_{\vec{x}} \psi_{\vec{x}}^{*}(t_{q}\ppr)\;\delta_{qp}\;\eta_{q}\;\bigg(-\im\hbar\;
\frac{\delta_{t_{q}\ppr\boldsymbol{,}t_{p}}-
\delta_{t_{q}\ppr\boldsymbol{,}t_{p}+\sdelta\!t_{p}}}{\sdelta\!t_{p}}+
\big(\hat{h}(\vec{x})-\im\;\hat{\ve}_{p}\big)\;
\delta_{t_{q}\ppr\boldsymbol{,}t_{p}+\sdelta\!t_{p}}\bigg)\;\psi_{\vec{x}}(t_{p})  \\ \no
&\hspace*{-3.3cm}=&\hspace*{-1.8cm} \int_{\breve{C}}dt_{p}\;dt_{q}\ppr\sum_{\vec{x}}
\psi_{\vec{x}}^{*}(t_{q}\ppr+\sdelta\!t_{q}\ppr)\;\delta_{qp}\;\eta_{q}\;\bigg(-\im\hbar\;
\frac{\delta_{t_{q}\ppr\boldsymbol{,}t_{p}-\sdelta\!t_{p}}-
\delta_{t_{q}\ppr\boldsymbol{,}t_{p}}}{\sdelta\!t_{p}}+
\big(\hat{h}(\vec{x})-\im\;\hat{\ve}_{p}\big)\;
\delta_{t_{q}\ppr\boldsymbol{,}t_{p}}\bigg)\;\psi_{\vec{x},s}(t_{p})\;;  \\  \lb{s2_16}
\hat{H}_{p}(\vec{x},t_{p}) &=& -\hat{E}_{p}-\im\;\hat{\ve}_{p}+\hat{h}(\vec{x}) =
-\im\hbar\frac{\pp}{\pp t_{p}}-\im\;\hat{\ve}_{p}+
\frac{\vec{p}^{\;2}}{2m}+u(\vec{x})-\mu_{0}\;; \\  \lb{s2_17}
\hat{H}_{\vec{x},\vec{x}\ppr}(t_{p},t_{q}\ppr)&=&\hat{H}_{p}(t_{p},\vec{x}) \;
\delta_{pq}\;\eta_{p}\;\delta(t_{p}-t_{q}\ppr)\;\delta_{\vec{x},\vec{x}\ppr}\;; 
\hspace*{0.46cm}\ve_{p}=\eta_{p}\;\ve\;\;;(\ve>0;\;\;\eta_{\pm}=\pm 1)  \;\;. 
\eeq
According to the anomalous doubling, one can attain order parameter matrices (\ref{s2_18}-\ref{s2_21})
which are nonlocal in time for case 'I' of static disorder and which are local in time for case 'II'
of dynamic disorder. This corresponds to the described picture of a 'return' probability in Anderson
localization for case 'I' without an additional white-noise distribution in time. 
We can use the given order parameter matrices (\ref{s2_18}-\ref{s2_21}) 
with their various forms for the anomalous doubled self-energy with 'equal time' doubling of fields
(\ref{s2_18},\ref{s2_20}) and for the 'Nambu'-terms extended density matrices with 'time shifted'
doubling of fields (\ref{s2_19},\ref{s2_21}). However, we can simplify to the spatially local
case due to the contact disorder of spatial Kronecker-deltas  with the second moments of the
two Gaussian distributions (\ref{s1_4},\ref{s1_5})
\beq\lb{s2_18}
\hat{\Phi}_{\vec{x},p;\vec{x}\ppr,q}^{(I)ab}(t,t\ppr)&=&
\Psi_{\vec{x}}^{a}(t_{p})\otimes \Psi_{\vec{x}\ppr}^{\dag b}(t_{q}\ppr) 
\Longrightarrow \mbox{ anomalous doubled self-energy of static disorder}  \;;  \\   \lb{s2_19}
\breve{\Phi}_{\vec{x},p;\vec{x}\ppr,q}^{(I)ab}(t,t\ppr)&=&
\breve{\Psi}_{\vec{x}}^{a}(t_{p})\otimes \breve{\Psi}_{\vec{x}\ppr}^{\sharp b}(t_{q}\ppr) 
\Longrightarrow \mbox{ density matrices with anomalous terms for static disorder}  \;; \\  \lb{s2_20}
\hat{\Phi}_{\vec{x},p;\vec{x}\ppr,q}^{(II)ab}(t)&=&
\Psi_{\vec{x}}^{a}(t_{p})\otimes \Psi_{\vec{x}\ppr}^{\dag b}(t_{q})
\Longrightarrow \mbox{ anomalous doubled self-energy of dynamic disorder}  \;;  \\   \lb{s2_21}
\breve{\Phi}_{\vec{x},p;\vec{x}\ppr,q}^{(II)ab}(t)&=&
\breve{\Psi}_{\vec{x}}^{a}(t_{p})\otimes \breve{\Psi}_{\vec{x}\ppr}^{\sharp b}(t_{q})   
\Longrightarrow \mbox{ density matrices with anomalous terms for dynamic disorder}\;.
\eeq
In order to achieve the precise, subsequent time steps at the integration boundaries, we have to
introduce a modified time contour integration '\(\breve{C}\)', '\(\int_{\breve{C}}dt_{p}\ldots\)' which
is extended to the parameter end points '\(\breve{0}_{p}=-\sdelta t\,\tfrac{1+\eta_{p}}{2}\)' and 
'\(\breve{T}_{p}=T_{0}+\sdelta t\,\tfrac{1-\eta_{p}}{2}\)'.
We thus take integrations with the extended doubled coherent state fields which are restricted
to a single field (\(a\equiv1\)) at the values '\(\breve{0}_{p}=-\sdelta t\,\tfrac{1+\eta_{p}}{2}\)' and 
'\(\breve{T}_{p}=T_{0}+\sdelta t\,\tfrac{1-\eta_{p}}{2}\)'. One therefore has the particular field values
\beq \lb{s2_22}
\breve{\Psi}_{\vec{x},+}^{(a\equiv1)}\bigl(t=-\sdelta t\bigr) &=&\psi_{\vec{x},+}^{*}(t=0)\;;\hspace*{1.62cm}
\breve{\Psi}_{\vec{x},-}^{(a\equiv1)}\bigl(t=0\bigr) =\psi_{\vec{x},-}(t=0)\;;
\\ \lb{s2_23} \breve{\Psi}_{\vec{x},+}^{\sharp,(a\equiv1)}\bigl(t=-\sdelta t)\bigr) &=&
\psi_{\vec{x},+}^{*}(t=0)\;; \hspace*{1.44cm}
\breve{\Psi}_{\vec{x},-}^{\sharp,(a\equiv1)}\bigl(t=0\bigr) =\psi_{\vec{x},-}(t=0)\;;\\ \lb{s2_24} 
\breve{\Psi}_{\vec{x},+}^{(a\equiv1)}\bigl(t=+T_{0}\bigr) &=&\psi_{\vec{x},+}(t=+T_{0})\;;\;\;
\breve{\Psi}_{\vec{x},-}^{(a\equiv1)}\bigl(t=T_{0}+\sdelta t\bigr) =\psi_{\vec{x},-}^{*}(t=+T_{0})\;;
\\ \lb{s2_25}  \breve{\Psi}_{\vec{x},+}^{\sharp,(a\equiv1)}\bigl(t=+T_{0}\bigr) &=&
\psi_{\vec{x},+}(t=+T_{0})\;;  
\breve{\Psi}_{\vec{x},-}^{\sharp,(a\equiv1)}\bigl(t=T_{0}+\sdelta t\bigr) =\psi_{\vec{x},-}^{*}(t=+T_{0})\;;  \\ \no
\psi_{\vec{x},+}(t=+T_{0}) &=& \psi_{\vec{x},-}(t=+T_{0}) \;;   \hspace*{1.1cm}
\psi_{\vec{x},+}^{*}(t=+T_{0}) = \psi_{\vec{x},-}^{*}(t=+T_{0}) \;\;\;,
\eeq
which have to be combined with anomalous doubled matrices \(\hat{M}_{\vec{x}\ppr,\vec{x}}^{ba}(t_{q}\ppr,t_{p})\)
(\ref{s2_26}) for the proper exact sequence of time steps in order to obtain the correct propagation with the
ensemble averaged generating functions (\ref{s2_9}-\ref{s2_11}). We list the various notations 
(\ref{s2_26},\ref{s2_27}) of extended
contour time integrations for the presented cases of disorder which are straightforwardly generalized from the
solely hermitian cases of interactions with the already described precise time steps and intervals in one of our earlier
references
\beq \lb{s2_26}\hspace*{-0.66cm}
\int_{\breve{C}}d t_{p}\;d t_{q}\ppr\sum_{\vec{x},\vec{x}\ppr}\sum_{a,b=1,2}
\breve{\Psi}_{\vec{x}\ppr}^{\sharp b}(t_{q}\ppr)\;\hat{M}_{\vec{x}\ppr,\vec{x}}^{ba}(t_{q}\ppr,t_{p})\;
\breve{\Psi}_{\vec{x}}^{a}(t_{p}) \hspace*{-0.28cm} &=&\hspace*{-0.28cm} 
\int_{\breve{0}}^{+\breve{T}}\hspace*{-0.36cm}d t\;d t\ppr\sum_{\vec{x},\vec{x}\ppr}
\sum_{p,q=\pm}^{a,b=1,2}
\breve{\Psi}_{\vec{x}\ppr\!,q}^{\sharp b}(t\ppr)\;\eta_{q}\;
\hat{M}_{\vec{x}\ppr\!,q;\vec{x},p}^{ba}(t\ppr,t)\;\eta_{p}\;
\breve{\Psi}_{\vec{x},p}^{a}(t); \\   \lb{s2_27}  \hspace*{-0.66cm}
\int_{\breve{C}}d t_{p}\;d t_{q}\ppr\sum_{\vec{x},\vec{x}\ppr}\sum_{a,b=1,2}
\breve{\Psi}_{\vec{x}\ppr}^{\sharp b}(t_{q}\ppr)\;\hat{J}_{\psi\psi;\vec{x}\ppr\!,\vec{x}}^{ba}(t_{q}\ppr,t_{p})\;
\breve{\Psi}_{\vec{x}}^{a}(t_{p}) \hspace*{-0.28cm} &=& \hspace*{-0.28cm}
\int_{\breve{0}}^{+\breve{T}}\hspace*{-0.36cm}d t\;d t\ppr\sum_{\vec{x},\vec{x}\ppr}\sum_{p,q=\pm}^{a,b=1,2}
\breve{\Psi}_{\vec{x}\ppr\!,q}^{\sharp b}(t\ppr)\;\eta_{q}\;\hat{J}_{\psi\psi;\vec{x}\ppr\!,q;
\vec{x},p}^{ba}(t\ppr,t)\;\eta_{p}\;
\breve{\Psi}_{\vec{x},p}^{a}(t)  \\ \no \hspace*{-0.66cm}
\hspace*{-0.28cm} &=&\hspace*{-0.28cm}  \int_{\breve{C}}d t_{p}\sum_{\vec{x}}\sum_{a,b=1,2}
\breve{\Psi}_{\vec{x}}^{\sharp b}(t_{p})\;\;\hat{J}_{\psi\psi;\vec{x}}^{ba}(t_{p})\;\;
\breve{\Psi}_{\vec{x}}^{a}(t_{p})\;\;\;.
\eeq
Concerning presentation and notation, the case 'II' of dynamic disorder turns out to have simpler relations
than the case 'I' because the static disorder involves the combination of two distinct time parameters
\(t\), \(t\ppr\) in the self-energy matrices as in the already given order parameter matrices
\(\hat{\Phi}_{\vec{x},p;\vec{x},q}^{(I)ab}(t,t\ppr)\) (\ref{s2_18}) and 
\(\breve{\Phi}_{\vec{x},p;\vec{x},q}^{(I)ab}(t,t\ppr)\) (\ref{s2_19}). The dynamic disorder with a white-noise
time distribution restricts to a single time parameter in the self-energy matrices, as in the order parameter
matrices \(\hat{\Phi}_{\vec{x},p;\vec{x},q}^{(II)ab}(t)\) (\ref{s2_20}),  
\(\breve{\Phi}_{\vec{x},p;\vec{x},q}^{(II)ab}(t)\) (\ref{s2_21}), but still incorporates the two metric sign
labels \(p,q=\pm\) as additional field degree of freedom for a 'disorder' quasi-particle. However, both cases
'I', 'II' of disorder allow for analogous treatment of HST's and coset decompositions of nonlinear sigma
models, as one only considers the relevant reduction to stationary time states of fields with a single
frequency '\(\omega_{p}\)' in case 'I' of static disorder.

Solely stationary states of time in case 'I' result in the analogous transformations and derivations to
nonlinear sigma models of case 'II', if one replaces the single time parameter '\(t\)' in
\(\hat{\Phi}_{\vec{x},p;\vec{x},q}^{(II)ab}(t)\) (\ref{s2_20}),  
\(\breve{\Phi}_{\vec{x},p;\vec{x},q}^{(II)ab}(t)\) (\ref{s2_21}) with a single frequency parameter '\(\omega\)'
in case 'I' of static disorder. After the corresponding Fourier transformation to a frequency (or energy)
contour \(\breve{C}_{\omega}\) according to (\ref{s2_28},\ref{s2_29}), we achieve for (\ref{s2_9}) the
generating function \(\ovv{Z_{I}[\hat{\mscr{J}}]}\) (\ref{s2_30}) which contains two independent,
non-stationary energy contour integrations '\(\hbar\omega_{p}\)' and '\(\hbar\omega_{q}\ppr\)' in the
ensemble averaged Gaussian part. As we reduce to stationary states or to a single frequency parameter
\(\omega=\omega\ppr\), we introduce the approximation (\ref{s2_31}) with various contour labels \(p,q\)
very similar as in \(\ovv{Z_{II}[\hat{\mscr{J}}]}\) (\ref{s2_10})
\beq\lb{s2_28}
\Omega_{max} &=&\frac{1}{\sdelta t}\;\;\;;\;\;\;0<t_{p}<+T_{0}\;\;\;;\;\;\;T_{0}=T_{max} \;; \\ \no
0<&\omega_{p}&<+\Omega_{0}\;\;\;;\;\;\;\Omega_{0}=\Omega_{max}=\frac{1}{\sdelta t}\;\;\;;\; \;\;
\sdelta\omega=\frac{2\pi}{T_{max}}\;\;\;;\;\;\mscr{N}_{t}=T_{max}/\sdelta t\;\;\;;  \\  \lb{s2_29} 
\int_{\breve{C}_{\omega}}\frac{d\omega_{p}}{\sdelta\omega}\ldots\hspace*{-0.28cm} &=&
\int_{-\sdelta\omega}^{+\Omega_{0}}\frac{d\omega_{+}}{(\frac{2\pi}{T_{max}})}\;\ldots+
\int_{+(\Omega_{0}+\sdelta\omega)}^{0}\frac{d\omega_{-}}{(\frac{2\pi}{T_{max}})}\;\ldots  =  \\  \no &=&
\int_{-\sdelta\omega}^{+\Omega_{0}}\frac{d\omega_{+}}{(\frac{2\pi}{T_{max}})}\;\ldots-
\int_{0}^{+(\Omega_{0}+\sdelta\omega)}\frac{d\omega_{-}}{(\frac{2\pi}{T_{max}})}\;\ldots  =
\sum_{p=\pm}\int_{\breve{0}_{p}}^{+\breve{\Omega}_{p}}\frac{d\omega_{p}}{(\frac{2\pi}{T_{max}})}\;\eta_{p}\;\ldots ; 
\\  \no && \breve{0}_{+}=-\sdelta\omega\;;\;\;\breve{\Omega}_{+}=\Omega_{0}\;;\;\;\breve{0}_{-}=0\;;\;\;
\breve{\Omega}_{-}=\Omega_{0}+\sdelta\omega\;\;\;;  \\  \lb{s2_30}
\ovv{Z_{I}[\hat{\mscr{J}}]}&=&\boldsymbol{\bigg\langle}
F[\psi^{*},\psi;J_{\psi},\hat{J}_{\psi\psi};\hat{\mscr{J}}] \;\times \\ \no &\times&\hspace*{-0.28cm}
\exp\bigg\{\hspace*{-0.19cm}
-\frac{R_{I}^{2}}{2\hbar^{2}}\frac{\mscr{N}_{t}^{2}}{\mcal{N}_{x}}\sum_{\vec{x}}\hspace*{-0.1cm}
\bigg(\int_{\breve{C}_{\omega}}\hspace*{-0.19cm}\frac{d\omega_{p}}{(\frac{2\pi}{T_{max}})}\,
\psi_{\vec{x}}^{*}(\omega_{p})\;e^{\im\:\omega_{p}\:\Delta t_{p}}\;\psi_{\vec{x}}(\omega_{p})\bigg)
\bigg(\int_{\breve{C}_{\omega}}\hspace*{-0.19cm}\frac{d\omega_{q}\ppr}{(\frac{2\pi}{T_{max}})}\,
\psi_{\vec{x}}^{*}(\omega_{q}\ppr)\;e^{\im\:\omega_{q}\ppr\:\Delta t_{q}}\;
\psi_{\vec{x}}(\omega_{q}\ppr)\bigg)\bigg\}
\boldsymbol{\hspace*{-0.1cm}\bigg\rangle_{\hspace*{-0.2cm}F[\psi^{*},\psi];}}  \\ \lb{s2_31}
\ovv{Z_{I}[\hat{\mscr{J}}]}&\simeq&\boldsymbol{\bigg\langle}
F[\psi^{*},\psi;J_{\psi},\hat{J}_{\psi\psi};\hat{\mscr{J}}] \;\times \\ \no &\times&\hspace*{-0.28cm}
\exp\bigg\{\hspace*{-0.19cm}
-\frac{R_{I}^{2}}{2\hbar^{2}}\frac{\mscr{N}_{t}^{2}}{\mcal{N}_{x}}
\int_{\breve{0}}^{+\breve{\Omega}}
\frac{d\omega}{(\frac{2\pi}{T_{max}})}\sum_{\vec{x}}\sum_{p,q=\pm}\hspace*{-0.19cm}
\bigg(\psi_{\vec{x},p}^{*}(\omega)\;
e^{\im\:\omega\:\Delta t_{p}}\;\eta_{p}\;\psi_{\vec{x},p}(\omega)\bigg)
\bigg(\psi_{\vec{x},q}^{*}(\omega)\;e^{\im\:\omega\:\Delta t_{q}}\;\eta_{q}\;
\psi_{\vec{x},q}(\omega)\bigg)\bigg\}
\boldsymbol{\hspace*{-0.1cm}\bigg\rangle_{\hspace*{-0.2cm}F[\psi^{*},\psi].}} 
\eeq
The 'time-shift' correction of complex conjugated fields implies additional 
phases \(e^{\im\:\omega\cdot{\scrscr\Delta}t_{p}}\) in the static disorder case whose consideration
also leads to the exact, proper frequency or energy steps within the propagation of the time
development operators as in case 'II' of dynamic disorder. It is even possible to substitute
the parameter \(R_{II}^{2}\), according to relation (\ref{s2_32}), and the contour time
integrals by contour frequency integrals \(\int_{\breve{C}_{\omega}}d\omega_{p}/(2\pi/T_{max})\ldots\)
(\ref{s2_29}) so that the nonlinear sigma model of case 'II' straightforwardly generalizes to case 'I'
of static disorder
\beq\lb{s2_32}
R_{II}^{2}&\simeq&R_{I}^{2}\;\frac{\mscr{N}_{t}^{2}\;\Omega_{max}}{\mcal{N}_{x}} \;\;\;.
\eeq
In the following sections \ref{s3}, \ref{s4} we can therefore concentrate on the dynamic disorder
case with a single time variable in the self-energy matrix and can then transfer the result of case
'II' to the stationary case with a single frequency variable '\(\omega\)' in the self-energy
matrices for case 'I' of static disorder. The approximation to a single frequency \(\omega_{p}\) for static
disorder can also be attained at very later steps of transformations to a nonlinear sigma model, e.g. as one
simplifies the two-time or two-frequency dependent, nonlocal disorder self-energy matrix
to a single dependence with contour frequency \(\omega_{p}\).

\section{HST for dynamic disorder and repulsive interaction with 'hinge' fields}  \lb{s3}

\subsection{Anomalous doubling of the one-particle part} \lb{s31}

We perform the anomalous doubling (\ref{s3_1}) of the
bosonic fields with inclusion of the contour time metric \(\eta_{p}\), due to the ensemble average
with a dynamic disorder. This defines a 'Nambu' metric tensor \(\hat{K}_{pq}^{ab}\) (\ref{s3_2}) for
the anomalous doubled fields \(\breve{\Psi}_{\vec{x},q}^{\sharp,b}(t)\),
\(\breve{\Psi}_{\vec{x},p}^{a}(t)\) whose dyadic product determines the density 
matrices \(\breve{R}_{\vec{x};pq}^{ab}(t)\) (\ref{s3_3}) with anomalous extended parts in the off-diagonal
blocks \(a\neq b\). The quartic disorder interaction of bosonic fields is therefore equivalent to the trace
relation (\ref{s3_4}) with the density matrices \(\breve{R}_{\vec{x};pq}^{ab}(t)\) (\ref{s3_3}) which are
modified by the 'Nambu' metric tensor (\ref{s3_2}) with a final multiplication of a factor '\(\tfrac{1}{4}\)'
corresponding to two dyadic product operations of (\ref{s3_1}) with anomalous doubling of relation (\ref{s2_21})
\beq\lb{s3_1}
\psi_{\vec{x}}^{*}(t_{p}\!\!+\!\!\sdelta t_{p})\;\eta_{p}\;\psi_{\vec{x}}(t_{p})&=&
\frac{1}{2}\Big(\psi_{\vec{x}}^{*}(t_{p}\!\!+\!\!\sdelta t_{p})\;\eta_{p}\;\psi_{\vec{x}}(t_{p})+
\psi_{\vec{x}}(t_{p})\;\eta_{p}\;\psi_{\vec{x}}^{*}(t_{p}\!\!+\!\!\sdelta t_{p})\Big)  =
\frac{1}{2}\;\breve{\Psi}_{\vec{x}}^{\sharp b}(t_{q})\;\hat{K}_{qp}^{ba}\;
\breve{\Psi}_{\vec{x}}^{a}(t_{p}) \;;  \\ \lb{s3_2}
\hat{K}_{pq}^{ab}&=&\Bigg(\bea{cc} \big(\hat{\eta}\big)_{pq}^{11} & \\ & 
\big(\hat{\eta}\big)_{pq}^{22} \eea\Bigg)_{pq}^{ab} =
\delta_{ab}\;\;\delta_{pq}\;\;\eta_{p} =\delta_{ab}\;\;\delta_{pq}\;\;
\mbox{diag}\Big\{\underbrace{+1\;,\;-1}_{a=1}\;;\;\underbrace{+1\;,\;-1}_{a=2}\Big\} \;;
\eeq
\beq\lb{s3_3}
\lefteqn{
\breve{R}_{\vec{x}}^{ab}(t_{p},t_{q}) = \breve{R}_{\vec{x};pq}^{ab}(t)=
\breve{\Psi}_{\vec{x}}^{a}(t_{p})\,\otimes\,\breve{\Psi}_{\vec{x}}^{\sharp b}(t_{q}) =
\breve{\Psi}_{\vec{x};p}^{a}(t)\otimes\breve{\Psi}_{\vec{x};q}^{\sharp b}(t)  } \\ \no &=&
\left(\bea{c}\Bigg(\bea{c} \psi_{\vec{x},+}\!(t) \\ \psi_{\vec{x},-}(t) \eea \Bigg)_{p}^{a=1} \\ 
\Bigg( \bea{c} \psi_{\vec{x},+}^{*}\!(t\!\!+\!\!\sdelta t) \\
\psi_{\vec{x},-}^{*}(t\!\!-\!\!\sdelta t) \eea\Bigg)_{p}^{a=2}\eea\right)_{p}^{a}\otimes
\bigg(\underbrace{\psi_{\vec{x},+}^{*}\!(t\!\!+\!\!\sdelta t)\,,\, 
\psi_{\vec{x},-}^{*}(t\!\!-\!\!\sdelta t)}_{b=1}\;;\;
\underbrace{\psi_{\vec{x},+}\!(t)\,,\, \psi_{\vec{x},-}(t)}_{b=2} \bigg)_{q}^{b}   \\ \no &=&
\left( \bea{cc}
\Bigg(\bea{cc} \psi_{\vec{x},+}\!(t)\;\psi_{\vec{x},+}^{*}\!(t\!\!+\!\!\sdelta t) &
\psi_{\vec{x},+}\!(t)\;\psi_{\vec{x},-}^{*}(t\!\!-\!\!\sdelta t) \\
\psi_{\vec{x},-}\!(t)\;\psi_{\vec{x},+}^{*}\!(t\!\!+\!\!\sdelta t) &
\psi_{\vec{x},-}\!(t)\;\psi_{\vec{x},-}^{*}(t\!\!-\!\!\sdelta t)  \eea\Bigg)_{pq}^{11} & 
\Bigg(\bea{cc}\psi_{\vec{x},+}\!(t)\;\psi_{\vec{x},+}\!(t) &
\psi_{\vec{x},+}\!(t)\;\psi_{\vec{x},-}(t)  \\  
\psi_{\vec{x},-}\!(t)\;\psi_{\vec{x},+}\!(t) &
\psi_{\vec{x},-}\!(t)\;\psi_{\vec{x},-}(t)   \eea\Bigg)_{pq}^{12} \\
\hspace*{-0.3cm}\Bigg(\bea{cc} \psi_{\vec{x},+}^{*}\!(t\!\!+\!\!\sdelta t)\;
\psi_{\vec{x},+}^{*}\!(t\!\!+\!\!\sdelta t) &
\psi_{\vec{x},+}^{*}\!(t\!\!+\!\!\sdelta t)\;\psi_{\vec{x},-}^{*}(t\!\!-\!\!\sdelta t) \\
\psi_{\vec{x},-}^{*}\!(t\!\!-\!\!\sdelta t)\;
\psi_{\vec{x},+}^{*}\!(t\!\!+\!\!\sdelta t) &
\psi_{\vec{x},-}^{*}\!(t\!\!-\!\!\sdelta t)\;\psi_{\vec{x},-}^{*}(t\!\!-\!\!\sdelta t)  \eea\Bigg)_{pq}^{21} &
\hspace*{-0.2cm}\Bigg(\bea{cc}\psi_{\vec{x},+}^{*}\!(t\!\!+\!\!\sdelta t)\;\psi_{\vec{x},+}\!(t) &
\psi_{\vec{x},+}^{*}\!(t\!\!+\!\!\sdelta t)\;\psi_{\vec{x},-}\!(t) \\
\psi_{\vec{x},-}^{*}\!(t\!\!-\!\!\sdelta t)\;\psi_{\vec{x},+}\!(t) &
\psi_{\vec{x},-}^{*}\!(t\!\!-\!\!\sdelta t)\;\psi_{\vec{x},-}\!(t) \eea\Bigg)_{pq}^{22}  \hspace*{-0.2cm}
\eea\right)_{pq\;\mbox{;}}^{ab}  
\eeq
\beq\lb{s3_4}
\Big(\psi_{\vec{x}}^{*}(t_{p}\!\!+\!\!\sdelta t_{p})\;\eta_{p}\;\psi_{\vec{x}}(t_{p})\Big)\;
\Big(\psi_{\vec{x}}^{*}(t_{q}\!\!+\!\!\sdelta t_{q})\;\eta_{q}\;\psi_{\vec{x}}(t_{q})\Big) &=&\frac{1}{4}
\TRAB\Big[\breve{R}_{\vec{x};pq}^{ab}(t)\;\hat{K}\;\breve{R}_{\vec{x};qp}^{ba}(t)\;\hat{K}\Big] \;.
\eeq
This anomalous doubling has also to be taken for the one-particle part (\ref{s2_15}) so that we
accomplish relation (\ref{s3_5}) with the anomalous doubled fields 
\(\breve{\Psi}_{\vec{x}\ppr,q}^{\sharp,b}(t\ppr)\), \(\breve{\Psi}_{\vec{x},p}^{a}(t)\) and the doubled
one-particle operator \(\breve{\mscr{H}}_{\vec{x}\ppr,q;\vec{x},p}^{ba}(t\ppr,t)\) (\ref{s3_6}) whose
lower block diagonal part \(\breve{\mscr{H}}_{\vec{x}\ppr,q;\vec{x},p}^{22}(t\ppr,t)\) (\ref{s3_8})
follows by transposition from the upper block diagonal part
\(\breve{\mscr{H}}_{\vec{x}\ppr,q;\vec{x},p}^{11}(t\ppr,t)\) (\ref{s3_7}). Corresponding to
propagation with the exact, precise time steps, we outline these two one-particle parts in Eqs.
(\ref{s3_7},\ref{s3_8}) with the exact Kronecker deltas of time and also describe a 'laxed'
kind of one-particle operators in (\ref{s3_9},\ref{s3_10}), only applicable in classical
approximations or equations
\beq \lb{s3_5} 
\lefteqn{\hspace*{-1.9cm}\int_{\breve{C}}dt_{p}\sum_{\vec{x}}
\psi_{\vec{x}}^{*}(t_{p}+\sdelta\!t_{p})\;\;\hat{H}_{p}(\vec{x},t_{p})\;\;
\psi_{\vec{x}}(t_{p})=\int_{\breve{C}}dt_{p}\sum_{\vec{x}}
\psi_{\vec{x}}(t_{p})\;\;\hat{H}_{p}^{T}(\vec{x},t_{p})\;\;
\psi_{\vec{x}}^{*}(t_{p}+\sdelta\!t_{p}) =}    \\  \no  &=&
\int_{\breve{C}}dt_{p}\;dt_{q}\ppr\sum_{\vec{x},\vec{x}\ppr}\frac{1}{2}\;
\breve{\Psi}_{\vec{x}\ppr}^{\sharp b}(t_{q}\ppr)\;\;
\breve{\mscr{H}}_{\vec{x}\ppr;\vec{x}}^{ba}(t_{q}\ppr,t_{p})\;\;
\breve{\Psi}_{\vec{x}}^{a}(t_{p})  \;\;\;; \\  \lb{s3_6}   
\breve{\mscr{H}}_{\vec{x}\ppr;\vec{x}}^{ba}(t_{q}\ppr,t_{p})&=&
\mbox{diag}\left(\breve{\mscr{H}}_{\vec{x}\ppr;\vec{x}}^{11}(t_{q}\ppr,t_{p})\;;\;
\breve{\mscr{H}}_{\vec{x}\ppr;\vec{x}}^{22}(t_{q}\ppr,t_{p})\right)  \;; \\  \lb{s3_7}
\breve{\mscr{H}}_{\vec{x}\ppr;\vec{x}}^{11}(t_{q}\ppr,t_{p}) &=&
\bigg(-\im\,\hbar\;\tfrac{\delta(t_{q}\ppr-(t_{p}-\sdelta\!t_{p})\,)-\delta(t_{q}\ppr-t_{p})}{\sdelta\!t_{p}}+
\big(\hat{h}(\vec{x}\ppr)-\im\;\hat{\ve}_{p}\big)\;\delta(t_{q}\ppr-t_{p})\bigg)\delta_{qp}\;\eta_{q}\;
\delta_{\vec{x}\ppr,\vec{x}} \;; \\ \lb{s3_8}
\breve{\mscr{H}}_{\vec{x}\ppr;\vec{x}}^{22}(t_{q}\ppr,t_{p}) &=&
\Big(\breve{\mscr{H}}_{\vec{x}\ppr;\vec{x}}^{11}(t_{q}\ppr,t_{p}) \Big)^{T}=
\bigg(-\im\,\hbar\;\tfrac{\delta((t_{q}\ppr-\sdelta\!t_{q}\ppr)-t_{p})-\delta(t_{q}\ppr-t_{p})}{\sdelta\!t_{p}}+
\big(\hat{h}^{T}(\vec{x}\ppr)-\im\;\hat{\ve}_{p}\big)\;\delta(t_{q}\ppr-t_{p})\bigg)\delta_{qp}\;\eta_{q}\;
\delta_{\vec{x}\ppr,\vec{x}}\;;  \\ \lb{s3_9}
\breve{\mscr{H}}_{\vec{x}\ppr;\vec{x}}^{ba}(t_{q}\ppr,t_{p}) &\simeq&
\hat{\mscr{H}}_{\vec{x}\ppr;\vec{x}}^{ba}(t_{q}\ppr,t_{p}) = 
\mbox{diag}\Big(\underbrace{\hat{H}_{+}(\vec{x}\ppr,t_{+}\ppr)\:,\:
-\hat{H}_{-}(\vec{x}\ppr,t_{-}\ppr)}_{a=1}\;;\;\underbrace{\hat{H}_{+}^{T}(\vec{x}\ppr,t_{+}\ppr)\:,\:
-\hat{H}_{-}^{T}(\vec{x}\ppr,t_{-}\ppr)}_{a=2}\Big)
\delta_{\vec{x}\ppr,\vec{x}}\;\delta_{qp}\;\delta(t_{q}\ppr-t_{p}) ; \\  \lb{s3_10}
\hat{H}_{p}(\vec{x},t_{p})&=&-\im\hbar\frac{\pp}{\pp t_{p}}-\im\;\ve_{p}+
\frac{\vec{p}^{\;2}}{2m}+u(\vec{x})-\mu_{0} \;; \hspace*{0.6cm}
\hat{H}_{p}^{T}(\vec{x},t_{p})=+\im\hbar\frac{\pp}{\pp t_{p}}-\im\;\ve_{p}+
\frac{\vec{p}^{\;2}}{2m}+u(\vec{x})-\mu_{0}\;.
\eeq

\subsection{Anomalous doubled self-energies and their coset decomposition} \lb{s32}

In the following we convey the results of our earlier articles to the presented case 'II'
of dynamic disorder with the additional contour time metric tensor \(\eta_{p}\) and have to use a
modified 'Nambu' metric tensor \(\wt{K}_{pq}^{ab}\), \(\wt{\kappa}^{ab}\) (\ref{s3_11},\ref{s3_12})
which changes the anomalous parts in the self-energy to anti-hermitian relations
\beq\lb{s3_11}
\wt{K}_{pq}^{ab} &=&\Bigg(\bea{cc} \big(\hat{\eta}\big)_{pq}^{11} & \\ & 
-\big(\hat{\eta}\big)_{pq}^{22} \eea\Bigg)_{pq}^{ab} =
\delta_{ab}\;\;\delta_{pq}\;\;\eta_{p}\;\;\wt{\kappa}^{ab} =\delta_{ab}\;\;\delta_{pq}\;\;
\mbox{diag}\Big\{\underbrace{+1\;,\;-1}_{a=1}\;;\;\underbrace{-1\;,\;+1}_{a=2}\Big\} \;;   \\  \lb{s3_12}
\wt{\kappa}^{ab} &=&\delta_{ab}\;\;\mbox{diag}\Big\{\underbrace{+1}_{a=1}\;;\;
\underbrace{-1}_{a=2}\Big\}_{\mbox{.}}
\eeq
The final HST is achieved by one half with the self-energy density variable \(\sigma_{R_{II}}^{(0)}(\vec{x},t)\)
(\ref{s3_13}) and by one half with the 'Nambu' parts \(\delta\hat{\Sigma}_{\vec{x};pq}^{12}(t)\),
\(\delta\hat{\Sigma}_{\vec{x};pq}^{21}(t)\) \((a\neq b)\) (\ref{s3_13}) of the anomalous doubled
self-energy matrix \(\delta\hat{\Sigma}_{\vec{x};pq}^{ab}(t)\) (\ref{s3_14}) whose block diagonal,
hermitian self-energy density parts \(\delta\hat{\Sigma}_{\vec{x};pq}^{11}(t)\),
\(\delta\hat{\Sigma}_{\vec{x};pq}^{22}(t)\) (\ref{s3_15}) are related by transposition and are only used
as 'hinge' fields in a SSB with a coset decomposition. The off-diagonal, 'Nambu' matrix parts
\(\delta\hat{\Sigma}_{\vec{x};pq}^{12}(t)\), \(\delta\hat{\Sigma}_{\vec{x};pq}^{21}(t)\) (\ref{s3_16})
of \(\delta\hat{\Sigma}_{\vec{x};pq}^{ab}(t)\) (\ref{s3_14}) are symmetric matrices and are related by
hermitian conjugation. One thus acquires following real and complex parameters
\(\delta B_{\vec{x};++}(t)\), \(\delta B_{\vec{x};--}(t)\) and \(\delta B_{\vec{x};+-}(t)\),
\(\delta B_{\vec{x};-+}(t)=\delta B_{\vec{x};+-}^{*}(t)\) and the solely complex field variables
\(\delta c_{\vec{x};++}(t)\), \(\delta c_{\vec{x};--}(t)\), \(\delta c_{\vec{x};-+}(t)=
\delta c_{\vec{x};+-}(t)\) in the corresponding anomalous doubled part (\ref{s3_16}) of the
total self-energy (\ref{s3_14})
\beq\lb{s3_13}
\sigma_{R_{II}}^{(0)}(\vec{x},t) \in\mathbb{R}&;&\delta\hat{\Sigma}_{\vec{x};pq}^{11}(t)\;;\;
\delta\hat{\Sigma}_{\vec{x};pq}^{22}(t)\;;\;\delta\hat{\Sigma}_{\vec{x};pq}^{12}(t)\;;\;
\delta\hat{\Sigma}_{\vec{x};pq}^{21}(t)\;\mbox{ in analogy to }\;
\breve{R}_{\vec{x};pq}^{ab}(t) \;;   \\  \lb{s3_14}
\delta\hat{\Sigma}_{\vec{x};pq}^{ab}(t)&=&\left(\bea{cc}
\delta\hat{\Sigma}_{\vec{x};pq}^{11}(t) & \delta\hat{\Sigma}_{\vec{x};pq}^{12}(t) \\
\delta\hat{\Sigma}_{\vec{x};pq}^{21}(t) & \delta\hat{\Sigma}_{\vec{x};pq}^{22}(t) \eea\right)\;;
\\  \lb{s3_15} \delta\hat{\Sigma}_{\vec{x};pq}^{11}(t)&=&\left(\bea{cc}
\delta B_{\vec{x};++}(t) & \delta B_{\vec{x};+-}(t) \\
\delta B_{\vec{x};+-}^{*}(t) & \delta B_{\vec{x};--}(t) \eea\right)\;\;;\hspace*{0.5cm}
\delta\hat{\Sigma}_{\vec{x};pq}^{22}(t)=\left(\bea{cc}
\delta B_{\vec{x};++}(t) & \delta B_{\vec{x};+-}^{*}(t) \\
\delta B_{\vec{x};+-}(t) & \delta B_{\vec{x};--}(t) \eea\right)\;;
\\ \no \delta\hat{\Sigma}_{\vec{x};pq}^{22,T}(t)=\delta\hat{\Sigma}_{\vec{x};pq}^{11}(t)
&;& \delta B_{\vec{x};++}(t)\;,\;\delta B_{\vec{x};--}\;\in\mathbb{R} \;\;;
\hspace*{0.3cm}\delta B_{\vec{x};+-}(t)\;\in\mathbb{C}\;\;;\;\;
\delta B_{\vec{x};-+}(t)=\delta B_{\vec{x};+-}^{*}(t)  \;;
\\ \lb{s3_16} \delta\hat{\Sigma}_{\vec{x};pq}^{12}(t)&=&  \left(\bea{cc}
\delta c_{\vec{x};++}(t) & \delta c_{\vec{x};+-}(t) \\
\delta c_{\vec{x};+-}(t) & \delta c_{\vec{x};--}(t) \eea\right)\;\;;\hspace*{0.5cm}
\delta\hat{\Sigma}_{\vec{x};pq}^{21}(t)=  \left(\bea{cc}
\delta c_{\vec{x};++}^{*}(t) & \delta c_{\vec{x};+-}^{*}(t) \\
\delta c_{\vec{x};+-}^{*}(t) & \delta c_{\vec{x};--}^{*}(t) \eea\right)\;;
\\ \no 
\delta\hat{\Sigma}_{\vec{x};pq}^{21}(t)=\delta\hat{\Sigma}_{\vec{x};pq}^{12,+}(t) &;&
\delta\hat{\Sigma}_{\vec{x};pq}^{12,T}= \delta\hat{\Sigma}_{\vec{x};pq}^{12}(t) \;\;;\;\;  
\delta c_{\vec{x};pq}(t)\in\mathbb{C}\;\;;\hspace*{0.4cm}
\delta c_{\vec{x};+-}(t)=\delta c_{\vec{x};-+}(t)  \;.
\eeq
We combine the self-energy density variable \(\sigma_{R_{II}}^{(0)}(\vec{x},t)\) (\ref{s3_13})
and the anomalous doubled self-energy matrix \(\delta\hat{\Sigma}_{\vec{x};pq}^{ab}(t)\) (\ref{s3_14})
into (\ref{s3_17}) with block diagonal parts \(\hat{\Sigma}_{\vec{x};pq}^{11}(t)\) (\ref{s3_18}),
\(\hat{\Sigma}_{\vec{x};pq}^{22}(t)\) (\ref{s3_19}) and introduce the modified self-energy matrix
\(\delta\wt{\Sigma}_{\vec{x};pq}^{ab}(t)\;\wt{K}\) (\ref{s3_20},\ref{s3_21}) with 'Nambu' metric
tensor '\(\wt{K}\)' (\ref{s3_11},\ref{s3_12}) and anti-hermitian related, off-diagonal, anomalous
doubled blocks \(\delta\wt{\Sigma}_{\vec{x};pq}^{a\neq b}(t)=\im\;\delta\hat{\Sigma}_{\vec{x};pq}^{a\neq b}(t)\).
This allows to perform a coset decomposition into densities and bosonic parts according to
\(\mbox{Sp}(4)\rightarrow\mbox{Sp}(4)/\mbox{U}(2)\,\otimes\,\mbox{U}(2)\)
\beq  \lb{s3_17}
\left(\bea{cc}
\hat{\Sigma}_{\vec{x};pq}^{11}(t) & \delta\hat{\Sigma}_{\vec{x};pq}^{12}(t) \\
\delta\hat{\Sigma}_{\vec{x};pq}^{21}(t) & -\hat{\Sigma}_{\vec{x};pq}^{22}(t) \eea\right)&=&
\sigma_{R_{II}}^{(0)}(\vec{x},t)\;\;\hat{K}+
\left(\bea{cc}
\delta\hat{\Sigma}_{\vec{x};pq}^{11}(t) & \delta\hat{\Sigma}_{\vec{x};pq}^{12}(t) \\
\delta\hat{\Sigma}_{\vec{x};pq}^{21}(t) & -\delta\hat{\Sigma}_{\vec{x};pq}^{22}(t) \eea\right)\;; \\ \lb{s3_18}
\hat{\Sigma}_{\vec{x};pq}^{11}(t) &=&
\sigma_{R_{II}}^{(0)}(\vec{x},t)\;\;\hat{\eta}_{p}\;\;\delta_{pq} +
\delta\hat{\Sigma}_{\vec{x};pq}^{11}(t) \;;  \\  \lb{s3_19}
\hat{\Sigma}_{\vec{x};pq}^{22}(t) &=&
-\sigma_{R_{II}}^{(0)}(\vec{x},t)\;\;\hat{\eta}_{p}\;\;\delta_{pq} +
\delta\hat{\Sigma}_{\vec{x};pq}^{22}(t)  \;;   \\    \lb{s3_20}
\delta\wt{\Sigma}_{\vec{x};pq}^{aa}(t)&=&\delta\hat{\Sigma}_{\vec{x};pq}^{aa}(t)
\hspace*{1.0cm}
\delta\wt{\Sigma}_{\vec{x};pq}^{a\neq b}(t)=\im\;\;\delta\hat{\Sigma}_{\vec{x};pq}^{a\neq b}(t) \;; \\  \lb{s3_21}
\delta\wt{\Sigma}_{\vec{x};pq}^{ab}(t)\;\;\wt{K}&=& \left(\bea{cc}
\left(\bea{cc} \delta B_{\vec{x};++}(t)   & -\delta B_{\vec{x};+-}(t) \\
\delta B_{\vec{x};+-}^{*}(t) & -\delta B_{\vec{x};--}(t) \eea\right)_{pq}^{11}  &
\im\;\left(\bea{cc} -\delta c_{\vec{x};++}(t) & \delta c_{\vec{x};+-}(t) \\
-\delta c_{\vec{x};+-}(t) & \delta c_{\vec{x};--}(t) \eea\right)_{pq}^{12}  \\
\im\;\left(\bea{cc} \delta c_{\vec{x};++}^{*}(t) & -\delta c_{\vec{x};+-}^{*}(t) \\
\delta c_{\vec{x};+-}^{*}(t) & -\delta c_{\vec{x};--}^{*}(t) \eea\right)_{pq}^{21} &
\left(\bea{cc} -\delta B_{\vec{x};++}(t) & \delta B_{\vec{x};+-}^{*}(t) \\
-\delta B_{\vec{x};+-}(t) & \delta B_{\vec{x};--}(t) \eea\right)_{pq}^{22}
\eea\right)_{\mbox{.}}
\eeq
As one includes a further, derived 'Nambu' metric tensor \(\hat{I}_{pq}^{ab}\) (\ref{s3_22}) so that
the various kinds of self-energy terms with hermitian or anti-hermitian off-diagonal block parts 
can be transformed into each other (\ref{s3_23}), one finally succeeds into the coset decomposition
(\ref{s3_24}) with block diagonal densities \(\delta\hat{\Sigma}_{D;pq}^{aa}(\vec{x},t)\;\wt{K}\)
(similar to (\ref{s3_15})) and coset matrices \(\hat{T}_{pq}^{ab}(\vec{x},t)\) (\ref{s3_25}).
The latter coset matrices (\ref{s3_25}) consist of the generator \(\hat{Y}_{pq}^{ab}(\vec{x},t)\)
(\ref{s3_26}) with sub-generators \(\hat{X}_{pq}(\vec{x},t)\), \(-\hat{X}_{pq}^{*}(\vec{x},t)\) (\ref{s3_27})
as the 'Nambu' or anomalous doubled field degrees of freedom
\beq\lb{s3_22}
\hat{I}_{pq}^{ab}&=&\delta_{ab}\;\;\delta_{pq}\;\;\mbox{diag}\Big\{\underbrace{+1\,,\,+1}_{a=1}\;;\;
\underbrace{+\im\,,\,+\im}_{a=2}\Big\}\;\;\;;  \\  \lb{s3_23}
\wt{\Sigma}_{\vec{x};pq}^{ab}(t) &=&\hat{I}\;
\left(\bea{cc}
\hat{\Sigma}_{\vec{x};pq}^{11}(t) & \delta\hat{\Sigma}_{\vec{x};pq}^{12}(t) \\
\delta\hat{\Sigma}_{\vec{x};pq}^{21}(t) & -\hat{\Sigma}_{\vec{x};pq}^{22}(t) \eea\right)\;\hat{I} =
\sigma_{R_{II}}^{(0)}(\vec{x},t)\;\;\wt{K}+\delta\wt{\Sigma}_{\vec{x};pq}^{ab}(t) \;;  \\  \lb{s3_24}
\wt{\Sigma}_{\vec{x};pq}^{ab}(t)\;\wt{K}&=&\sigma_{R_{II}}^{(0)}(\vec{x},t)\;\;\hat{1}_{4\times 4}+
\delta\wt{\Sigma}_{\vec{x};pq}^{ab}(t)\;\;\wt{K}=   \\ \no &=&
\Big(\hat{T}(\vec{x},t)\Big)_{pp\ppr}^{aa\ppr}\;
\Big(\sigma_{R_{II}}^{(0)}(\vec{x},t)\;\;\hat{1}_{4\times 4}+
\delta\hat{\Sigma}_{D;p\ppr q\ppr}^{a\ppr=b\ppr}(\vec{x},t)\;\;\wt{K}\Big)\;
\Big(\hat{T}^{-1}(\vec{x},t)\Big)_{q\ppr q}^{b\ppr b} \;; \\   \lb{s3_25}
\hat{T}(\vec{x},t) &=&\exp\Big\{-\hat{Y}_{pq}^{ab}(\vec{x},t)\Big\}\;\;;\hspace*{0.5cm}
\hat{T}^{-1}(\vec{x},t) = \exp\Big\{\hat{Y}_{pq}^{ab}(\vec{x},t)\Big\}  \;;  \\    \lb{s3_26}
\hat{Y}_{pq}^{ab}(\vec{x},t)&=&\left(\bea{cc} \Big(0\Big)_{pq}^{11} &
\Big(\hat{X}_{pq}(\vec{x},t)\Big)^{12}  \\
\Big(-\hat{X}_{pq}^{*}(\vec{x},t)\Big)^{21} & \Big(0\Big)_{pq}^{22} \eea\right)=  
\left(\bea{cc} \Big(0\Big)_{pq}^{11} &
\Big(\hat{X}_{pq}(\vec{x},t)\Big)^{12}  \\
\Big(-\eta_{p}\;\hat{X}_{pq}\pdag(\vec{x},t)\;\eta_{q}\Big)^{21} & \Big(0\Big)_{pq}^{22} \eea\right) \;; \\
   \lb{s3_27}   \hat{X}_{pq}(\vec{x},t) &=&
\left(\bea{cc} -\delta c_{D;++}(\vec{x},t) & \delta c_{D;+-}(\vec{x},t) \\
-\delta c_{D;+-}(\vec{x},t) & \delta c_{D;--}(\vec{x},t) \eea\right)\;\;;\;\;
-\eta_{p}\;\hat{X}_{pq}\pdag(\vec{x},t)\;\eta_{q} =
\left(\bea{cc} \delta c_{D;++}^{*}(\vec{x},t) & -\delta c_{D;+-}^{*}(\vec{x},t) \\
\delta c_{D;+-}^{*}(\vec{x},t) & -\delta c_{D;--}^{*}(\vec{x},t) \eea\right)_{\mbox{.}}
\eeq
It is further possible to diagonalize the various block diagonal density parts 
\(\delta\hat{\Sigma}_{D;pq}^{aa}(\vec{x},t)\;\wt{K}\) as in (\ref{s3_28}) to (\ref{s3_35}) and
the various 'Nambu' generators \(\hat{Y}_{pq}^{ab}(\vec{x},t)\), \(\hat{X}_{pq}(\vec{x},t)\)
as in (\ref{s3_36}) to (\ref{s3_40}). In the case of the density parts we have the diagonal
eigenvalue elements (\ref{s3_29}) with diagonalizing 'rotation' matrices (\ref{s3_30}) so that
the lower block diagonal '22' part is related by transposition to the upper '11' part
\beq \lb{s3_28}
\delta\hat{\Sigma}_{D;pq}^{aa}(\vec{x},t)\;\;\wt{K} &=&
\hat{Q}_{pp\ppr}^{-1,aa}(\vec{x},t)\;\;\delta\hat{\Lambda}_{p\ppr}^{a}(\vec{x},t)\;\;
\hat{Q}_{p\ppr q}^{aa}(\vec{x},t)  \\ \no &=& \left(\bea{cc}
\left(\bea{cc}
\delta B_{D;++}(\vec{x},t) & -\delta B_{D;+-}(\vec{x},t) \\
\delta B_{D;+-}^{*}(\vec{x},t) & -\delta B_{D;--}(\vec{x},t) \eea\right)_{pq}^{11} & \\ &\hspace*{-0.46cm}
\left(\bea{cc}
-\delta B_{D;++}(\vec{x},t) & \delta B_{D;+-}^{*}(\vec{x},t) \\
-\delta B_{D;+-}(\vec{x},t) & \delta B_{D;--}(\vec{x},t) \eea\right)_{pq}^{22} \eea\right) \;; \\   \lb{s3_29}
\delta\hat{\Lambda}_{p}^{a}(\vec{x},t)&=&
\mbox{diag}\Big\{\underbrace{\overbrace{\bigl(+\delta\lambda_{+}(\vec{x},t)\:,
\:-\delta\lambda_{-}(\vec{x},t)\,\bigr)}^{p\cdot\delta\hat{\lambda}_{p}(\vec{x},t)}}_{
a=1}\;;\;\underbrace{\overbrace{\bigl(-\delta\lambda_{+}(\vec{x},t)\:,\:+\delta\lambda_{-}(\vec{x},t)\,\bigr)}^{-p\cdot 
\delta\hat{\lambda}_{p}(\vec{x},t)}}_{a=2}\Big\}=
\delta\hat{\lambda}_{p}(\vec{x},t)\;\wt{K}_{pp}^{aa} \;;  \\  \no
\delta\hat{\lambda}_{p}(\vec{x},t)  &=&  \mbox{diag}\bigl\{\delta\lambda_{+}(\vec{x},t)\:,\:
\delta\lambda_{-}(\vec{x},t)\bigr\}\;;\;\;\; \delta\hat{\lambda}_{p}(\vec{x},t) \in\mathbb{R}\;; \\   \lb{s3_30}
\hat{Q}_{pq}^{11}(\vec{x},t) &=&\Big(\exp\Big\{\im\;\hat{\mscr{B}}_{D}(\vec{x},t)\Big\}\Big)_{pq}\;\;;\hspace*{0.5cm}
\hat{Q}_{pq}^{22}(\vec{x},t) = \Big(\exp\Big\{\im\;\hat{\mscr{B}}_{D}^{T}(\vec{x},t)\Big\}\Big)_{pq}  \;; \\  \lb{s3_31}
\hat{\mscr{B}}_{D}(\vec{x},t) &=&\left(\bea{cc} 0 & -\mscr{B}_{D;+-}(\vec{x},t)  \\
\mscr{B}_{D;+-}^{*}(\vec{x},t) & 0 \eea\right)  \;\;;\hspace*{0.5cm}
\hat{Q}_{pq}^{22,T}(\vec{x},t)=\hat{Q}_{pq}^{11}(\vec{x},t)  \;;  \\    \no 
\mscr{B}_{D;+-}(\vec{x},t) &:=& |\mscr{B}_{D}(\vec{x},t)|\;\;\exp\{\im\,\beta_{D}(\vec{x},t)\,\}  \;; \;\;\;
\mscr{B}_{D;+-}(\vec{x},t) \in\mathbb{C}\;;     \\  \lb{s3_32}
\delta\hat{\Sigma}_{D;pq}^{aa}(\vec{x},t)\;\wt{K}^{aa}_{qq} &=& \hat{Q}_{pp\ppr}^{aa;-1}(\vec{x},t)\;\;
\delta\hat{\Lambda}_{p\ppr}^{a}(\vec{x},t)\;\;\hat{Q}_{p\ppr q}^{aa}(\vec{x},t)\;\;;\hspace*{0.5cm}
\hat{Q}_{pq}^{aa}(\vec{x},t)=\left(\bea{cc} \hat{Q}_{pq}^{11}(\vec{x},t) & 0 \\ 0 & 
\hat{Q}_{pq}^{22}(\vec{x},t) \eea\right) \;; \\ \lb{s3_33}
\Big(\delta\hat{\Sigma}_{D}^{11}(\vec{x},t)\;\hat{\eta}\Big) &=&
\hat{Q}^{11,-1}(\vec{x},t)\;\;\big(p\cdot\delta\hat{\lambda}_{p}(\vec{x},t)\big)\;\;\hat{Q}^{11}(\vec{x},t) \;; \\  \lb{s3_34}
-\Big(\delta\hat{\Sigma}_{D}^{22}(\vec{x},t)\;\hat{\eta}\Big) &=&
\hat{Q}^{22,-1}(\vec{x},t)\;\;\big(-p\cdot\delta\hat{\lambda}_{p}(\vec{x},t)\big)\;\;\hat{Q}^{22}(\vec{x},t) \;; \\  \lb{s3_35}
\delta\hat{\Sigma}_{D;pq}^{22,T}(\vec{x},t) &=& \delta\hat{\Sigma}_{D;pq}^{11}(\vec{x},t) \;.
\eeq
A similar diagonalization (\ref{s3_36},\ref{s3_37}) 
is achieved for the off-diagonal block parts with sub-eigenvalue parts
\(\hat{X}_{D;pq}(\vec{x},t)\) (\ref{s3_38}) and 'eigenvector' matrices \(\hat{P}_{2\times 2}^{11}(\vec{x},t)\),
\(\hat{P}_{2\times 2}^{22}(\vec{x},t)\) (\ref{s3_39}) in such a manner that the symmetry relations of
(\ref{s3_25}) to (\ref{s3_27}) are still retained. This involves the complex parameters
\(\ovv{c}_{++}(\vec{x},t)\), \(\ovv{c}_{--}(\vec{x},t)\) as eigenvalues and \(\mscr{C}_{D;+-}(\vec{x},t)\),
\(\mscr{C}_{D;+-}^{*}(\vec{x},t)\) as the angular parameters of the rotation for the
off-diagonal elements \(\delta c_{D;+-}(\vec{x},t)\), \(\delta c_{D;+-}^{*}(\vec{x},t)\) and diagonal elements
\(\delta c_{D;++}(\vec{x},t)\), \(\delta c_{D;++}^{*}(\vec{x},t)\), 
\(\delta c_{D;--}(\vec{x},t)\), \(\delta c_{D;--}^{*}(\vec{x},t)\) within 
\(\hat{X}_{pq}(\vec{x},t)\), \(-\eta_{p}\;\hat{X}_{pq}\pdag(\vec{x},t)\;\eta_{q}\) (\ref{s3_27})
\beq\lb{s3_36}
\hat{Y}_{pq}^{ab}(\vec{x},t) &=&
\hat{P}_{4\times 4}^{-1}(\vec{x},t)\;\;\hat{Y}_{D;4\times 4}(\vec{x},t)\;\;\hat{P}_{4\times 4}(\vec{x},t)\;; \\ \lb{s3_37}
\hat{Y}_{D;4\times 4}(\vec{x},t) &=&\left(\bea{cc} \Big(0\Big)_{pq}^{11} &
\hat{X}_{D;pq}(\vec{x},t)  \\ -\hat{X}_{D;pq}\pdag(\vec{x},t) & \Big(0\Big)_{pq}^{22} \eea\right)
\;\;;\hspace*{0.3cm}\hat{P}_{4\times 4}(\vec{x},t)=\left(\bea{cc}\hat{P}_{2\times 2}^{11}(\vec{x},t) & 0 \\ 0 &
\hat{P}_{2\times 2}^{22}(\vec{x},t)\eea\right)  \;; \\  \lb{s3_38}
\hat{X}_{D;pq}(\vec{x},t) &=& \left(\bea{cc} -\ovv{c}_{++}(\vec{x},t)  &  0 \\
0 & \ovv{c}_{--}(\vec{x},t)   \eea\right)_{pq}  \;\;;\hspace*{0.3cm}
-\hat{X}_{D;pq}\pdag(\vec{x},t)=  \left(\bea{cc} \ovv{c}_{++}^{*}(\vec{x},t)  &  0 \\
0 & -\ovv{c}_{--}^{*}(\vec{x},t)   \eea\right)_{pq}  \;; \\  \no    \ovv{c}_{++}(\vec{x},t) &:=&
|\ovv{c}_{+}(\vec{x},t)|\;\exp\{\im\,\varphi_{+}(\vec{x},t)\}\;;\;\;\;\;   \ovv{c}_{--}(\vec{x},t) :=
|\ovv{c}_{-}(\vec{x},t)|\;\exp\{\im\,\varphi_{-}(\vec{x},t)\}\;;    \\   \lb{s3_39}
\hat{P}_{2\times 2}^{11}(\vec{x},t) &=& \exp\Big\{\im\;\hat{\mscr{C}}_{D;2\times 2}(\vec{x},t)\Big\}\;\;;\hspace*{0.3cm}
\hat{P}_{2\times 2}^{22}(\vec{x},t) = \exp\Big\{\im\;\hat{\mscr{C}}_{D;2\times 2}^{T}(\vec{x},t)\Big\} \;; \\
   \lb{s3_40}   \hat{\mscr{C}}_{D;2\times 2}(\vec{x},t)&=&\hat{\mscr{C}}_{D;pq}(\vec{x},t)=\left(\bea{cc}
0 & -\mscr{C}_{D;+-}(\vec{x},t)  \\ \mscr{C}_{D;+-}^{*}(\vec{x},t) \eea\right)_{pq}  \;\;;\;\;
\hat{P}_{2\times 2}^{22,T}(\vec{x},t)=\hat{P}_{2\times 2}^{11}(\vec{x},t)  \;;     \\    \no 
\mscr{C}_{D;+-}(\vec{x},t) &:=&  |\mscr{C}_{D}(\vec{x},t)|\;\exp\{\im\,\gamma_{D}(\vec{x},t)\}\;\;\;.
\eeq

\subsection{HST transformations with 'hinge'-fields}\lb{s33}

Eventually, we can collect the various parameters \(\delta B_{\vec{x};++}(t)\),
\(\delta B_{\vec{x};--}(t)\), \(\delta B_{\vec{x};+-}(t)\), \(\delta B_{\vec{x};+-}^{*}(t)\)
for the density parts (\ref{s3_43},\ref{s3_44}) and \(\delta c_{\vec{x};pq}(t)\),
\(\delta c_{\vec{x};pq}^{*}(t)\) within the 'Nambu' terms (\ref{s3_45}) in order to determine
the HST (\ref{s3_46})
\beq \hspace*{-1.0cm}\lb{s3_41}
\sigma_{R_{II}}^{(0)}(\vec{x},t)&\in&\mathbb{R}  \hspace*{0.64cm}\mbox{'hinge' functions :}\;
\delta\hat{\Sigma}_{\vec{x};pq}^{11}(t),\;\delta\hat{\Sigma}_{\vec{x};pq}^{22}(t)   \\ \lb{s3_42}
\delta c_{\vec{x};pq}(t)&\in&\mathbb{C};\hspace*{0.64cm}
\delta B_{\vec{x};++}(t),\;\;\delta B_{\vec{x};--}(t)\;\in\;\mathbb{R}\;;\;
\delta B_{\vec{x};+-}(t)\;\in\;\mathbb{C}   \\ \lb{s3_43}
\delta\hat{\Sigma}_{\vec{x};pq}^{11}(t) &=&
\left(\bea{cc}
\delta B_{\vec{x};++}(t) & \delta B_{\vec{x};+-}(t) \\
\delta B_{\vec{x};+-}^{*}(t) & \delta B_{\vec{x};--}(t)
\eea\right)  \\ \lb{s3_44}
\delta\hat{\Sigma}_{\vec{x};pq}^{22}(t) &=&
\left(\bea{cc}
\delta B_{\vec{x};++}(t) & \delta B_{\vec{x};+-}^{*}(t) \\
\delta B_{\vec{x};+-}(t) & \delta B_{\vec{x};--}(t)
\eea\right)\hspace*{0.55cm}
\Big(\delta\hat{\Sigma}_{\vec{x};pq}^{22}(t)\Big)^{T}=
\delta\hat{\Sigma}_{\vec{x};pq}^{11}(t)  \\ \lb{s3_45}
\delta\hat{\Sigma}_{\vec{x};pq}^{12}(t) &=&
\left(\bea{cc}
\delta c_{\vec{x};++}(t) & \delta c_{\vec{x};+-}(t) \\
\delta c_{\vec{x};+-}(t) & \delta c_{\vec{x};--}(t)
\eea\right)\hspace*{0.55cm}
\Big(\delta\hat{\Sigma}_{\vec{x};pq}^{21}(t)\Big)\pdag=
\delta\hat{\Sigma}_{\vec{x};pq}^{12}(t)\;\;\;.
\eeq
The HST of a quartic interaction of fields can be taken in various manners;
we consider the case where one half of the quartic interaction (with correspondingly reduced
pre-factor \(R_{II}^{2}/(4\hbar^{2})\) in the exponent) is transformed by the real self-energy
density variable \(\sigma_{R_{II}}^{(0)}(\vec{x},t)\) as invariant vacuum or ground state in a SSB
and where the other half of the quartic interaction (also with pre-factor \(R_{II}^{2}/(4\hbar^{2})\) in the exponent)
remains within the anomalous or off-diagonal blocks of the Gaussian transformations. This can
be verified as one decomposes the trace inside the exponential of the last line of (\ref{s3_46}).
As we remove the trace and dyadic product inside the exponent of the last line in (\ref{s3_46}),
one notes that terms with the block diagonal self-energy density \(\delta\hat{\Sigma}_{\vec{x};pq}^{11}(t)\),
\(\delta\hat{\Sigma}_{\vec{x};pq}^{22}(t)\) cancel and only the self-energy variable
\(\sigma_{R_{II}}^{(0)}(\vec{x},t)\) and anomalous parts \(\delta\hat{\Sigma}_{\vec{x};pq}^{12}(t)\),
\(\delta\hat{\Sigma}_{\vec{x};pq}^{21}(t)\) couple to the anomalous doubled, bilinear fields
\(\breve{\Psi}_{\vec{x};q}^{\sharp,b}(t)\ldots\breve{\Psi}_{\vec{x};p}^{a}(t)\)
\beq \lb{s3_46}
\lefteqn{\exp\bigg\{-\frac{R_{II}^{2}}{2\hbar^{2}}\int_{\breve{0}}^{+\breve{T}}d t
\sum_{p,q=\pm}\sum_{\vec{x}}\Big(\psi_{\vec{x}}^{*}(t_{p}+\sdelta\!t_{p})\;\eta_{p}\;
\psi_{\vec{x}}(t_{p})\Big)\;\;
\Big(\psi_{\vec{x}}^{*}(t_{q}+\sdelta\!t_{q})\;\eta_{q}\;\psi_{\vec{x}}(t_{q})\Big)\bigg\}= } \\ \no &=&
\int d[\sigma_{R_{II}}^{(0)}(\vec{x},t)]\;\;
\exp\bigg\{-\frac{1}{4}\frac{1}{R_{II}^{2}}\int_{\breve{0}}^{+\breve{T}}d t\sum_{\vec{x}}
\sigma_{R_{II}}^{(0)}(\vec{x},t)\;\;\sigma_{R_{II}}^{(0)}(\vec{x},t)\bigg\}\;\; \times \\ \no &\times&
\int d[\wt{\Sigma}_{\vec{x};pq}^{ab}(t)\;\wt{K}]\;\;
\exp\bigg\{-\frac{1}{8}\frac{1}{R_{II}^{2}}\int_{\breve{0}}^{+\breve{T}} dt
\sum_{\vec{x}}\TRAB\Big[\delta\wt{\Sigma}_{\vec{x};pq}^{ab}(t)\;
\wt{K}\;\delta\wt{\Sigma}_{\vec{x};qp}^{ba}(t)\;\wt{K}\Big]\bigg\}\;\times
\\ \no &\times &\exp\Bigg\{-\frac{\im}{4\hbar}\int_{\breve{0}}^{+\breve{T}} dt\sum_{\vec{x}}
\TRAB\Bigg[\Bigg(\bea{cc}
\breve{R}_{\vec{x};qp}^{11}(t) & \breve{R}_{\vec{x};qp}^{12}(t) \\
\breve{R}_{\vec{x};qp}^{21}(t) & \breve{R}_{\vec{x};qp}^{22}(t)
\eea\Bigg)\underbrace{\Bigg(\bea{cc} \hat{\eta} & 0 \\ 0 & \hat{\eta} \eea\Bigg)}_{\hat{K}} 
\Bigg(\bea{cc}
\hat{\Sigma}_{\vec{x};pq}^{11}(t) & \delta\hat{\Sigma}_{\vec{x};pq}^{12}(t) \\
\delta\hat{\Sigma}_{\vec{x};pq}^{21}(t) & -\hat{\Sigma}_{\vec{x};pq}^{22}(t)
\eea\Bigg)\underbrace{\Bigg(\bea{cc} \hat{\eta} & 0 \\ 0 & \hat{\eta} \eea\Bigg)}_{\hat{K}}
\Bigg]\Bigg\}_{\mbox{.}}
\eeq
In analogy the HST is obtained for the repulsive, quartic contact interaction of Bose fields;
one half of the HST follows from a Gaussian identity with the real self-energy variable
\(\sigma_{V_{0}}^{(0)}(\vec{x},t_{p})\) and the other half is given by a Gaussian identity of
self-energy matrices \(\delta\wt{\sigma}_{\vec{x}}^{ab}(t_{p})\),
\(\delta\sigma_{\vec{x}}^{ab}(t_{p})\) which are shifted by the anomalous doubled density
matrix \(\breve{R}_{\vec{x};pp}^{ab}(t)\). We emphasize the missing couplings
of the different branches '\(\pm\)' of the time contour, due to the hermitian property
of the repulsive contact interaction, so that one has to adapt the corresponding symmetries
of \(\delta\wt{\sigma}_{\vec{x}}^{ab}(t_{p})\), \(\delta\sigma_{\vec{x}}^{ab}(t_{p})\)
to that of a single time branch with \(\delta\sigma_{\vec{x}}^{ab}(t_{+})\) and 
\(\delta\sigma_{\vec{x}}^{ab}(t_{-})\) being unrelated self-energy field variables
\beq \lb{s3_47}
\lefteqn{\exp\bigg\{-\frac{\im}{\hbar}\int_{\breve{C}}d t_{p}\sum_{\vec{x}}V_{0}\;\;
\Big(\psi_{\vec{x}}^{*}(t_{p}+\sdelta t_{p})\Big)^{2}\;\;\Big(\psi_{\vec{x}}(t_{p})\Big)^{2}\bigg\} = } \\ \no &=&
\int d[\sigma_{V_{0}}^{(0)}(\vec{x},t_{p})]\;\;
\exp\bigg\{\frac{\im}{2\hbar}\int_{\breve{C}}d t_{p}\sum_{\vec{x}}
\frac{\sigma_{V_{0}}^{(0)}(\vec{x},t_{p})\;\;\sigma_{V_{0}}^{(0)}(\vec{x},t_{p})}{V_{0}-\im\:\ve_{p}}\bigg\}\;\times  \\ \no &\times&
\int d[\delta\wt{\sigma}_{\vec{x}}^{ab}(t_{p})\;\wt{\kappa}]\;\;
\exp\bigg\{\frac{\im}{4\hbar}\int_{\breve{C}}d t_{p}\sum_{\vec{x}}\trab
\bigg[\frac{\delta\wt{\sigma}_{\vec{x}}^{ab}(t_{p})\;\wt{\kappa}\;
\wt{\sigma}_{\vec{x}}^{ba}(t_{p})\;\wt{\kappa}}{V_{0}-\im\:\ve_{p}}\bigg]\bigg\}\;\times \\ \no &\times&
\exp\Bigg\{-\frac{\im}{2\hbar}\int_{\breve{C}}d t_{p}\sum_{\vec{x}}\trab\Bigg[
\Bigg(\bea{cc} \breve{R}_{\vec{x};pp}^{11}(t) & \breve{R}_{\vec{x};pp}^{12}(t) \\
\breve{R}_{\vec{x};pp}^{21}(t) & \breve{R}_{\vec{x};pp}^{22}(t)
\eea\Bigg)\Bigg(\bea{cc}
\sigma_{\vec{x}}^{11}(t_{p}) & \delta\sigma_{\vec{x}}^{12}(t_{p}) \\
\delta\sigma_{\vec{x}}^{21}(t_{p}) & -\sigma_{\vec{x}}^{22}(t_{p})
\eea\Bigg)\Bigg]\Bigg\}  \;;   \\             \lb{s3_48}
\sigma_{\vec{x}}^{11}(t_{p})&=&\sigma_{V_{0}}^{(0)}(\vec{x},t_{p}) +
\delta\sigma_{\vec{x}}^{11}(t_{p})\;\;;\hspace*{0.64cm}
\delta\sigma_{\vec{x}}^{11}(t_{p})\;,\;\;\delta\sigma_{\vec{x}}^{22}(t_{p})\;\in\;\mathbb{R} \;; \\ \lb{s3_49}
\sigma_{\vec{x}}^{22}(t_{p})&=&-\sigma_{V_{0}}^{(0)}(\vec{x},t_{p}) +
\delta\sigma_{\vec{x}}^{22}(t_{p})\;\;;\hspace*{1.81cm}
\sigma_{V_{0}}^{(0)}(\vec{x},t_{p})\;\in\;\mathbb{R} \;; \\ \lb{s3_50}
\delta\wt{\sigma}_{\vec{x}}^{aa}(t_{p})&=&\delta\sigma_{\vec{x}}^{aa}(t_{p})\;\;;\hspace*{4.33cm}
\delta\wt{\sigma}_{\vec{x}}^{ab}(t_{p})=\im\;\delta\sigma_{\vec{x}}^{ab}(t_{p})\;(a\neq b) \;; \\ \lb{s3_51}
\delta\sigma_{\vec{x}}^{12}(t_{p})&\in&\mathbb{C}\;\;;\hspace*{0.46cm}
\Big(\delta\sigma_{\vec{x}}^{21}(t_{p})\Big)^{*}=\delta\sigma_{\vec{x}}^{12}(t_{p})\;\;;\hspace*{1.0cm}
\delta\sigma_{\vec{x}}^{11}(t_{p})=\delta\sigma_{\vec{x}}^{22}(t_{p})\;\;\;.
\eeq

\subsection{Removal of the 'hinge'-fields from the generating function}\lb{s34}

As we insert the two HST's (\ref{s3_46},\ref{s3_47}) of the quartic, non-hermitian disorder term and
the repulsive contact interaction into (\ref{s2_10},\ref{s2_11},\ref{s2_31}), we achieve the
ensemble averaged path integral \(\ovv{Z_{II}[\hat{\mscr{J}}]}\) (\ref{s3_52}) with only linear
and bilinear anomalous doubled fields \(\breve{\Psi}_{\vec{x}\ppr,q}^{\sharp b}(t\ppr)\),
\(\breve{\Psi}_{\vec{x},p}^{a}(t)\). Aside from the linear coupling to the doubled source fields
\(J_{\psi;\vec{x}}^{a}(t_{p})\), \(J_{\psi;\vec{x}}^{\dag a}(t_{p})\), we abbreviate the bilinear
term of fields \(\breve{\Psi}_{\vec{x}\ppr,q}^{\sharp b}(t\ppr)\;\ldots\;
\breve{\Psi}_{\vec{x},p}^{a}(t)\) by introducing the matrix
\(\hat{\mscr{M}}_{\vec{x}\ppr,\vec{x}}^{ba}(t_{q}\ppr,t_{p})\) (\ref{s3_53}) which consists of the
one-particle part \(\breve{H}_{\vec{x}\ppr,\vec{x}}(t_{q}\ppr,t_{p})\) (\ref{s3_5}-\ref{s3_10}), the
source matrix \(\hat{\mscr{J}}_{\vec{x}\ppr,\vec{x}}^{ba}(t_{q}\ppr,t_{p})\) for generating bilinear
observables of bosonic fields, the condensate seed field \(\hat{J}_{\psi\psi;\vec{x}}^{ba}(t_{p})\)
(\ref{s2_13},\ref{s2_14},\ref{s2_27}) and the various self-energy 
variables \(\sigma_{R_{II}}^{(0)}(\vec{x},t)\), \(\sigma_{V_{0}}^{(0)}(\vec{x},t_{p})\) and
matrices \(\delta\hat{\Sigma}_{\vec{x};qp}^{ba}(t)\), \(\delta\sigma_{\vec{x}}^{ba}(t_{p})\),
defined in previous sections \ref{s32}, \ref{s33}
\beq \lb{s3_52}
\lefteqn{\ovv{Z_{II}[\hat{\mscr{J}}]}=\int d[\delta\wt{\Sigma}_{\vec{x};pq}^{ab}(t)\;\wt{K}]\;\;
\exp\bigg\{-\frac{1}{8}\frac{1}{R_{II}^{2}}\int_{\breve{0}}^{+\breve{T}}dt\sum_{\vec{x}}
\TRAB\Big[\delta\wt{\Sigma}_{\vec{x};pq}^{ab}(t)\;\wt{K}\;
\delta\wt{\Sigma}_{\vec{x};qp}^{ba}(t)\;\wt{K}\Big]\bigg\} } \\ \no &\times&
\int d[\delta\wt{\sigma}_{\vec{x}}^{ab}(t_{p})\;\wt{\kappa}]\;\;
\exp\bigg\{\frac{\im}{4\hbar}\int_{\breve{C}}dt_{p}\sum_{\vec{x}}
\trab\bigg[\frac{\delta\wt{\sigma}_{\vec{x}}^{ab}(t_{p})\;\wt{\kappa}\;
\delta\wt{\sigma}_{\vec{x}}^{ba}(t_{p})\;\wt{\kappa}}{V_{0}-\im\:\ve_{p}}\bigg]
\bigg\}  \\ \no &\times&
\int d[\sigma_{V_{0}}^{(0)}(\vec{x},t_{p})]\;\;
\exp\bigg\{\frac{\im}{2\hbar}\int_{\breve{C}}dt_{p}\sum_{\vec{x}}
\frac{\sigma_{V_{0}}^{(0)}(\vec{x},t_{p})\;\;\sigma_{V_{0}}^{(0)}(\vec{x},t_{p})}{V_{0}-\im\:\ve_{p}}
\bigg\}   \\ \no &\times&
\int d[\sigma_{R_{II}}^{(0)}(\vec{x},t)]\;\;
\exp\bigg\{-\frac{1}{4}\frac{1}{R_{II}^{2}}\int_{\breve{0}}^{+\breve{T}}dt\sum_{\vec{x}}
\sigma_{R_{II}}^{(0)}(\vec{x},t)\;\;\sigma_{R_{II}}^{(0)}(\vec{x},t)\bigg\}  \\ \no &\times &
\int d[\psi_{\vec{x}}(t_{p})]\;\;
\exp\bigg\{-\frac{\im}{2\hbar}\int_{\breve{0}}^{+\breve{T}}dt\;dt\ppr\sum_{\vec{x},\vec{x}\ppr}
\mcal{N}_{x}\sum_{p,q=\pm} \breve{\Psi}_{\vec{x}\ppr,q}^{\sharp b}(t\ppr)\;
\hat{\mscr{M}}_{\vec{x}\ppr,q;\vec{x},p}^{ba}(t\ppr,t)\;
\breve{\Psi}_{\vec{x},p}^{a}(t)  \bigg\}   \\ \no &\times&
\exp\bigg\{-\frac{\im}{2\hbar}\int_{\breve{C}}dt_{p}\sum_{\vec{x}}\Big(J_{\psi;\vec{x}}^{\dag a}(t_{p})\;\;
\breve{\Psi}_{\vec{x}}^{a}(t_{p})+\breve{\Psi}_{\vec{x}}^{\sharp a}(t_{p})\;\;J_{\psi;\vec{x}}^{a}(t_{p})\Big)\bigg\}
\;;   \\  \lb{s3_53}
\lefteqn{\hat{\mscr{M}}_{\vec{x}\ppr,\vec{x}}^{ba}(t_{q}\ppr,t_{p}) =
\breve{\mscr{H}}_{\vec{x}\ppr,\vec{x}}^{ba}(t_{q}\ppr,t_{p}) +
\hat{K}\;\;\frac{\hat{\mscr{J}}_{\vec{x}\ppr,\vec{x}}^{ba}(t_{q}\ppr,t_{p})}{\mcal{N}_{x}}\;\;
\hat{K} + }   \\ \no &+&\delta(t-t\ppr)\;\delta_{\vec{x},\vec{x}\ppr}\;\delta_{pq}\;\eta_{p}\;
\bigg(\hat{J}_{\psi\psi;\vec{x}}^{ba}(t_{p})+\delta_{ab}\;\Big(\sigma_{V_{0}}^{(0)}(\vec{x},t_{p})+
\frac{1}{2}\;\sigma_{R_{II}}^{(0)}(\vec{x},t)\Big)\bigg)+   \\ \no &+&
\delta(t-t\ppr)\;\delta_{\vec{x},\vec{x}\ppr}\;\frac{1}{2}\;\hat{K}\;\left(\bea{cc}
\delta\hat{\Sigma}_{\vec{x};qp}^{11}(t)+2\;\delta_{pq}\;\eta_{p}\;\delta\sigma_{\vec{x}}^{11}(t_{p}) &
\delta\hat{\Sigma}_{\vec{x};qp}^{12}(t)+2\;\delta_{pq}\;\eta_{p}\;\delta\sigma_{\vec{x}}^{12}(t_{p})  \\
\delta\hat{\Sigma}_{\vec{x};qp}^{21}(t)+2\;\delta_{pq}\;\eta_{p}\;\delta\sigma_{\vec{x}}^{21}(t_{p}) &
-\Big(\delta\hat{\Sigma}_{\vec{x};qp}^{22}(t)+2\;\delta_{pq}\;\eta_{p}\;\delta\sigma_{\vec{x}}^{22}(t_{p})\Big)
\eea\right)_{qp}^{ba}\;\hat{K}\;\;\;.
\eeq
After integration over the bilinear, anomalous doubled fields in (\ref{s3_52}), one attains the inverse
square root of the determinant of \(\hat{\mscr{M}}_{\vec{x}\ppr,\vec{x}}^{ba}(t_{q}\ppr,t_{p})\), which
is transformed to a '\(-\tfrac{1}{2}\mbox{trace}\ln [\ldots]\)' in an exponential with normalizing term
\(\hbar\Omega_{max}\,\mcal{N}_{x}\) and to a propagator part
\(\hat{\mscr{M}}_{\vec{x}\ppr,q;\vec{x},p}^{-1;ba}(t\ppr,t)\) between bilinear, 'Nambu' doubled source
fields \(J_{\psi;\vec{x}\ppr}^{\dag b}(t_{q}\ppr)\;\ldots\; J_{\psi;\vec{x}}^{a}(t_{p})\)
\beq\lb{s3_54}
\lefteqn{\ovv{Z_{II}[\hat{\mscr{J}}]}=\int d[\delta\wt{\Sigma}_{\vec{x};pq}^{ab}(t)\;\wt{K}]\;\;
\exp\bigg\{-\frac{1}{8}\frac{1}{R_{II}^{2}}\int_{\breve{0}}^{+\breve{T}}dt\sum_{\vec{x}}
\TRAB\Big[\delta\wt{\Sigma}_{\vec{x};pq}^{ab}(t)\;\wt{K}\;
\delta\wt{\Sigma}_{\vec{x};qp}^{ba}(t)\;\wt{K}\Big]\bigg\} } \\ \no &\times&
\int d[\delta\wt{\sigma}_{\vec{x}}^{ab}(t_{p})\;\wt{\kappa}]\;\;
\exp\bigg\{\frac{\im}{4\hbar}\int_{\breve{C}}dt_{p}\sum_{\vec{x}}
\trab\bigg[\frac{\delta\wt{\sigma}_{\vec{x}}^{ab}(t_{p})\;\wt{\kappa}\;
\delta\wt{\sigma}_{\vec{x}}^{ba}(t_{p})\;\wt{\kappa}}{V_{0}-\im\:\ve_{p}}\bigg]
\bigg\}  \\ \no &\times&
\int d[\sigma_{V_{0}}^{(0)}(\vec{x},t_{p})]\;\;
\exp\bigg\{\frac{\im}{2\hbar}\int_{\breve{C}}dt_{p}\sum_{\vec{x}}
\frac{\sigma_{V_{0}}^{(0)}(\vec{x},t_{p})\;\;\sigma_{V_{0}}^{(0)}(\vec{x},t_{p})}{V_{0}-\im\:\ve_{p}}
\bigg\}   \\ \no &\times&
\int d[\sigma_{R_{II}}^{(0)}(\vec{x},t)]\;\;
\exp\bigg\{-\frac{1}{4}\frac{1}{R_{II}^{2}}\int_{\breve{0}}^{+\breve{T}}dt\sum_{\vec{x}}
\sigma_{R_{II}}^{(0)}(\vec{x},t)\;\;\sigma_{R_{II}}^{(0)}(\vec{x},t)\bigg\}  \\ \no &\times &
\exp\bigg\{-\frac{1}{2}\int_{\breve{C}}\frac{dt_{p}}{\hbar}\eta_{p}\sum_{\vec{x}}\hbar\Omega_{max}
\mcal{N}_{x}\;\;
\TRAB\ln\Big[\hat{\mscr{M}}_{\vec{x}\ppr,\vec{x}}^{ba}(t_{q}\ppr,t_{p})\Big]\bigg\}  \\  \no &\times &
\exp\bigg\{\frac{\im}{2}\frac{\Omega_{max}^{2}}{\hbar}\int_{\breve{0}}^{+\breve{T}}
dt\;\;dt\ppr\sum_{\vec{x},\vec{x}\ppr}
\mcal{N}_{x}\sum_{p,q=\pm}J_{\psi;\vec{x}\ppr}^{\dag b}(t_{q}\ppr)\;\;
\hat{\mscr{M}}_{\vec{x}\ppr,q;,\vec{x},p}^{-1;ba}(t\ppr,t)\;\;J_{\psi;\vec{x}}^{a}(t_{p})\bigg\}\;\;\;.
\eeq
In order to simplify the self-energy matrices in a coset decomposition, we shift the matrix
\(\delta\hat{\Sigma}_{\vec{x};pq}^{ab}(t)\) and self-energy variable \(\sigma_{V_{0}}^{(0)}(\vec{x},t_{p})\),
according to (\ref{s3_55},\ref{s3_56}) and (\ref{s3_57}), respectively, and perform the transformations
(\ref{s3_58}) of \(\hat{\mscr{M}}_{\vec{x},\vec{x}\ppr}^{ab}(t_{p},t_{q}\ppr)\) (\ref{s3_53}) with the
'Nambu' metric tensors (\ref{s3_2},\ref{s3_11},\ref{s3_12},\ref{s3_22}) to a modified matrix
\(\wt{\mscr{M}}_{\vec{x},\vec{x}\ppr}^{ab}(t_{p},t_{q}\ppr)\) (\ref{s3_59},\ref{s3_60}). This does not alter
the value of the determinant (\ref{s3_61}) and allows for the factorization (\ref{s3_63}-\ref{s3_66})
into density-related self-energy matrices and coset matrices \(\hat{T}_{pq}^{ab}(\vec{x},t)\)
(\ref{s3_25}-\ref{s3_27}). Similar transformations (\ref{s3_62}) are also accomplished for the
propagator part from \(\hat{\mscr{M}}_{\vec{x},\vec{x}\ppr}^{-1;ab}(t_{p},t_{q}\ppr)\) to
\(\wt{\mscr{M}}_{\vec{x},\vec{x}\ppr}^{-1;ab}(t_{p},t_{q}\ppr)\) with 
\(\delta\wt{\Sigma}_{\vec{x};pq}^{ab}(t)\;\wt{K}\) (\ref{s3_63}) for the coset decomposition
\(\mbox{Sp}(4)/\mbox{U}(2)\,\otimes\mbox{U}(2)\)
\beq\lb{s3_55}
\delta\hat{\Sigma}_{\vec{x};pq}^{ab}(t)&\to&\delta\hat{\Sigma}_{\vec{x};pq}^{ab}(t)-2\;\delta_{pq}\;
\eta_{p}\;\delta\hat{\sigma}_{\vec{x}}^{ab}(t_{p}) \;;   \\ \lb{s3_56}
\delta\hat{\Sigma}_{\vec{x};pq}^{ab}(t)&\to&\delta\hat{\Sigma}_{\vec{x};pq}^{ab}(t)-2\;\delta_{pq}\;
\eta_{p}\;\hat{J}_{\psi\psi;\vec{x}}^{ab}(t_{p}) \;;    \\    \lb{s3_57}
\sigma_{V_{0}}^{(0)}(\vec{x},t_{p})&\to&\sigma_{V_{0}}^{(0)}(\vec{x},t_{p})-\frac{1}{2}\;\;
\sigma_{R_{II}}^{(0)}(\vec{x},t) \;\;\;;   \\       \lb{s3_58}
\hat{\mscr{M}}_{\vec{x},\vec{x}\ppr}^{ab}(t_{p},t_{q}\ppr) &\longrightarrow &
\hat{K}\;\hat{I}^{-1}\;\Big(\hat{I}\;\hat{K}\;
\hat{\mscr{M}}_{\vec{x},\vec{x}\ppr}^{ab}(t_{p},t_{q}\ppr)\;\hat{K}\;\hat{I}\Big)\;\hat{I}^{-1}\;\hat{K}
\\ \no    &\longrightarrow &
\hat{K}\;\hat{I}^{-1}\;\underbrace{\Big(\hat{I}\;\hat{K}\;
\hat{\mscr{M}}_{\vec{x},\vec{x}\ppr}^{ab}(t_{p},t_{q}\ppr)\;
\hat{K}\;\hat{I}\;\wt{K}\Big)}_{\wt{\mscr{M}}_{\vec{x},\vec{x}\ppr}^{ab}(t_{p},t_{q}\ppr)}
\;\wt{K}\;\hat{I}^{-1}\;\hat{K} \;;
\\       \lb{s3_59}   \wt{\mscr{M}}_{\vec{x},\vec{x}\ppr}^{ab}(t_{p},t_{q}\ppr) &=&\Big(\hat{I}\;\hat{K}\;
\hat{\mscr{M}}_{\vec{x},\vec{x}\ppr}^{ab}(t_{p},t_{q}\ppr)\;
\hat{K}\;\hat{I}\;\wt{K}\Big)  \;;   \\   \lb{s3_60}
\wt{\mscr{M}}_{\vec{x},\vec{x}\ppr}^{ab}(t_{p},t_{q}\ppr) &=& 
\Big[\hat{H}_{p}^{a}(t_{p})+\sigma_{V_{0}}^{(0)}(\vec{x},t_{p})\Big] \delta_{ab}\;\delta_{pq}\;
\delta(t_{p}-t_{q}\ppr)\;\delta_{\vec{x},\vec{x}\ppr}+   \\  \no &+&
\hat{I}\;\frac{\hat{\mscr{J}}_{\vec{x},\vec{x}\ppr}^{ab}(t_{p},t_{q}\ppr)}{\mcal{N}_{x}}\;
\hat{I}\;\wt{K}+\frac{1}{2}\;\delta(t-t\ppr)\;\delta_{\vec{x},\vec{x}\ppr}\;
\left(\bea{cc} \delta\hat{\Sigma}_{\vec{x};pq}^{11}(t) & \im\;\delta\hat{\Sigma}_{\vec{x};pq}^{12}(t) \\
\im\;\delta\hat{\Sigma}_{\vec{x};pq}^{21}(t) & \delta\hat{\Sigma}_{\vec{x};pq}^{22}(t) \eea\right)\;\;
\wt{K}   \;; \\  \no 
\breve{\mfrak{H}}_{\vec{x},\vec{x}\ppr}^{ab}(t_{p},t_{q}\ppr)=
\breve{\mscr{H}}_{\vec{x},\vec{x}\ppr}^{ab}(t_{p},t_{q}\ppr)\,\hat{K} &=& 
\hat{H}_{p}^{a}(t_{p})\;\delta_{ab}\;\delta_{pq}\;
\delta(t_{p}-t_{q}\ppr)\;\delta_{\vec{x},\vec{x}\ppr}\;\;\;;\;\;\;
\hat{H}_{p}^{a=2}(t_{p}) = \Big(\hat{H}_{p}^{a=1}(t_{p})\Big)^{T}\;;   \\   \lb{s3_61}
\mbox{DET}\Big\{\hat{\mscr{M}}_{\vec{x},\vec{x}\ppr}^{ab}(t_{p},t_{q}\ppr)\Big\} &=&\mbox{DET}\Big\{
\hat{K}\;\hat{I}^{-1}\;\wt{\mscr{M}}_{\vec{x},\vec{x}\ppr}^{ab}(t_{p},t_{q}\ppr)\;
\wt{K}\;\hat{I}^{-1}\;\hat{K}\Big\} =
\mbox{DET}\Big\{\wt{\mscr{M}}_{\vec{x},\vec{x}\ppr}^{ab}(t_{p},t_{q}\ppr)\Big\} \;; \\  \lb{s3_62}
\hat{\mscr{M}}_{\vec{x},\vec{x}\ppr}^{-1;ab}(t_{p},t_{q}\ppr)  &=&
\hat{K}\;\hat{I}\;\wt{K}\;\wt{\mscr{M}}_{\vec{x},\vec{x}\ppr}^{-1;ab}(t_{p},t_{q}\ppr)\;
\hat{I}\;\hat{K} \;;    \\     \lb{s3_63}
\left(\bea{cc} \delta\hat{\Sigma}_{\vec{x};pq}^{11}(t) & \im\;\delta\hat{\Sigma}_{\vec{x};pq}^{12}(t) \\
\im\;\delta\hat{\Sigma}_{\vec{x};pq}^{21}(t) & \delta\hat{\Sigma}_{\vec{x};pq}^{22}(t) \eea\right)^{ab}\;\;
\wt{K} &=&\hat{T}_{pp\ppr}^{aa\ppr}(\vec{x},t)\;\;
\delta\hat{\Sigma}_{D;p\ppr q\ppr}^{a\ppr=b\ppr}(\vec{x},t)\;\wt{K}\;\;
\hat{T}_{q\ppr q}^{-1;b\ppr b}(\vec{x},t)\;\;\;; \\ \lb{s3_64}
\delta\hat{\Sigma}_{D;pq}^{aa}(\vec{x},t) &=&\left(\bea{cc} \delta\hat{\Sigma}_{D;pq}^{11}(\vec{x},t) &     \\
    & \delta\hat{\Sigma}_{D;pq}^{22}(\vec{x},t)  \eea\right)_{pq}^{aa}  \;\;\;; \\   \lb{s3_65}
\delta\hat{\Sigma}_{D;pq}^{11}(\vec{x},t)\;\;\eta_{q} &=& \left(\bea{cc}
\delta B_{D;++}(\vec{x},t) & -\delta B_{D;+-}(\vec{x},t)  \\
\delta B_{D;+-}^{*}(\vec{x},t) & -\delta B_{D;--}(\vec{x},t) \eea\right)_{pq} \;\;\;;  \\   \lb{s3_66}
-\delta\hat{\Sigma}_{D;pq}^{22}(\vec{x},t)\;\;\eta_{q} &=& \left(\bea{cc}
-\delta B_{D;++}(\vec{x},t) & \delta B_{D;+-}^{*}(\vec{x},t)  \\
-\delta B_{D;+-}(\vec{x},t) & \delta B_{D;--}(\vec{x},t) \eea\right)_{pq}\;\;\;.
\eeq
Since the shift of the matrix \(\delta\hat{\Sigma}_{\vec{x};pq}^{ab}(t)\) (\ref{s3_55}) and self-energy
variable \(\sigma_{V_{0}}^{(0)}(\vec{x},t_{p})\) (\ref{s3_57}) has removed the self-energy matrix
\(\delta\hat{\sigma}_{\vec{x}}^{ab}(t_{p})\) and variable \(\sigma_{R_{II}}^{(0)}(\vec{x},t)\) from
\(\hat{\mscr{M}}_{\vec{x},\vec{x}\ppr}^{ab}(t_{p},t_{q}\ppr)\) or 
\(\wt{\mscr{M}}_{\vec{x},\vec{x}\ppr}^{ab}(t_{p},t_{q}\ppr)\) (\ref{s3_59},\ref{s3_60}) (compare with (\ref{s3_53})),
we can completely eliminate latter self-energies by integration of Gaussian identities which involves the
appearance of a new parameter \(\mu_{p}^{(II)}\) (\ref{s3_69}) with the ratio \(R_{II}^{2}/(\hbar V_{0})\)
in the denominator
\beq \no
\lefteqn{\hspace*{-0.6cm}\int d[\sigma_{R_{II}}^{(0)}(\vec{x},t)]\;\;
\exp\bigg\{-\frac{1}{4}\frac{1}{R_{II}^{2}}\int_{\breve{0}}^{+\breve{T}}dt\sum_{\vec{x}}
\sigma_{R_{II}}^{(0)}(\vec{x},t)\;\;\sigma_{R_{II}}^{(0)}(\vec{x},t)\bigg\} \;
\exp\bigg\{\frac{\im}{2\hbar}\int_{\breve{C}}dt_{p}\sum_{\vec{x}}
\frac{\Big(\sigma_{V_{0}}^{(0)}(\vec{x},t_{p})-\frac{1}{2}\;
\sigma_{R_{II}}^{(0)}(\vec{x},t)\Big)^{2}}{V_{0}-\im\;\ve_{p}}\bigg\} =}
\\ \lb{s3_67} &=&
\exp\bigg\{-\frac{1}{4}\frac{R_{II}^{2}}{(\hbar V_{0})^{2}}\int_{\breve{0}}^{+\breve{T}}dt\sum_{\vec{x}}
\Big(\sigma_{V_{0}}^{(0)}(\vec{x},t_{+})-\sigma_{V_{0}}^{(0)}(\vec{x},t_{-})\Big)^{2}\bigg\} \;
\exp\bigg\{\frac{\im}{2\hbar}\int_{\breve{C}}dt_{p}\sum_{\vec{x}}
\frac{\sigma_{V_{0}}^{(0)}(\vec{x},t_{p})\;\;\sigma_{V_{0}}^{(0)}(\vec{x},t_{p})}{V_{0}-\im\;\ve_{p}} \bigg\}; 
  \\    \lb{s3_68}     \lefteqn{\int d[\delta\wt{\sigma}_{\vec{x}}^{ab}(t_{p})\;\wt{\kappa}]\;\;
\exp\bigg\{\frac{\im}{4\hbar}\int_{\breve{C}}dt_{p}\sum_{\vec{x}}
\trab\bigg[\frac{\delta\wt{\sigma}_{\vec{x}}^{ab}(t_{p})\;\wt{\kappa}\;
\delta\wt{\sigma}_{\vec{x}}^{ba}(t_{p})\;\wt{\kappa}}{V_{0}-\im\;\ve_{p}}\bigg]\bigg\}
\;\;\times }  \\ \no &&
\exp\bigg\{-\frac{1}{8}\frac{1}{R_{II}^{2}}\int_{\breve{0}}^{+\breve{T}}dt\sum_{\vec{x}}
\TRAB\bigg[\Big(\delta\wt{\Sigma}_{\vec{x};pq}^{ab}(t)-2\;\delta_{pq}\;\eta_{p}\;
\Big(\delta\wt{\sigma}_{\vec{x}}^{ab}(t_{p})+\wt{J}_{\psi\psi;\vec{x}}^{ab}(t_{p})\Big)\Big)\;
\wt{K}\;\times  \\ \no &\times& \Big(\delta\wt{\Sigma}_{\vec{x};qp}^{ba}(t)-2\;\delta_{pq}\;\eta_{p}\;
\Big(\delta\wt{\sigma}_{\vec{x}}^{ba}(t_{p})+\wt{J}_{\psi\psi;\vec{x}}^{ba}(t_{p})\Big)\Big)\;
\wt{K}\bigg]\bigg\} =  \\ \no &=& 
\exp\bigg\{-\frac{1}{8}\frac{1}{R_{II}^{2}}\int_{\breve{0}}^{+\breve{T}}dt\sum_{\vec{x}}\sum_{p,q=\pm}
\Big(1-\delta_{pq}\;\mu_{p}^{(II)}\Big)\;
\trab\Big[\delta\wt{\Sigma}_{\vec{x};pq}^{ab}(t)\;\wt{K}\;
\delta\wt{\Sigma}_{\vec{x};qp}^{ba}(t)\;\wt{K}\Big]\bigg\}    \times  \\ \no &\times&
\exp\bigg\{\frac{1}{2\;R_{II}^{2}}\int_{\breve{0}}^{+\breve{T}}dt
\sum_{\vec{x}}\sum_{p=\pm}\eta_{p}\;
\Big(1-\mu_{p}^{(II)}\Big)\;\trab\Big[\delta\wt{\Sigma}_{\vec{x};pp}^{ab}(t)\;\wt{\kappa}\;
\wt{J}_{\psi\psi;\vec{x}}^{ba}(t_{p})\;\wt{\kappa}\Big]\bigg\} \times  \\ \no &\times&
\exp\bigg\{-\frac{1}{2\;R_{II}^{2}}\int_{\breve{0}}^{+\breve{T}}dt\sum_{\vec{x}}\sum_{p=\pm}
\Big(1-\mu_{p}^{(II)}\Big)\;\trab\Big[\wt{J}_{\psi\psi;\vec{x}}^{ab}(t_{p})\;\wt{\kappa}\;
\wt{J}_{\psi\psi;\vec{x}}^{ba}(t_{p})\;\wt{\kappa}\Big]\bigg\}  \;\;\;; \\  \lb{s3_69}  &&
\mu_{p}^{(II)}=\frac{1}{\Big(1-\big(\im/2\big)\;\eta_{p}\;\Big(R_{II}^{2}/(\hbar V_{0})\Big)\Big)}\;\;\;.
\eeq
After insertion of the Gaussian identities (\ref{s3_67},\ref{s3_68}) into the path integral (\ref{s3_54}),
one obtains \(\ovv{Z_{II}[\hat{\mscr{J}}]}\) (\ref{s3_70}) which is determined by the self-energy variable
\(\sigma_{V_{0}}^{(0)}(\vec{x},t_{p})\) and the self-energy matrix term 
\(\delta\wt{\Sigma}_{\vec{x};pq}^{ab}(t)\;\wt{K}\) for the coset decomposition as remaining field degrees
of freedom
\beq \lb{s3_70}
\lefteqn{\ovv{Z_{II}[\hat{\mscr{J}}]} = \exp\bigg\{-\frac{1}{2\;R_{II}^{2}}\int_{\breve{0}}^{+\breve{T}}
dt\sum_{\vec{x}}\sum_{p=\pm}\Big(1-\mu_{p}^{(II)}\Big)\;\trab\Big[\wt{J}_{\psi\psi;\vec{x}}^{ab}(t_{p})\;
\wt{\kappa}\;\wt{J}_{\psi\psi;\vec{x}}^{ba}(t_{p})\;\wt{\kappa}\Big]\bigg\} \times } \\ \no &\times&
\int d[\sigma_{V_{0}}^{(0)}(\vec{x},t_{p})]\;\;
\exp\bigg\{-\frac{1}{4}\frac{R_{II}^{2}}{(\hbar V_{0})^{2}}\int_{\breve{0}}^{+\breve{T}}dt\sum_{\vec{x}}
\Big(\sigma_{V_{0}}^{(0)}(\vec{x},t_{+})-\sigma_{V_{0}}^{(0)}(\vec{x},t_{-})\Big)^{2}\bigg\} \times \\ \no
&\times& \exp\bigg\{\frac{\im}{2\hbar}\int_{\breve{C}}dt_{p}\sum_{\vec{x}}
\frac{\sigma_{V_{0}}^{(0)}(\vec{x},t_{p})\;\;\sigma_{V_{0}}^{(0)}(\vec{x},t_{p})}
{V_{0}-\im\;\ve_{p}}\bigg\} \times
\int d[\delta\wt{\Sigma}_{\vec{x};pq}^{ab}(t)\;\wt{K}] \;\;\times  \\ \no &\times&
\exp\bigg\{-\frac{1}{8}\frac{1}{R_{II}^{2}}\int_{\breve{0}}^{+\breve{T}}dt
\sum_{\vec{x}}\sum_{p,q=\pm}
\Big(1-\delta_{pq}\;\mu_{p}^{(II)}\Big)\;
\trab\Big[\delta\wt{\Sigma}_{\vec{x};pq}^{ab}(t)\;\wt{K}\;
\delta\wt{\Sigma}_{\vec{x};qp}^{ba}(t)\;\wt{K}\Big]\bigg\} \times \\ \no &\times&
\exp\bigg\{\frac{1}{2\;R_{II}^{2}}\int_{\breve{0}}^{+\breve{T}}dt
\sum_{\vec{x}}\sum_{p=\pm}\eta_{p}\;
\Big(1-\mu_{p}^{(II)}\Big)\;\trab\Big[\delta\wt{\Sigma}_{\vec{x};pp}^{ab}(t)\;\wt{\kappa}\;
\wt{J}_{\psi\psi;\vec{x}}^{ba}(t_{p})\;\wt{\kappa}\Big]\bigg\} \times \\ \no &\times&
\exp\bigg\{-\frac{1}{2}\int_{\breve{C}}
\frac{dt_{p}}{\hbar}\eta_{p}\sum_{\vec{x}}\hbar\Omega_{max}\;\mcal{N}_{x}\;
\trab\ln\Big[\wt{\mscr{M}}_{\vec{x}\ppr,\vec{x}}^{ba}(t_{q}\ppr,t_{p})\Big]\bigg\} \times \\ \no &\times&
\exp\bigg\{\frac{\im}{2}\frac{\Omega_{max}^{2}}{\hbar}\int_{\breve{0}}^{+\breve{T}}
dt\;dt\ppr\sum_{p,q=\pm}\sum_{\vec{x},\vec{x}\ppr}\mcal{N}_{x}\sum_{a,b=1,2}
J_{\psi;\vec{x}\ppr}^{+b}(t_{q}\ppr)\;\hat{I}\;\wt{K}\;
\wt{\mscr{M}}_{\vec{x}\ppr,\vec{x}}^{-1;ba}(t_{q}\ppr,t_{p})\;\hat{I}\;J_{\psi;\vec{x}}^{a}(t_{p})
\bigg\}\;\;\;.
\eeq
Finally, we factorize the self-energy matrix term \(\delta\wt{\Sigma}_{\vec{x};pq}^{ab}(t)\;\wt{K}\) 
inside \(\wt{\mscr{M}}_{\vec{x},\vec{x}\ppr}^{ab}(t_{p},t_{q}\ppr)\) (\ref{s3_71}) corresponding to
the coset decomposition \(\mbox{Sp}(4)/\mbox{U}(2)\,\otimes\mbox{U}(2)\)
(\ref{s3_22}-\ref{s3_27}) and apply this factorization inside the action
\(\mscr{A}_{DET}[\hat{T},\delta\hat{\Sigma}_{D},\sigma_{V_{0}}^{(0)};\hat{\mscr{J}}]\) (\ref{s3_73}) from
the determinant and within
\(\mscr{A}_{J_{\psi}\pdag,J_{\psi}}[\hat{T},\delta\hat{\Sigma}_{D},\sigma_{V_{0}}^{(0)};\hat{\mscr{J}}]\)
(\ref{s3_74}) from the propagator part of the source fields \(J_{\psi;\vec{x}}^{a}(t_{p})\),
\(J_{\psi;\vec{x}\ppr}^{\dag b}(t_{q}\ppr)\)
\beq\lb{s3_71}
\wt{\mscr{M}}_{\vec{x},\vec{x}\ppr}^{ab}(t_{p},t_{q}\ppr) &=& 
\Big[\hat{H}_{p}^{a}(t_{p})+\sigma_{V_{0}}^{(0)}(\vec{x},t_{p})\Big] \;\delta_{ab}\;\delta_{pq}\;
\delta(t_{p}-t_{q}\ppr)\;\delta_{\vec{x},\vec{x}\ppr}+   
\hat{I}\;\frac{\hat{\mscr{J}}_{\vec{x},\vec{x}\ppr}^{ab}(t_{p},t_{q}\ppr)}{\mcal{N}_{x}}\;
\hat{I}\;\wt{K}+    \\   \no &+&  \frac{1}{2}\;\delta(t-t\ppr)\;\delta_{\vec{x},\vec{x}\ppr}\;\;
\hat{T}(\vec{x},t)\;\hat{Q}_{4\times 4}^{-1}(\vec{x},t)\;\delta\hat{\Lambda}_{4\times 4}(\vec{x},t)\;
\Big(\hat{T}(\vec{x},t)\;\hat{Q}_{4\times 4}^{-1}(\vec{x},t)\Big)^{-1}_{\mbox{;}}    \\  \lb{s3_72}
\hat{T}(\vec{x},t)\;\;\hat{Q}_{4\times 4}^{-1}(\vec{x},t)&=&\hat{T}_{0}(\vec{x},t)\;\;\;;\hspace*{0.7cm}
\hat{T}_{0}^{-1}(\vec{x},t)=\hat{Q}_{4\times 4}(\vec{x},t)\;\;\hat{T}^{-1}(\vec{x},t)\;;   \\  \lb{s3_73}
\mscr{A}_{DET}[\hat{T},\delta\hat{\Sigma}_{D},\sigma_{V_{0}}^{(0)};\hat{\mscr{J}}]&=&\frac{1}{2}
\int_{\breve{C}}\frac{dt_{p}}{\hbar}\eta_{p}\sum_{\vec{x}}\hbar\Omega_{max}\;\mcal{N}_{x}\;
\trab\ln\Big[\wt{\mscr{M}}_{\vec{x}\ppr,\vec{x}}^{ba}(t_{q}\ppr,t_{p})\Big]  \;;   \\  \lb{s3_74}
\mscr{A}_{J_{\psi}\pdag,J_{\psi}}[\hat{T},\delta\hat{\Sigma}_{D},\sigma_{V_{0}}^{(0)};\hat{\mscr{J}}]
&=&\frac{\Omega^{2}}{2\hbar}\int_{\breve{0}}^{+\breve{T}}dt\;dt\ppr
\sum_{\vec{x},\vec{x}\ppr}\mcal{N}_{x}\sum_{p,q=\pm}^{a,b=1,2}
J_{\psi;\vec{x}\ppr}^{+b}(t_{q}\ppr)\;\hat{I}\;\wt{K}\;
\wt{\mscr{M}}_{\vec{x}\ppr,\vec{x}}^{-1;ba}(t_{q}\ppr,t_{p})\;\hat{I}\;J_{\psi;\vec{x}}^{a}(t_{p})\;\;.
\eeq
In subsequent steps we straightforwardly outline how to factorize the matrix
\(\wt{\mscr{M}}_{\vec{x},\vec{x}\ppr}^{ab}(t_{p},t_{q}\ppr)\) (\ref{s3_75},\ref{s3_76}) with the
coset matrices \(\hat{T}_{pq}^{ab}(\vec{x},t)\) (\ref{s3_22}-\ref{s3_27}) and with the various
'Nambu' metric tensors (\ref{s3_2},\ref{s3_11},\ref{s3_12},\ref{s3_22}) into the modified matrices
\(\wt{\mscr{N}}_{\vec{x},\vec{x}\ppr}^{ab}(t_{p},t_{q}\ppr;\delta\hat{\Sigma}_{D}\:\wt{K})\) (\ref{s3_77}),
\(\wt{N}_{\vec{x},\vec{x}\ppr}^{ab}(t_{p},t_{q}\ppr;\hat{I}^{-1}\:\delta\hat{\Sigma}_{D}\:\hat{I}^{-1})\)
(\ref{s3_78}), \(\wt{N}_{\vec{x},\vec{x}\ppr}^{ab}(t_{p},t_{q}\ppr)\) (\ref{s3_79}),
\(\hat{\mscr{O}}_{\vec{x},\vec{x}\ppr}^{ab}(t_{p},t_{q}\ppr)\) (\ref{s3_80}) where the first two kinds
of matrices (\ref{s3_77},\ref{s3_78}) do still contain the block diagonal self-energy density
matrices \(\delta\hat{\Sigma}_{D;pq}^{11}(\vec{x},t)\), \(\delta\hat{\Sigma}_{D;pq}^{22}(\vec{x},t)\)
as 'hinge' fields in a SSB with a coset decomposition \(\mbox{Sp}(4)/\mbox{U}(2)\,\otimes\mbox{U}(2)\)
\beq\lb{s3_75}
\wt{\mscr{M}}_{\vec{x},\vec{x}\ppr}^{ab}(t_{p},t_{q}\ppr) &=&
\Big(\hat{T}(\vec{x},t)\Big)_{pp\ppr}^{aa\ppr}\;\;
\wt{\mscr{N}}_{\vec{x},\vec{x}\ppr}^{a\ppr b\ppr}(t_{p\ppr},t_{q\ppr}\ppr;\delta\hat{\Sigma}_{D}\;
\wt{K})\;\;\Big(\hat{T}^{-1}(\vec{x}\ppr,t\ppr)\Big)_{q\ppr q}^{b\ppr b}  \;;  \\  \lb{s3_76}
\wt{\mscr{M}}_{\vec{x},\vec{x}\ppr}^{ab}(t_{p},t_{q}\ppr) &=&\hat{T}_{pp\ppr}^{aa\ppr}(\vec{x},t)\;\hat{I}\;\hat{K}\;
\underbrace{\Big(\hat{K}\;\hat{I}^{-1}\;
\wt{\mscr{N}}_{\vec{x},\vec{x}\ppr}^{a\ppr b\ppr}(t_{p\ppr},t_{q\ppr}\ppr;\delta\hat{\Sigma}_{D}\;
\wt{K})\;\wt{K}\;\hat{I}^{-1}\;\hat{K}\Big)}_{
\wt{N}_{\vec{x},\vec{x}\ppr}^{a\ppr b\ppr}(t_{p\ppr},t_{q\ppr}\ppr;
\hat{I}^{-1}\;\delta\hat{\Sigma}_{D}\;\hat{I}^{-1})}\;\hat{K}\;\hat{I}\;\wt{K}\;
\hat{T}_{q\ppr q}^{-1;b\ppr b}(\vec{x}\ppr,t\ppr) ;   \\         \lb{s3_77}
\wt{N}_{\vec{x},\vec{x}\ppr}^{ab}(t_{p},t_{q}\ppr;
\hat{I}^{-1}\;\delta\hat{\Sigma}_{D}\;\hat{I}^{-1})  &=&
\Big(\hat{K}\;\hat{I}^{-1}\;
\wt{\mscr{N}}_{\vec{x},\vec{x}\ppr}^{ab}(t_{p},t_{q}\ppr;\delta\hat{\Sigma}_{D}\;
\wt{K})\;\wt{K}\;\hat{I}^{-1}\;\hat{K}\Big)   \\   \lb{s3_78}
\wt{N}_{\vec{x},\vec{x}\ppr}^{ab}(t_{p},t_{q}\ppr;
\hat{I}^{-1}\;\delta\hat{\Sigma}_{D}\;\hat{I}^{-1}) &=&
\wt{N}_{\vec{x},\vec{x}\ppr}^{ab}(t_{p},t_{q}\ppr)+\delta(t-t\ppr)\;\delta_{\vec{x},\vec{x}\ppr}\;\;
\hat{K}\left(\bea{cc} \delta\hat{\Sigma}_{D;pq}^{11}(\vec{x}\ppr,t\ppr) & \\
  & -\delta\hat{\Sigma}_{D;pq}^{22}(\vec{x}\ppr,t\ppr) \eea\right)^{ab}\hat{K}\;\;;  \\    \lb{s3_79}
\wt{N}_{\vec{x},\vec{x}\ppr}^{ab}(t_{p},t_{q}\ppr) &=&\hat{K}\;
\bigg\{\bigg(\Big[\hat{H}_{p}^{a}(t_{p})+\sigma_{V_{0}}^{(0)}(\vec{x},t_{p})\Big]
\delta_{ab}\,\delta_{pq}\delta_{\vec{x},\vec{x}\ppr}\:\delta(t_{p}-t_{q}\ppr) +  \\ \no &+&
\Big[\Big(\hat{T}(\vec{x},t)\;\hat{I}\Big)_{pq\ppr}^{-1;ab\ppr}\;\hat{H}_{q\ppr}^{b\ppr}(t_{q\ppr}\ppr)\;
\Big(\hat{T}(\vec{x}\ppr,t\ppr)\;\hat{I}\Big)_{q\ppr q}^{b\ppr b}-
\hat{H}_{p}^{a}(t_{p})\delta_{ab}\,\delta_{pq}
\delta_{\vec{x},\vec{x}\ppr}\:\delta(t_{p}-t_{q}\ppr) \Big]\bigg)\bigg\}_{\vec{x},\vec{x}\ppr}^{ab}
\hspace*{-0.5cm}(t_{p},t_{q}\ppr) + \\ \no &+&\hat{K}\; \bigg\{
\Big(\hat{T}\;\hat{I}\Big)^{-1}\;\hat{I}\;\frac{\hat{\mscr{J}}}{\mcal{N}_{x}}\;\hat{I}\;\wt{K}\;
\Big(\hat{T}\;\hat{I}\Big)\bigg\}_{\vec{x},\vec{x}\ppr}^{ab}\hspace*{-0.5cm}(t_{p},t_{q}\ppr) = 
\hat{K}\;\;\hat{I}^{-1}\;\;\hat{\mscr{O}}_{\vec{x},\vec{x}\ppr}^{ab}(t_{p},t_{q}\ppr)\;\;\hat{I}\;; \\  \lb{s3_80}
\hat{\mscr{O}}_{\vec{x}\ppr,\vec{x}}^{ba}(t_{q}\ppr,t_{p})  &=&
\Big(\hat{H}_{q}^{b}(t_{q}\ppr)+\sigma_{V_{0}}^{(0)}(\vec{x}\ppr,t_{q}\ppr)\Big)\;
\delta_{ba}\;\delta_{qp}\;\delta_{\vec{x}\ppr,\vec{x}}\;\delta(t_{q}\ppr-t_{p})+
\Big(\hat{T}^{-1}\;\hat{I}\;\frac{\mscr{J}}{\mcal{N}_{x}}\;\hat{I}\;\wt{K}\;\hat{T}
\Big)_{\hspace*{-0.1cm}\vec{x}\ppr,\vec{x}}^{\hspace*{-0.1cm}ba}\hspace*{-0.1cm}(t_{q}\ppr,t_{p}) +  
\\ \no &+&\underbrace{
\Big(\hat{T}_{qp\ppr}^{-1;ba\ppr}(\vec{x}\ppr,t\ppr)\;\hat{H}_{p\ppr}^{a\ppr}(t_{p\ppr})\;
\hat{T}_{p\ppr p}^{a\ppr a}(\vec{x},t)-
\hat{H}_{q}^{b}(t_{q}\ppr)\;\delta_{ba}\;\delta_{qp}\;\delta_{\vec{x}\ppr,\vec{x}}\;
\delta(t_{q}\ppr-t_{p})\;\Big)}_{\delta\breve{\mscr{H}}(\hat{T}^{-1},\hat{T})}\;.
\eeq
The factorization with the coset matrices \(\hat{T}_{pq}^{ab}(\vec{x},t)\) (\ref{s3_22}-\ref{s3_27})
has isolated the 'hinge' fields for the density related terms and has introduced a new gradient term
\(\delta\breve{\mscr{H}}(\hat{T}^{-1},\hat{T})\) for the anomalous doubled parts
\be\lb{s3_81}
\delta\breve{\mscr{H}}_{\vec{x}\ppr,\vec{x}}^{ba}(\hat{T}^{-1},\hat{T};t_{q}\ppr,t_{p}) =
\Big(\hat{T}_{qp\ppr}^{-1;ba\ppr}(\vec{x}\ppr,t\ppr)\;\hat{H}_{p\ppr}^{a\ppr}(t_{p\ppr})\;
\hat{T}_{p\ppr p}^{a\ppr a}(\vec{x},t)-
\hat{H}_{q}^{b}(t_{q}\ppr)\;\delta_{ba}\;\delta_{qp}\;\delta_{\vec{x}\ppr,\vec{x}}\;\delta(t_{q}\ppr-t_{p})\;\Big) \;,
\ee
so that we can use this separation into 'hinge' density-related self-energy terms and 'Nambu' gradient terms for
the actions \(\mscr{A}_{DET}[\hat{T},\delta\hat{\Sigma}_{D},\sigma_{V_{0}}^{(0)};\hat{\mscr{J}}]\) (\ref{s3_73}) and
\(\mscr{A}_{J_{\psi}\pdag,J_{\psi}}[\hat{T},\delta\hat{\Sigma}_{D},\sigma_{V_{0}}^{(0)};\hat{\mscr{J}}]\)
(\ref{s3_74}). Since the determinant and propagator of the matrices (\ref{s3_79},\ref{s3_80}) without 'hinge'
fields are related by
\beq\lb{s3_82}
\mbox{DET}\Big(\wt{N}_{\vec{x},\vec{x}\ppr}^{ab}(t_{p},t_{q}\ppr)\Big) &=&
\mbox{DET}\Big(\hat{\mscr{O}}_{\vec{x},\vec{x}\ppr}^{ab}(t_{p},t_{q}\ppr)\Big) \;, \\   \lb{s3_83}
\wt{N}_{\vec{x},\vec{x}\ppr}^{-1;ab}(t_{p},t_{q}\ppr) &=&
\hat{I}^{-1}\;\hat{\mscr{O}}_{\vec{x},\vec{x}\ppr}^{-1;ab}(t_{p},t_{q}\ppr)\;\hat{I}\;\hat{K} \;,
\eeq
we can considerably simplify the actions (\ref{s3_73},\ref{s3_74}) by following transformations
\beq\lb{s3_84}
\mbox{DET}\Big(\wt{\mscr{M}}_{\vec{x},\vec{x}\ppr}^{ab}(t_{p},t_{q}\ppr)\Big) &=&
\mbox{DET}\Big(\wt{N}_{\vec{x},\vec{x}\ppr}^{ab}(t_{p},t_{q}\ppr;
\hat{I}^{-1}\;\delta\hat{\Sigma}_{D}\;\hat{I}^{-1})\Big) \;;     \\  \lb{s3_85}
\mscr{A}_{DET}[\hat{T},\delta\hat{\Sigma}_{D},\sigma_{V_{0}}^{(0)};\hat{\mscr{J}}] &=& \frac{1}{2}
\int_{\breve{C}}\frac{dt_{p}}{\hbar}\eta_{p}\sum_{\vec{x}}\hbar\Omega_{max}\;\mcal{N}_{x}\;\times \\ \no &\times&
\trab\ln\bigg[\wt{N}_{\vec{x},\vec{x}\ppr}^{ab}(t_{p},t_{q}\ppr)+\delta(t-t\ppr)\delta_{\vec{x},\vec{x}\ppr}
\hat{K}\;\bigg(\hspace*{-0.19cm}\bea{cc} \delta\hat{\Sigma}_{D;pq}^{11}(\vec{x}\ppr,t\ppr) &   \\  
& \hspace*{-0.46cm}-\delta\hat{\Sigma}_{D;pq}^{22}(\vec{x}\ppr,t\ppr)\eea\hspace*{-0.19cm}
\bigg)\;\hat{K}\bigg]_{\mbox{;}}     \\           \lb{s3_86}
\wt{\mscr{M}}_{\vec{x},\vec{x}\ppr}^{-1;ab}(t_{p},t_{q}\ppr) &=&
\hat{T}_{pp\ppr}^{aa\ppr}(\vec{x},t)\;\wt{K}\;\hat{I}^{-1}\;\hat{K}\;
\wt{N}_{\vec{x},\vec{x}\ppr}^{-1;a\ppr b\ppr}(t_{p\ppr},t_{q\ppr}\ppr;
\hat{I}^{-1}\;\delta\hat{\Sigma}_{D}\;\hat{I}^{-1})\;\hat{K}\;\hat{I}^{-1}\;
\hat{T}_{q\ppr q}^{-1;b\ppr b}(\vec{x}\ppr,t\ppr) \;;
\\ \lb{s3_87} \mscr{A}_{J_{\psi}\pdag,J_{\psi}}[\hat{T},\delta\hat{\Sigma}_{D},\sigma_{V_{0}}^{(0)};\hat{\mscr{J}}]
&=& \frac{\Omega_{max}^{2}}{2\hbar}\int_{\breve{0}}^{+\breve{T}}
dt\;dt\ppr\sum_{\vec{x},\vec{x}\ppr}
\mcal{N}_{x}\sum_{p,q=\pm}^{a,b=1,2}  J_{\psi;\vec{x}\ppr}^{+b}(t_{q}\ppr)\;
\Big(\hat{I}\;\wt{K}\;\hat{T}(\vec{x}\ppr,t\ppr)\;\wt{K}\;\hat{I}^{-1}\Big)_{qq\ppr}^{bb\ppr}
\times \\ \no  &\times &  \hat{K}\;
\wt{N}_{\vec{x}\ppr,\vec{x}}^{-1;b\ppr a\ppr}(t_{q\ppr}\ppr,t_{p\ppr};
\hat{I}^{-1}\;\delta\hat{\Sigma}_{D}\;\hat{I}^{-1})\;\hat{K}\;
\Big(\hat{I}^{-1}\;\hat{T}^{-1}(\vec{x},t)\;\hat{I}\Big)_{p\ppr p}^{a\ppr a}\;J_{\psi;\vec{x}}^{a}(t_{p})\;;  \\  \lb{s3_88}
\Big[\Big(\hat{I}^{-1}\;\hat{T}^{-1}(\vec{x},t)\;\hat{I}\Big)\;J_{\psi;\vec{x}}^{a}(t_{p})\Big]\pdag  &=&
J_{\psi;\vec{x}}^{+a}(t_{p})\;\Big(\hat{I}^{-1}\;\hat{T}^{-1}(\vec{x},t)\;\hat{I}\Big)\pdag=
J_{\psi;\vec{x}}^{+a}(t_{p})\;\Big(\hat{I}\;\wt{K}\;\hat{T}(\vec{x},t)\;\wt{K}\;\hat{I}^{-1}\Big) \;;       \\      \lb{s3_89}
\Big(\hat{I}^{-1}\;\hat{T}^{-1}(\vec{x},t)\;\hat{I}\Big)\pdag&=&\hat{I}^{-1}\;
\exp\Big\{\hat{Y}\pdag(\vec{x},t)\Big\}\;\hat{I} = 
\hat{I}^{-1}\;\;\exp\left(\bea{cc} 0 & +\hat{\eta}\;\hat{X}\;\hat{\eta} \\
-\hat{X}\pdag   & 0 \eea\right)\;\;\hat{I}=   \\  \no &=&   \hat{I}^{-1}\;\hat{K}\;\;
\exp\left(\bea{cc}   0  & \hat{X}  \\ -\hat{\eta}\;\hat{X}\pdag\;\hat{\eta}  & 0 \eea\right)\;\;
\hat{K}\;\hat{I} =   \hat{I}\;\;\wt{K}\;\;\hat{T}(\vec{x},t)\;\;\wt{K}\;\;\hat{I}^{-1}\;.
\eeq
After substitution of above separating relations into self-energy 
densities \(\delta\hat{\Sigma}_{D;pq}^{aa}(\vec{x},t)\) and gradient term (\ref{s3_81}) of coset matrices 
into the path integral (\ref{s3_70}), we can 're-introduce' integrals of bosonic, anomalous doubled fields
\(\breve{\Psi}_{\vec{x}\ppr}^{\sharp b}(t_{q}\ppr)\;\ldots\;\breve{\Psi}_{\vec{x}}^{a}(t_{p})\) so that
the determinant and propagator part are changed back to the bilinear, 'Nambu' doubled term of bosonic fields
and to the linear coupling with the condensate seed fields \(J_{\psi;\vec{x}}^{a}(t_{p})\),
\(J_{\psi;\vec{x}}^{\dag a}(t_{p})\). In this manner one performs a projection onto the anomalous doubled field
degrees of freedom
\beq\lb{s3_90}
\lefteqn{\ovv{Z_{II}[\hat{\mscr{J}}]}=
\exp\bigg\{-\frac{1}{2\;R_{II}^{2}}\int_{\breve{0}}^{+\breve{T}}
dt\sum_{\vec{x}}\sum_{p=\pm}
\Big(1-\mu_{p}^{(II)}\Big)\;
\trab\Big[\wt{J}_{\psi\psi;\vec{x}}^{ab}(t_{p})\;\wt{\kappa}\;
\wt{J}_{\psi\psi;\vec{x}}^{ba}(t_{p})\Big]\bigg\} \times }  \\ \no &\times&
\int d[\sigma_{V_{0}}^{(0)}(\vec{x},t_{p})]\;\;
\exp\bigg\{-\frac{1}{4}\frac{R_{II}^{2}}{(\hbar V_{0})^{2}}\int_{\breve{0}}^{+\breve{T}}dt\sum_{\vec{x}}
\Big(\sigma_{V_{0}}^{(0)}(\vec{x},t_{+})-\sigma_{V_{0}}^{(0)}(\vec{x},t_{-})\Big)^{2}\bigg\} \times
\\ \no &\times & \exp\bigg\{\frac{\im}{2\;\hbar}\int_{\breve{C}}dt_{p}\sum_{\vec{x}}
\frac{\sigma_{V_{0}}^{(0)}(\vec{x},t_{p})\;\;\sigma_{V_{0}}^{(0)}(\vec{x},t_{p})}{V_{0}-\im\;\ve_{p}}\bigg\} \times
\int d[\delta\wt{\Sigma}_{\vec{x};pq}^{ab}(t)\;\wt{K}] \;\;\times  \\ \no &\times&
\exp\bigg\{-\frac{1}{8}\frac{1}{R_{II}^{2}}\int_{\breve{0}}^{+\breve{T}}dt\sum_{\vec{x}}\sum_{p,q=\pm}
\Big(1-\delta_{pq}\;\mu_{p}^{(II)}\Big)\;
\trab\Big[\delta\wt{\Sigma}_{\vec{x};pq}^{ab}(t)\;\wt{K}\;
\delta\wt{\Sigma}_{\vec{x};qp}^{ba}(t)\;\wt{K}\Big]\bigg\} \times \\ \no &\times&
\exp\bigg\{\frac{1}{2\;R_{II}^{2}}\int_{\breve{0}}^{+\breve{T}}dt\sum_{\vec{x}}\sum_{p=\pm}\eta_{p}\;
\Big(1-\mu_{p}^{(II)}\Big)\;
\trab\Big[\delta\wt{\Sigma}_{\vec{x};pp}^{ab}(t)\;\wt{\kappa}\;
\wt{J}_{\psi\psi;\vec{x}}^{ba}(t_{p})\;\wt{\kappa}\Big]\bigg\} \times
\int d[\psi_{\vec{x}}(t_{p})]\;\;\times \\ \no &\times&
\exp\bigg\{-\frac{\im}{2\;\hbar}\int_{\breve{C}}dt_{p}\;dt_{q}\ppr
\sum_{\vec{x},\vec{x}\ppr}\mcal{N}_{x}
\Psi_{\vec{x}\ppr}^{+b}(t_{q}\ppr)\;\bigg[\hat{K}\;
\wt{N}_{\vec{x}\ppr,\vec{x}}^{ba}(t_{q}\ppr,t_{p})\;\hat{K}+\bigg(\bea{cc}
\delta\hat{\Sigma}_{D;2\times 2}^{11}  & 0 \\ 0  & -\delta\hat{\Sigma}_{D;2\times 2}^{22}
\eea\bigg)\bigg]_{\vec{x}\ppr,\vec{x}}^{ba}\hspace*{-0.4cm}(t_{q}\ppr,t_{p})
\;\;\;\Psi_{\vec{x}}^{a}(t_{p})\bigg\} \times  \\ \no &\times&
\exp\bigg\{-\frac{\im}{2\hbar}\int_{\breve{C}}dt_{p}\sum_{\vec{x}}
\bigg(\underbrace{J_{\psi;\vec{x}}^{+b}(t_{p\ppr})\;\Big(\hat{I}\;\wt{K}\;
\hat{T}(\vec{x},t)\;\wt{K}\;\hat{I}^{-1}\Big)_{p}^{a}}_{
\wt{J}_{\psi;\vec{x}}^{+a}(t_{p})}\;\Psi_{\vec{x}}^{a}(t_{p}) +
\Psi_{\vec{x}}^{+a}(t_{p})\;\underbrace{^{a}_{p}\Big(\hat{I}^{-1}\;\hat{T}^{-1}(\vec{x},t)\;\hat{I}\Big)\;
J_{\psi;\vec{x}}^{b}(t_{p\ppr})}_{\wt{J}_{\psi;\vec{x}}^{a}(t_{p})} \bigg)\bigg\}_{\mbox{.}}
\eeq
However, the presented factorization of \(\wt{\mscr{M}}_{\vec{x},\vec{x}\ppr}^{ab}(t_{p},t_{q}\ppr)\)
with the various 'Nambu' metric tensors and coset matrices has modified the original 
path integral (\ref{s3_70},\ref{s3_71}) or similarly (\ref{s3_73},\ref{s3_74}) to (\ref{s3_85},\ref{s3_87})
or (\ref{s3_90}) in such a manner that the bilinear coupling of doubled, Bose fields
\(\breve{\Psi}_{\vec{x}\ppr}^{\sharp b}(t_{q}\ppr)\;\ldots\;\breve{\Psi}_{\vec{x}}^{a}(t_{p})\) with
the self-energy densities \(\delta\hat{\Sigma}_{D;pq}^{11}(\vec{x},t)\),
\(\delta\hat{\Sigma}_{D;pq}^{22}(\vec{x},t)\) cancels in (\ref{s3_90}) and restricts the corresponding
exponential (\ref{s3_91}) to the value of unity
\be\lb{s3_91}
\exp\bigg\{-\frac{\im}{2\;\hbar}\int_{\breve{C}}dt_{p}\;dt_{q}\ppr
\sum_{\vec{x},\vec{x}\ppr}\mcal{N}_{x}\;\;
\Psi_{\vec{x}\ppr}^{+b}(t_{q}\ppr)\;\bigg(\bea{cc}
\delta\hat{\Sigma}_{D;2\times 2}^{11}  & 0 \\ 0  & -\delta\hat{\Sigma}_{D;2\times 2}^{22}
\eea\bigg)_{\vec{x}\ppr,\vec{x}}^{ba}\hspace*{-0.4cm}(t_{q}\ppr,t_{p})
\;\;\;\Psi_{\vec{x}}^{a}(t_{p})\bigg\} \;\;\equiv \;\;1\;\;\;.
\ee
The above transformations with 'Nambu' metric tensors and factorization of the total self-energy
\(\delta\wt{\Sigma}_{\vec{x};pq}^{ab}(t)\:\wt{K}\) have resulted into a projection onto the
anomalous doubled field degrees of freedom with the coset matrices \(\hat{T}_{pq}^{ab}(\vec{x},t)\)
and the self-energy variable \(\sigma_{V_{0}}^{(0)}(\vec{x},t_{p})\) as the invariant vacuum or ground
state in a SSB. We thus obtain the path integral (\ref{s3_92}) with the operator
\(\hat{\mscr{O}}_{\vec{x}\ppr,\vec{x}}^{ba}(t_{q}\ppr,t_{p})\) (\ref{s3_93}), being composed of
gradient \(\delta\breve{\mscr{H}}(\hat{T}^{-1},\hat{T})\) (\ref{s3_81}) and density part 
\(\hat{H}_{p}^{a}(t_{p})+\sigma_{V_{0}}^{(0)}(\vec{x},t_{p})\),
where the block diagonal 'hinge' densities \(\delta\hat{\Sigma}_{D;pq}^{aa}(\vec{x},t)\) are only
contained within Gaussian factors and the traces of \(\delta\wt{\Sigma}_{\vec{x};pq}^{ab}(t)\:\wt{K}\) or
\(\delta\wt{\Sigma}_{\vec{x};pp}^{ab}(t)\:\wt{\kappa}\) so that these 'hinge' fields
can be removed by Gaussian like integrations
\beq\lb{s3_92}
\ovv{Z_{II}[\hat{\mscr{J}}]}&=&\exp\bigg\{-\frac{1}{2\;R_{II}^{2}}
\int_{\breve{0}}^{+\breve{T}}d t\sum_{p=\pm}\sum_{\vec{x}}
\Big(1-\mu_{p}^{(II)}\Big)\;
\trab\Big[\wt{J}_{\psi\psi;\vec{x}}^{ab}(t_{p})\;\wt{\kappa}\;
\wt{J}_{\psi\psi;\vec{x}}^{ba}(t_{p})\;\wt{\kappa}\Big]\bigg\}  \\ \no &\times&
\int d[\sigma_{V_{0}}^{(0)}(\vec{x},t_{p})]\;\;
\exp\bigg\{-\frac{1}{4}\frac{R_{II}^{2}}{(\hbar\;V_{0})^{2}}\int_{\breve{0}}^{+\breve{T}}dt\sum_{\vec{x}}
\Big(\sigma_{V_{0}}^{(0)}(\vec{x},t_{+})-\sigma_{V_{0}}^{(0)}(\vec{x},t_{-})\Big)^{2}\bigg\} \\ \no &\times&
\exp\bigg\{\frac{\im}{2\hbar}\int_{\breve{C}}d t_{p}\sum_{\vec{x}}
\frac{\sigma_{V_{0}}^{(0)}(\vec{x},t_{p})\;\;\sigma_{V_{0}}^{(0)}(\vec{x},t_{p})}
{V_{0}-\im\;\ve_{p}}\bigg\} \hspace*{0.46cm}
\int d[\delta\wt{\Sigma}_{\vec{x};pq}^{ab}(t)\;\wt{K}]  \\ \no &\times&
\exp\bigg\{-\frac{1}{8}\frac{1}{R_{II}^{2}}\int_{\breve{0}}^{+\breve{T}}dt\sum_{p,q=\pm}\sum_{\vec{x}}
\Big(1-\delta_{pq}\;\mu_{p}^{(II)}\Big)\;
\trab\Big[\delta\wt{\Sigma}_{\vec{x};pq}^{ab}(t)\;\wt{K}\;
\delta\wt{\Sigma}_{\vec{x};qp}^{ba}(t)\;\wt{K}\Big]\bigg\}  \\ \no &\times&
\exp\bigg\{\frac{1}{2R_{II}^{2}}\int_{\breve{0}}^{+\breve{T}}dt\sum_{p=\pm}\sum_{\vec{x}}
\eta_{p}\;\Big(1-\mu_{p}^{(II)}\Big)
\trab\Big[\delta\wt{\Sigma}_{\vec{x};pp}^{ab}(t)\;\wt{\kappa}\;
\wt{J}_{\psi\psi;\vec{x}}^{ba}(t_{p})\;\wt{\kappa}\Big]\bigg\}  \\ \no &\times&
\exp\bigg\{-\frac{1}{2}\int_{\breve{C}}\frac{dt_{p}}{\hbar}\eta_{p}\sum_{\vec{x}}\hbar\Omega_{max}\mcal{N}_{x}
\trab\ln\Big[\hat{\mscr{O}}_{\vec{x}\ppr,\vec{x}}^{ba}(t_{q}\ppr,t_{p})\Big]\bigg\}  \\ \no &\times&
\exp\bigg\{\frac{\im}{2}\frac{\Omega_{max}^{2}}{\hbar}\int_{\breve{0}}^{+\breve{T}}\hspace*{-0.46cm}dt\;dt\ppr\hspace*{-0.2cm}
\sum_{p\ppr,q\ppr=\pm}^{a\ppr,b\ppr=1,2}\sum_{p,q=\pm}^{a,b=1,2}
\sum_{\vec{x},\vec{x}\ppr}\mcal{N}_{x}\;J_{\psi;\vec{x}\ppr}^{+b\ppr}(t_{q\ppr}\ppr)\;\hat{I}\;\wt{K}\;
\hat{T}_{q\ppr q}^{b\ppr b}(\vec{x}\ppr,t\ppr) \;
\hat{\mscr{O}}_{\vec{x}\ppr,\vec{x}}^{-1;ba}(t_{q}\ppr,t_{p})\;
\hat{T}_{pp\ppr}^{-1;aa\ppr}(\vec{x},t)\;\hat{I}\;J_{\psi;\vec{x}}^{a\ppr}(t_{p\ppr})\bigg\}_{\mbox{;}}  \\  \lb{s3_93}
\lefteqn{\hat{\mscr{O}}_{\vec{x}\ppr,\vec{x}}^{ba}(t_{q}\ppr,t_{p}) =
\Big(\hat{H}_{q}^{b}(t_{q}\ppr)+\sigma_{V_{0}}^{(0)}(\vec{x}\ppr,t_{q}\ppr)\Big)\;\delta_{ba}\;\delta_{qp}\;
\delta_{\vec{x}\ppr,\vec{x}}\;\delta(t_{q}\ppr-t_{p})+
\Big(\hat{T}^{-1}\;\hat{I}\;\frac{\mscr{J}}{\mcal{N}_{x}}\;\hat{I}\;\wt{K}\;\hat{T}
\Big)_{\vec{x}\ppr,\vec{x}}^{ba}(t_{q}\ppr,t_{p}) + }      \\ \no &+&\underbrace{
\Big(\hat{T}_{qp\ppr}^{-1;ba\ppr}(\vec{x}\ppr,t\ppr)\;\hat{H}_{p\ppr}^{a\ppr}(t_{p\ppr})\;
\hat{T}_{p\ppr p}^{a\ppr a}(\vec{x},t)-
\hat{H}_{q}^{b}(t_{q}\ppr)\;\delta_{ba}\;\delta_{qp}\;\delta_{\vec{x}\ppr,\vec{x}}\;
\delta(t_{q}\ppr-t_{p})\;\Big)}_{\delta\breve{\mscr{H}}(\hat{T}^{-1},\hat{T})}\;.
\eeq

\subsection{The change of integration measure for anomalous and density-related parts}\lb{s35}

In order to integrate over the density parts of the Gaussian like factors, we determine the
change of integrate measure from the 'flat' Euclidean, total self-energy \(\delta\wt{\Sigma}_{\vec{x};pq}^{ab}(t)\:\wt{K}\)
to the density-related and anomalous doubled parts; we use the factorization of latter two different kinds
of various blocks '\(a=b\)' and '\(a\neq b\)' and furthermore separate a background averaging functional
\(\langle\ldots\rangle_{\boldsymbol{\hat{\sigma}_{V_{0}}^{(0)}}}\)
from the path integral (\ref{s3_92},\ref{s3_93}) with the coset field degrees of freedom
\beq\lb{s3_94}
\lefteqn{d[\delta\wt{\Sigma}_{\vec{x};pq}^{ab}(t)\,\wt{K}] = \mcal{N}_{\delta\wt{\Sigma}\,\wt{K}}
\bigg(\prod_{\{\vec{x},t\}}d\bigl(\delta\lambda_{+}(\vec{x},t)\bigr)\;
d\bigl(\delta\lambda_{-}(\vec{x},t)\bigr)\;\Bigl(\delta\lambda_{+}(\vec{x},t)\Bigr)^{2}\;
\Bigl(\delta\lambda_{-}(\vec{x},t)\Bigr)^{2}\times } \\   \no &\times&
\Bigl(\bigl(\delta\lambda_{+}(\vec{x},t)\bigr)^{2}-\bigl(\delta\lambda_{-}(\vec{x},t)\bigr)^{2}\Bigr)^{2}\times
d\bigl(|\mscr{B}_{D}(\vec{x},t)|\bigr)\;\sinh\bigl(4|\mscr{B}_{D}(\vec{x},t)|\bigr)\;d\beta_{D}(\vec{x},t)\bigg)\times \\ \no &\times&
\bigg(\prod_{\{\vec{x},t\}}d\bigl(|\ovv{c}_{+}(\vec{x},t)|\bigr)\;\Bigl(\sin\bigl(2|\ovv{c}_{+}(\vec{x},t)|\bigr)\Bigr)^{2}\;d\varphi_{+}(\vec{x},t)\;
\times\;d\bigl(|\ovv{c}_{-}(\vec{x},t)|\bigr)\;\Bigl(\sin\bigl(2|\ovv{c}_{-}(\vec{x},t)|\bigr)\Bigr)^{2}\;d\varphi_{-}(\vec{x},t)\;\times \\ \no &\times&
d\bigl(|\mscr{C}_{D}(\vec{x},t)|\bigr)\;\sinh\bigl(2|\mscr{C}_{D}(\vec{x},t)|\bigr)\;d\gamma_{D}(\vec{x},t) \bigg)=  
\mcal{N}_{\delta\wt{\Sigma}\,\wt{K}}
\bigg(\prod_{\{\vec{x},t\}}d\bigl(\delta\lambda_{+}(\vec{x},t)\bigr)\,
d\bigl(\delta\lambda_{-}(\vec{x},t)\bigr)\;
\mbox{det}^{\!\boldsymbol{2}}\hspace*{-0.06cm}\Bigg[\hspace*{-0.2cm}
\bea{cc}\delta\lambda_{+}(\vec{x},t) \hspace*{-0.1cm}& \delta\lambda_{-}(\vec{x},t)  \\
\delta\lambda_{+}^{3}(\vec{x},t)\hspace*{-0.1cm} & \delta\lambda_{-}^{3}(\vec{x},t) \eea\hspace*{-0.2cm}
\Bigg]\hspace*{-0.16cm}\times  \\ \no &\times&\tfrac{1}{4}\;
d\Bigl(\cosh\bigl(4|\mscr{B}_{D}(\vec{x},t)|\bigr)\Bigr)\;d\beta_{D}(\vec{x},t)\bigg) \times \bigg(\prod_{\{\vec{x},t\}}
\tfrac{1}{2}\;d\Bigl(\cosh\bigl(2|\mscr{C}_{D}(\vec{x},t)|\bigr)\Bigr)\;d\gamma_{D}(\vec{x},t) \;\times \\ \no &\times&
\tfrac{1}{2}\;d\Bigl(|\ovv{c}_{+}(\vec{x},t)|-\tfrac{1}{4}\sin\bigl(4|\ovv{c}_{+}(\vec{x},t)|\bigr)\Bigr)\;d\varphi_{+}(\vec{x},t)\; \times\;
\tfrac{1}{2}\;d\Bigl(|\ovv{c}_{-}(\vec{x},t)|-\tfrac{1}{4}\sin\bigl(4|\ovv{c}_{-}(\vec{x},t)|\bigr)\Bigr)\;d\varphi_{-}(\vec{x},t)\bigg)\;.
\eeq 
\beq\lb{s3_95}
\lefteqn{\ovv{Z_{II}[\hat{\mscr{J}}]}=\bigg\langle\exp\bigg\{-\frac{1}{2\;R_{II}^{2}}
\int_{\breve{0}}^{+\breve{T}}d t\sum_{p=\pm}\sum_{\vec{x}}
\Big(1-\mu_{p}^{(II)}\Big)\;
\trab\Big[\wt{J}_{\psi\psi;\vec{x}}^{ab}(t_{p})\;\wt{\kappa}\;
\wt{J}_{\psi\psi;\vec{x}}^{ba}(t_{p})\;\wt{\kappa}\Big]\bigg\}  
\int d[\delta\wt{\Sigma}_{\vec{x};pq}^{ab}(t)\;\wt{K}]     }   \\ \no &\times&
\exp\bigg\{-\frac{1}{8}\frac{1}{R_{II}^{2}}\int_{\breve{0}}^{+\breve{T}}dt\sum_{p,q=\pm}\sum_{\vec{x}}
\Big(1-\delta_{pq}\;\mu_{p}^{(II)}\Big)\;
\trab\Big[\delta\wt{\Sigma}_{\vec{x};pq}^{ab}(t)\;\wt{K}\;
\delta\wt{\Sigma}_{\vec{x};qp}^{ba}(t)\;\wt{K}\Big]\bigg\}  \\ \no &\times&
\exp\bigg\{\frac{1}{2R_{II}^{2}}\int_{\breve{0}}^{+\breve{T}}dt\sum_{p=\pm}\sum_{\vec{x}}
\eta_{p}\;\Big(1-\mu_{p}^{(II)}\Big)
\trab\Big[\delta\wt{\Sigma}_{\vec{x};pp}^{ab}(t)\;\wt{\kappa}\;
\wt{J}_{\psi\psi;\vec{x}}^{ba}(t_{p})\;\wt{\kappa}\Big]\bigg\}  \\ \no &\times&
\exp\bigg\{-\frac{1}{2}\int_{\breve{C}}\frac{dt_{p}}{\hbar}\eta_{p}\sum_{\vec{x}}\hbar\Omega_{max}\mcal{N}_{x}
\trab\ln\bigg[\hat{1}+\delta\breve{\mscr{H}}(\hat{T}^{-1},\hat{T})\;\bigl(\breve{\mfrak{H}}+\hat{\sigma}_{V_{0}}^{(0)}\bigr)^{-1}+
\bigl(\hat{T}^{-1}\hat{I}\tfrac{1}{\mcal{N}_{x}}\hat{\mscr{J}}\hat{I}\wt{K}\hat{T}\bigr)\;
\bigl(\breve{\mfrak{H}}+\hat{\sigma}_{V_{0}}^{(0)}\bigr)^{-1}\bigg]_{\vec{x}=\vec{x}\ppr;p=q}^{a=b}\hspace*{-1.3cm}(t,t\ppr=t)\bigg\} 
 \\ \no &\times&
\exp\bigg\{\frac{\im}{2}\frac{\Omega_{max}^{2}}{\hbar}\int_{\breve{0}}^{+\breve{T}}\hspace*{-0.46cm}dt\,dt\pppr\hspace*{-0.2cm}
\sum_{p\ppr,q\pppr=\pm}^{a\ppr,b\pppr=1,2}
\sum_{\vec{x},\vec{x}\pppr}\mcal{N}_{x}\;J_{\psi;\vec{x}\pppr}^{+b\pppr}(t_{q\pppr}\pppr)\;\hat{I}\;\wt{K}\;  \\  \no &\times&
\bigg[\int_{\breve{0}}^{+\breve{T}}\hspace*{-0.46cm}dt\ppr\sum_{p,q,q\ppr=\pm}^{a,b,b\ppr=1,2}
\sum_{\vec{x}\ppr}\mcal{N}_{x}\bigg\{\hat{T}_{q\pppr q\ppr}^{b\pppr b\ppr}(\vec{x}\pppr,t\pppr) \;
\bigl(\breve{\mfrak{H}}+\hat{\sigma}_{V_{0}}^{(0)}\bigr)_{\vec{x}\pppr,\vec{x}\ppr}^{-1;b\ppr=b}(t_{q\ppr}\pppr,t_{q}\ppr)\Big[\hat{1}+
\delta\breve{\mscr{H}}(\hat{T}^{-1},\hat{T})\;\bigl(\breve{\mfrak{H}}+\hat{\sigma}_{V_{0}}^{(0)}\bigr)^{-1} +  \\     \no &+&
\bigl(\hat{T}^{-1}\hat{I}\tfrac{1}{\mcal{N}_{x}}\hat{\mscr{J}}\hat{I}\wt{K}\hat{T}\bigr)\;
\bigl(\breve{\mfrak{H}}+\hat{\sigma}_{V_{0}}^{(0)}\bigr)^{-1}\Big]_{\vec{x}\ppr,\vec{x}}^{-1;ba}(t_{q}\ppr,t_{p})\;
\hat{T}_{pp\ppr}^{-1;aa\ppr}(\vec{x},t)\bigg\}_{\vec{x}\pppr,\vec{x}}^{b\pppr a\ppr}\hspace{-0.6cm}(t_{q\pppr}\pppr,t_{p\ppr})
-\bigl(\breve{\mfrak{H}}+\hat{\sigma}_{V_{0}}^{(0)}\bigr)_{\vec{x}\pppr,\vec{x}}^{-1;b\pppr a\ppr}(t_{q\pppr}\pppr,t_{p\ppr})\bigg]
\hat{I}\;J_{\psi;\vec{x}}^{a\ppr}(t_{p\ppr})\bigg\}\bigg\rangle_{\boldsymbol{\hat{\sigma}_{V_{0}}^{(0)}};}
\eeq
\beq\lb{s3_96}
\lefteqn{\bigg\langle\Big(\mbox{coset fields and }\hat{\sigma}_{V_{0}}^{(0)}\Big)
\bigg\rangle_{\boldsymbol{\hat{\sigma}_{V_{0}}^{(0)}}}=
\int d[\sigma_{V_{0}}^{(0)}(\vec{x},t_{p})]\;\;
\exp\bigg\{-\frac{1}{4}\frac{R_{II}^{2}}{(\hbar\;V_{0})^{2}}\int_{\breve{0}}^{+\breve{T}}dt\sum_{\vec{x}}
\Big(\sigma_{V_{0}}^{(0)}(\vec{x},t_{+})-\sigma_{V_{0}}^{(0)}(\vec{x},t_{-})\Big)^{2}\bigg\}   }  \\ \no &\times&
\exp\bigg\{\frac{\im}{2\hbar}\int_{\breve{C}}d t_{p}\sum_{\vec{x}}
\frac{\sigma_{V_{0}}^{(0)}(\vec{x},t_{p})\;\;\sigma_{V_{0}}^{(0)}(\vec{x},t_{p})}
{V_{0}-\im\;\ve_{p}}-\int_{\breve{C}}\frac{dt_{p}}{\hbar}\,\eta_{p}\sum_{\vec{x}}\hbar\Omega_{max}\mcal{N}_{x}
\trpq\ln\Big[\Big(\breve{\mfrak{H}}^{11}+\hat{\sigma}_{V_{0}}^{(0)}\Big)_{\vec{x}=\vec{x}\ppr;p=q}^{11}(t,t\ppr=t)\Big]\bigg\} 
\\   \no &\times& \exp\bigg\{\im\frac{\Omega_{max}^{2}}{\hbar}\int_{\breve{0}}^{+\breve{T}}dt\;dt\ppr\sum_{p,q=\pm}\sum_{\vec{x},\vec{x}\ppr}
\mcal{N}_{x}\;j_{\psi;\vec{x}\ppr}\pdag(t_{q}\ppr)\;\Big[\bigl(\breve{\mfrak{H}}^{11}+\hat{\sigma}_{V_{0}}^{(0)}\bigr)
\hat{\eta}\Big]_{\vec{x}\ppr,\vec{x}}^{-1}(t_{q}\ppr,t_{p})\;j_{\psi;\vec{x}}(t_{p})\bigg\}  \times
\Big(\mbox{coset fields and }\hat{\sigma}_{V_{0}}^{(0)}\Big)_{\mbox{.}}
\eeq
We hint again to following abbreviations, already specified in (\ref{s3_60},\ref{s3_81})
\beq\lb{s3_97}
\Big(\breve{\mfrak{H}}+\hat{\sigma}_{V_{0}}^{(0)}\Big)_{\vec{x},\vec{x}\ppr}^{ab}(t_{p},t_{q}\ppr) &=&
\Big(\hat{H}_{p}^{a}(t_{p})+\sigma_{V_{0}}^{(0)}(\vec{x},t_{p})\Big)\delta_{ab}\;\delta_{pq}\;\delta_{\vec{x},\vec{x}\ppr}\;
\delta(t_{p}-t_{q}\ppr)\;;  \\ \lb{s3_98}
\delta\breve{\mscr{H}}_{\vec{x},\vec{x}\ppr}^{ab}(\hat{T}^{-1},\hat{T};t_{p},t_{q}\ppr) &=&
\hat{T}_{pp\ppr}^{-1;aa\ppr}(\vec{x},t)\;\breve{\mfrak{H}}_{\vec{x},\vec{x}\ppr}^{a\ppr=b\ppr}(t_{p\ppr},t_{q\ppr}\ppr)\;
\hat{T}_{q\ppr q}^{b\ppr b}(\vec{x}\ppr,t\ppr)-\breve{\mfrak{H}}_{\vec{x},\vec{x}\ppr}^{a=b}(t_{p},t_{q}\ppr)\;.
\eeq
Further simplification of above path integrals (\ref{s3_94},\ref{s3_95}) with coset field degrees of freedom arises from taking a separate
saddle point approximation of the background averaging functional (\ref{s3_96}) with respect to a variation with \(\delta\sigma_{V_{0}}^{(0)}(\vec{x},t_{p})\).
This results in definite, complex-valued functions \(\langle\sigma_{V_{0}}^{(0)}(\vec{x},t_{p})\rangle\) of space and time where the imaginary part
of  \(\langle\sigma_{V_{0}}^{(0)}(\vec{x},t_{p})\rangle\) has to comply with the original, already introduced, convergence generating '\(-\im\,\hat{\ve}_{p}\)'
terms for proper convergence and analytic properties of Green functions.

\section{Summary and conclusion}\lb{s4}

\subsection{Transformation to the case 'I' with static disorder reduced to stationary states} \lb{s41}

At the end of section \ref{s2} we have already specified the path integral \(\ovv{Z_{I}[\hat{\mscr{J}}]}\) (\ref{s2_31}) with the approximation
to stationary states in frequency space which can also be attained from simplifying the two frequency dependence of the corresponding total
disorder self-energy \(\delta\wt{\Sigma}_{\vec{x};pq}^{ab}(\omega,\omega\ppr)\rightarrow\delta_{\omega,\omega\ppr}\:
\delta\wt{\Sigma}_{\vec{x};pq}^{ab}(\omega)\) at very later steps of transformations to a nonlinear sigma model 
\beq \lb{s4_1}
\ovv{Z_{I}[\hat{\mscr{J}}]}&\simeq&\boldsymbol{\bigg\langle}
F[\psi^{*},\psi;J_{\psi},\hat{J}_{\psi\psi};\hat{\mscr{J}}] \;\times \\ \no &\times&\hspace*{-0.28cm}
\exp\bigg\{\hspace*{-0.19cm}
-\frac{R_{I}^{2}}{2\hbar^{2}}\frac{\mscr{N}_{t}^{2}}{\mcal{N}_{x}}
\int_{\breve{0}}^{+\breve{\Omega}}
\frac{d\omega}{(\frac{2\pi}{T_{max}})}\sum_{\vec{x}}\sum_{p,q=\pm}\hspace*{-0.19cm}
\bigg(\psi_{\vec{x},p}^{*}(\omega)\;
e^{\im\:\omega\:\Delta t_{p}}\;\eta_{p}\;\psi_{\vec{x},p}(\omega)\bigg)
\bigg(\psi_{\vec{x},q}^{*}(\omega)\;e^{\im\:\omega\:\Delta t_{q}}\;\eta_{q}\;
\psi_{\vec{x},q}(\omega)\bigg)\bigg\}
\boldsymbol{\hspace*{-0.1cm}\bigg\rangle_{\hspace*{-0.2cm}F[\psi^{*},\psi].}} \;;  \\   \lb{s4_2}
\Omega_{max} &=&\frac{1}{\sdelta t}\;\;\;;\;\;\;0<t_{p}<+T_{0}\;\;\;;\;\;\;T_{0}=T_{max} \;; \\ \no
0<&\omega_{p}&<+\Omega_{0}\;\;\;;\;\;\;\Omega_{0}=\Omega_{max}=\frac{1}{\sdelta t}\;\;\;;\; \;\;
\sdelta\omega=\frac{2\pi}{T_{max}}\;\;\;;\;\;\mscr{N}_{t}=T_{max}/\sdelta t\;\;\;;  \\  \lb{s4_3} 
\int_{\breve{C}_{\omega}}\frac{d\omega_{p}}{\sdelta\omega}\ldots\hspace*{-0.28cm} &=&
\int_{-\sdelta\omega}^{+\Omega_{0}}\frac{d\omega_{+}}{(\frac{2\pi}{T_{max}})}\;\ldots+
\int_{+(\Omega_{0}+\sdelta\omega)}^{0}\frac{d\omega_{-}}{(\frac{2\pi}{T_{max}})}\;\ldots  =  \\  \no &=&
\int_{-\sdelta\omega}^{+\Omega_{0}}\frac{d\omega_{+}}{(\frac{2\pi}{T_{max}})}\;\ldots-
\int_{0}^{+(\Omega_{0}+\sdelta\omega)}\frac{d\omega_{-}}{(\frac{2\pi}{T_{max}})}\;\ldots  =
\sum_{p=\pm}\int_{\breve{0}_{p}}^{+\breve{\Omega}_{p}}\frac{d\omega_{p}}{(\frac{2\pi}{T_{max}})}\;\eta_{p}\;\ldots ; 
\\  \no && \breve{0}_{+}=-\sdelta\omega\;;\;\;\breve{\Omega}_{+}=\Omega_{0}\;;\;\;\breve{0}_{-}=0\;;\;\;
\breve{\Omega}_{-}=\Omega_{0}+\sdelta\omega\;\;\;;    \\       \lb{s4_4}
R_{II}^{2}&\simeq&R_{I}^{2}\;\frac{\mscr{N}_{t}^{2}\;\Omega_{max}}{\mcal{N}_{x}} \;\;\;.
\eeq
Corresponding to (\ref{s4_1}-\ref{s4_4}), we can exchange the parameter \(\mu_{p}^{(II)}\) (\ref{s3_69}) by \(\mu_{p}^{(I)}\) in (\ref{s3_95})
\be\lb{s4_5}
\mu_{p}^{(II)}=\frac{1}{1-\frac{\im}{2}\eta_{p}\frac{R_{II}^{2}}{\hbar V_{0}}}\rightarrow\mu_{p}^{(I)}=
\frac{1}{1-\frac{\im}{2}\eta_{p}\frac{R_{I}^{2}\;\Omega_{max}}{\hbar V_{0}}\frac{\mscr{N}_{t}^{2}}{\mcal{N}_{x}}}\;\;\;,
\ee
and can perform the Fourier transformation to frequency space for the anomalous doubled one-particle part according to following relations
\beq \lb{s4_6}
\int_{\breve{0}}^{+\breve{T}}\!\!\!dt\;\ldots &=&T_{max}\int_{\breve{0}}^{+\breve{T}}\!\!\!\frac{dt}{T_{max}}\;\ldots \rightarrow
T_{max}\int_{\breve{0}}^{+\breve{\Omega}}\!\!\!\frac{d\omega}{(\frac{2\pi}{T_{max}})}\;\ldots\;; \\   \lb{s4_7}
\lefteqn{\hspace*{-1.9cm}\int_{\breve{C}_{\omega}}d\omega_{p}\sum_{\vec{x}}\psi_{\vec{x},p}^{*}(\omega)\;
e^{\im\:\omega\:\sdelta t_{p}}\;\;\hat{H}_{p}(\vec{x},\omega_{p})\;\;
\psi_{\vec{x},p}(\omega)=\int_{\breve{C}_{\omega}}d\omega_{p}\sum_{\vec{x}}
\psi_{\vec{x},p}(\omega)\;\;\hat{H}_{p}^{T}(\vec{x},\omega_{p})\;\;\psi_{\vec{x},p}^{*}(\omega)\;e^{\im\:\omega\:\Delta t_{p}} =}    \\  \no  &=&
\int_{\breve{C}_{\omega}}d\omega_{p}\;d\omega_{q}\ppr\sum_{\vec{x},\vec{x}\ppr}\frac{1}{2}\;
\breve{\Psi}_{\vec{x}\ppr}^{\sharp b}(\omega_{q}\ppr)\;\;
\breve{\mscr{H}}_{\vec{x}\ppr;\vec{x}}^{ba}(\omega_{q}\ppr,\omega_{p})\;\;
\breve{\Psi}_{\vec{x}}^{a}(\omega_{p})  \;\;\;; \\  \lb{s4_8}  
\breve{\mscr{H}}_{\vec{x}\ppr;\vec{x}}^{ba}(\omega_{q}\ppr,\omega_{p})&=&
\mbox{diag}\left(\breve{\mscr{H}}_{\vec{x}\ppr;\vec{x}}^{11}(\omega_{q}\ppr,\omega_{p})\;;\;
\breve{\mscr{H}}_{\vec{x}\ppr;\vec{x}}^{22}(\omega_{q}\ppr,\omega_{p})\right)  \;; \\  \lb{s4_9}
\breve{\mscr{H}}_{\vec{x}\ppr;\vec{x}}^{11}(\omega_{q}\ppr,\omega_{p}) &=&
\bigg(-\im\,\hbar\;\tfrac{\exp\{-\im\,\omega_{q}\ppr\;\sdelta t\}-1}{\sdelta t}+
\big(\hat{h}(\vec{x}\ppr)-\im\;\hat{\ve}_{p}\big)\bigg)\delta_{qp}\;\eta_{q}\;\delta(\omega_{q}\ppr-\omega_{p})\;
\delta_{\vec{x}\ppr,\vec{x}} \;; \\ \lb{s4_10}
\breve{\mscr{H}}_{\vec{x}\ppr;\vec{x}}^{22}(\omega_{q}\ppr,\omega_{p}) &=&
\Big(\breve{\mscr{H}}_{\vec{x}\ppr;\vec{x}}^{11}(\omega_{q}\ppr,\omega_{p}) \Big)^{T}=
\bigg(-\im\,\hbar\;\tfrac{\exp\{-\im\,\omega_{q}\ppr\;\sdelta t\}-1}{\sdelta t}+
\big(\hat{h}^{T}(\vec{x}\ppr)-\im\;\hat{\ve}_{p}\big)\bigg)\delta_{qp}\;\eta_{q}\;\delta(\omega_{q}\ppr-\omega_{p})\;
\delta_{\vec{x}\ppr,\vec{x}}\;\;   \;;  \\ \lb{s4_11}
\breve{\mscr{H}}_{\vec{x}\ppr;\vec{x}}^{ba}(\omega_{q}\ppr,\omega_{p}) &\simeq&
\hat{\mscr{H}}_{\vec{x}\ppr;\vec{x}}^{ba}(\omega_{q}\ppr,\omega_{p}) =   \\  \no &=&
\mbox{diag}\Big(\underbrace{\hat{H}_{+}(\vec{x}\ppr,\omega_{+}\ppr)\:,\:
-\hat{H}_{-}(\vec{x}\ppr,\omega_{-}\ppr)}_{a=1}\;;\;\underbrace{\hat{H}_{+}^{T}(\vec{x}\ppr,\omega_{+}\ppr)\:,\:
-\hat{H}_{-}^{T}(\vec{x}\ppr,\omega_{-}\ppr)}_{a=2}\Big)\delta_{qp}\;
\delta_{\vec{x}\ppr,\vec{x}}\;\delta(\omega_{q}\ppr-\omega_{p}) ; \\  \lb{s4_12}
\hat{H}_{p}(\vec{x},\omega_{p})&=&-\hbar\;\omega_{p}-\im\;\ve_{p}+
\frac{\vec{p}^{\;2}}{2m}+u(\vec{x})-\mu_{0} \;; \hspace*{0.6cm}
\hat{H}_{p}^{T}(\vec{x},\omega_{p})=-\hbar\;\omega_{p}-\im\;\ve_{p}+
\frac{\vec{p}^{\;2}}{2m}+u(\vec{x})-\mu_{0}\;,
\eeq
where we additionally define doubled, shifted fields \(\breve{\Psi}_{\vec{x}}^{a}(\omega_{p})\),  \(\breve{\Psi}_{\vec{x}\ppr}^{\sharp b}(\omega_{q}\ppr)\)
with analogous boundary conditions as in relations (\ref{s2_22}-\ref{s2_25}) for the case of dynamic disorder (cf. also relations (\ref{s2_1}-\ref{s2_8}))
\be\lb{s4_13}
\breve{\Psi}_{\vec{x}}^{a}(\omega_{p})  =\left(\bea{l} (a=1)\;\psi_{\vec{x},p}(\omega) \\  (a=2)\;\psi_{\vec{x},p}^{*}(\omega)\;
e^{\im\:\omega\:\Delta t_{p}}\eea\right)^{a}\;\;;\;\;\;\breve{\Psi}_{\vec{x}\ppr}^{\sharp b}(\omega_{q}\ppr)=\Big(\underbrace{
\psi_{\vec{x}\ppr,q}^{*}(\omega\ppr)\;e^{\im\:\omega\ppr\:\Delta t_{q}\ppr}}_{b=1}\;\boldsymbol{;}\;
\underbrace{\psi_{\vec{x}\ppr,q}(\omega\ppr)}_{b=2} \Big)^{b}\;.
\ee
Hence, under restriction to stationary states 
the total disorder self-energy \(\delta\wt{\Sigma}_{\vec{x};pq}(t)\) simply changes to \(\delta\wt{\Sigma}_{\vec{x};pq}(\omega)\) with
analogous changes within the the coset decomposition \(\mbox{Sp}(4)/\mbox{U}(2)\,\otimes\,\mbox{U}(2)\) of the coset matrices 
\(\hat{T}_{pq}^{ab}(\vec{x},\omega)\) and the involved generators \(\hat{Y}_{pq}^{ab}(\omega)\), \(\hat{X}_{pq}(\omega)\) so that the
factorization into eigenvalues \(\ovv{c}_{+}(\vec{x},\omega)\), \(\ovv{c}_{-}(\vec{x},\omega)\) and angular parameter \(\mscr{C}_{D}(\vec{x},\omega)\)
similarly keeps its form as in (\ref{s3_25}-\ref{s3_27}) and (\ref{s3_36}-\ref{s3_40}) 
\beq \lb{s4_14}
\delta\wt{\Sigma}_{\vec{x};pq}^{ab}(t)\;\wt{K}&\rightarrow&\delta\wt{\Sigma}_{\vec{x};pq}^{ab}(\omega)\;\wt{K}\;;  \\   \lb{s4_15}
\hat{T}(\vec{x},\omega) &=&\exp\Big\{-\hat{Y}_{pq}^{ab}(\vec{x},\omega)\Big\}\;\;;\hspace*{0.5cm}
\hat{T}^{-1}(\vec{x},\omega) = \exp\Big\{\hat{Y}_{pq}^{ab}(\vec{x},\omega)\Big\}  \;;  \\    \lb{s4_16}
\hat{Y}_{pq}^{ab}(\vec{x},\omega)&=&\left(\bea{cc} \Big(0\Big)_{pq}^{11} &
\Big(\hat{X}_{pq}(\vec{x},\omega)\Big)^{12}  \\
\Big(-\hat{X}_{pq}^{*}(\vec{x},\omega)\Big)^{21} & \Big(0\Big)_{pq}^{22} \eea\right)=  
\left(\bea{cc} \Big(0\Big)_{pq}^{11} &
\Big(\hat{X}_{pq}(\vec{x},\omega)\Big)^{12}  \\
\Big(-\eta_{p}\;\hat{X}_{pq}\pdag(\vec{x},\omega)\;\eta_{q}\Big)^{21} & \Big(0\Big)_{pq}^{22} \eea\right) \;; \\
   \lb{s4_17}   \hat{X}_{pq}(\vec{x},\omega) &=&
\left(\bea{cc} -\delta c_{D;++}(\vec{x},\omega) & \delta c_{D;+-}(\vec{x},\omega) \\
-\delta c_{D;+-}(\vec{x},\omega) & \delta c_{D;--}(\vec{x},\omega) \eea\right);
-\eta_{p}\;\hat{X}_{pq}\pdag(\vec{x},\omega)\;\eta_{q} =
\left(\bea{cc} \delta c_{D;++}^{*}(\vec{x},\omega) & -\delta c_{D;+-}^{*}(\vec{x},\omega) \\
\delta c_{D;+-}^{*}(\vec{x},\omega) & -\delta c_{D;--}^{*}(\vec{x},\omega) \eea\right)_{\!\!\mbox{;}}   \\         \lb{s4_18}
\hat{Y}_{pq}^{ab}(\vec{x},\omega) &=&
\hat{P}_{4\times 4}^{-1}(\vec{x},\omega)\;\;\hat{Y}_{D;4\times 4}(\vec{x},\omega)\;\;\hat{P}_{4\times 4}(\vec{x},\omega)\;; \\ \lb{s4_19}
\hat{Y}_{D;4\times 4}(\vec{x},\omega) &=&\left(\bea{cc} \Big(0\Big)_{pq}^{11} &
\hat{X}_{D;pq}(\vec{x},\omega)  \\ -\hat{X}_{D;pq}\pdag(\vec{x},\omega) & \Big(0\Big)_{pq}^{22} \eea\right)
\;\;;\hspace*{0.3cm}\hat{P}_{4\times 4}(\vec{x},\omega)=\left(\bea{cc}\hat{P}_{2\times 2}^{11}(\vec{x},\omega) & 0 \\ 0 &
\hat{P}_{2\times 2}^{22}(\vec{x},\omega)\eea\right)  \;; \\  \lb{s4_20}
\hat{X}_{D;pq}(\vec{x},\omega) &=& \left(\bea{cc} -\ovv{c}_{++}(\vec{x},\omega)  &  0 \\
0 & \ovv{c}_{--}(\vec{x},\omega)   \eea\right)_{pq}  \;\;;\hspace*{0.3cm}
-\hat{X}_{D;pq}\pdag(\vec{x},\omega)=  \left(\bea{cc} \ovv{c}_{++}^{*}(\vec{x},\omega)  &  0 \\
0 & -\ovv{c}_{--}^{*}(\vec{x},\omega)   \eea\right)_{pq}  \;; \\  \no    \ovv{c}_{++}(\vec{x},\omega) &:=&
|\ovv{c}_{+}(\vec{x},\omega)|\;\exp\{\im\,\varphi_{+}(\vec{x},\omega)\}\;;\;\;\;\;   \ovv{c}_{--}(\vec{x},\omega) :=
|\ovv{c}_{-}(\vec{x},\omega)|\;\exp\{\im\,\varphi_{-}(\vec{x},\omega)\}\;;    \\   \lb{s4_21}
\hat{P}_{2\times 2}^{11}(\vec{x},\omega) &=& \exp\Big\{\im\;\hat{\mscr{C}}_{D;2\times 2}(\vec{x},\omega)\Big\}\;\;;\hspace*{0.3cm}
\hat{P}_{2\times 2}^{22}(\vec{x},\omega) = \exp\Big\{\im\;\hat{\mscr{C}}_{D;2\times 2}^{T}(\vec{x},\omega)\Big\} \;; \\
   \lb{s4_22}   \hat{\mscr{C}}_{D;2\times 2}(\vec{x},\omega)&=&\hat{\mscr{C}}_{D;pq}(\vec{x},\omega)=\left(\bea{cc}
0 & -\mscr{C}_{D;+-}(\vec{x},\omega)  \\ \mscr{C}_{D;+-}^{*}(\vec{x},\omega) \eea\right)_{pq}  \;\;;\;\;
\hat{P}_{2\times 2}^{22,T}(\vec{x},\omega)=\hat{P}_{2\times 2}^{11}(\vec{x},\omega)  \;;     \\    \no 
\mscr{C}_{D;+-}(\vec{x},\omega) &:=&  |\mscr{C}_{D}(\vec{x},\omega)|\;\exp\{\im\,\gamma_{D}(\vec{x},\omega)\}\;\;\;.
\eeq
These similarities also hold for the block diagonal density-related parts so that one can immediately state following relations
in place of (\ref{s3_28}-\ref{s3_35}) 
\beq \lb{s4_23}
\delta\hat{\Sigma}_{D;pq}^{aa}(\vec{x},\omega)\;\;\wt{K} &=&
\hat{Q}_{pp\ppr}^{-1,aa}(\vec{x},\omega)\;\;\delta\hat{\Lambda}_{p\ppr}^{a}(\vec{x},\omega)\;\;
\hat{Q}_{p\ppr q}^{aa}(\vec{x},\omega)  \\ \no &=& \left(\bea{cc}
\left(\bea{cc}
\delta B_{D;++}(\vec{x},\omega) & -\delta B_{D;+-}(\vec{x},\omega) \\
\delta B_{D;+-}^{*}(\vec{x},\omega) & -\delta B_{D;--}(\vec{x},\omega) \eea\right)_{pq}^{11} & \\ &\hspace*{-0.46cm}
\left(\bea{cc}
-\delta B_{D;++}(\vec{x},\omega) & \delta B_{D;+-}^{*}(\vec{x},\omega) \\
-\delta B_{D;+-}(\vec{x},\omega) & \delta B_{D;--}(\vec{x},\omega) \eea\right)_{pq}^{22} \eea\right) \;; \\   \lb{s4_24}
\delta\hat{\Lambda}_{p}^{a}(\vec{x},\omega)&=&
\mbox{diag}\Big\{\underbrace{\overbrace{\bigl(+\delta\lambda_{+}(\vec{x},\omega)\:,
\:-\delta\lambda_{-}(\vec{x},\omega)\,\bigr)}^{p\cdot\delta\hat{\lambda}_{p}(\vec{x},\omega)}}_{
a=1}\;;\;\underbrace{\overbrace{\bigl(-\delta\lambda_{+}(\vec{x},\omega)\:,\:+\delta\lambda_{-}(\vec{x},\omega)\,\bigr)}^{-p\cdot 
\delta\hat{\lambda}_{p}(\vec{x},\omega)}}_{a=2}\Big\}=
\delta\hat{\lambda}_{p}(\vec{x},\omega)\;\wt{K}_{pp}^{aa} \;;  \\  \no
\delta\hat{\lambda}_{p}(\vec{x},\omega)  &=&  \mbox{diag}\bigl\{\delta\lambda_{+}(\vec{x},\omega)\:,\:
\delta\lambda_{-}(\vec{x},\omega)\bigr\}\;;\;\;\; \delta\hat{\lambda}_{p}(\vec{x},\omega) \in\mathbb{R}\;; \\   \lb{s4_25}
\hat{Q}_{pq}^{11}(\vec{x},\omega) &=&\Big(\exp\Big\{\im\;\hat{\mscr{B}}_{D}(\vec{x},\omega)\Big\}\Big)_{pq}\;\;;\hspace*{0.5cm}
\hat{Q}_{pq}^{22}(\vec{x},\omega) = \Big(\exp\Big\{\im\;\hat{\mscr{B}}_{D}^{T}(\vec{x},\omega)\Big\}\Big)_{pq}  \;; \\  \lb{s4_26}
\hat{\mscr{B}}_{D}(\vec{x},\omega) &=&\left(\bea{cc} 0 & -\mscr{B}_{D;+-}(\vec{x},\omega)  \\
\mscr{B}_{D;+-}^{*}(\vec{x},\omega) & 0 \eea\right)  \;\;;\hspace*{0.5cm}
\hat{Q}_{pq}^{22,T}(\vec{x},\omega)=\hat{Q}_{pq}^{11}(\vec{x},\omega)  \;;  \\    \no 
\mscr{B}_{D;+-}(\vec{x},\omega) &:=& |\mscr{B}_{D}(\vec{x},\omega)|\;\;\exp\{\im\,\beta_{D}(\vec{x},\omega)\,\}  \;; \;\;\;
\mscr{B}_{D;+-}(\vec{x},\omega) \in\mathbb{C}\;;     \\  \lb{s4_27}
\delta\hat{\Sigma}_{D;pq}^{aa}(\vec{x},\omega)\;\wt{K}^{aa}_{qq} &=& \hat{Q}_{pp\ppr}^{aa;-1}(\vec{x},\omega)\;\;
\delta\hat{\Lambda}_{p\ppr}^{a}(\vec{x},\omega)\;\;\hat{Q}_{p\ppr q}^{aa}(\vec{x},\omega)\;\;;\hspace*{0.5cm}
\hat{Q}_{pq}^{aa}(\vec{x},\omega)=\left(\bea{cc} \hat{Q}_{pq}^{11}(\vec{x},\omega) & 0 \\ 0 & 
\hat{Q}_{pq}^{22}(\vec{x},\omega) \eea\right) \;; \\ \lb{s4_28}
\Big(\delta\hat{\Sigma}_{D}^{11}(\vec{x},\omega)\;\hat{\eta}\Big) &=&
\hat{Q}^{11,-1}(\vec{x},\omega)\;\;\big(p\cdot\delta\hat{\lambda}_{p}(\vec{x},\omega)\big)\;\;\hat{Q}^{11}(\vec{x},\omega) \;; \\  \lb{s4_29}
-\Big(\delta\hat{\Sigma}_{D}^{22}(\vec{x},\omega)\;\hat{\eta}\Big) &=&
\hat{Q}^{22,-1}(\vec{x},\omega)\;\;\big(-p\cdot\delta\hat{\lambda}_{p}(\vec{x},\omega)\big)\;\;\hat{Q}^{22}(\vec{x},\omega) \;; \\  \lb{s4_30}
\delta\hat{\Sigma}_{D;pq}^{22,T}(\vec{x},\omega) &=& \delta\hat{\Sigma}_{D;pq}^{11}(\vec{x},\omega) \;.
\eeq
However, the real self-energy variable \(\sigma_{V_{0}}^{(0)}(\vec{x},t_{p})\) for the hermitian contact interaction takes a dependence on the
difference of two frequencies \(\sdelta\omega_{p}=\omega_{p}-\omega_{p}\ppr\) which therefore reduces in the limit 
\(\omega_{p}\rightarrow\omega_{p}\ppr\) for stationary states to the zero frequency mode of a real-valued self-energy variable
\(\sigma_{V_{0}}^{(0)}(\vec{x},\sdelta\omega_{p}\equiv0)\)
\be\lb{s4_31}
\sigma_{V_{0}}^{(0)}(\vec{x},t_{p})\rightarrow\sigma_{V_{0}}^{(0)}(\vec{x},\sdelta\omega_{p}=\omega_{p}-\omega_{p}\ppr)
\stackrel{\sdelta\omega_{p}=0}{\rightarrow}\ovv{\sigma}_{V_{0};p}^{(0)}(\vec{x})\in\mathbb{R}\;.
\ee
Consequently, we can directly convey the integration measure (\ref{s3_94}) and path integral \(\ovv{Z_{II}[\hat{\mscr{J}}]}\) (\ref{s3_95})
with background averaging functional (\ref{s3_96}) to the case \(\ovv{Z_{I}[\hat{\mscr{J}}]}\) of static disorder for the restriction to
stationary states; we do not outline the Fourier transformations of source fields and source matrices in detail and simply introduce their results
for brevity
\beq\lb{s4_32}
\lefteqn{d[\delta\wt{\Sigma}_{\vec{x};pq}^{ab}(\omega)\,\wt{K}] = \mcal{N}_{\delta\wt{\Sigma}\,\wt{K}}
\bigg(\prod_{\{\vec{x},t\}}d\bigl(\delta\lambda_{+}(\vec{x},\omega)\bigr)\;
d\bigl(\delta\lambda_{-}(\vec{x},\omega)\bigr)\;\Bigl(\delta\lambda_{+}(\vec{x},\omega)\Bigr)^{2}\;
\Bigl(\delta\lambda_{-}(\vec{x},\omega)\Bigr)^{2}\times } \\   \no &\times&
\Bigl(\bigl(\delta\lambda_{+}(\vec{x},\omega)\bigr)^{2}-\bigl(\delta\lambda_{-}(\vec{x},\omega)\bigr)^{2}\Bigr)^{2}\times
d\bigl(|\mscr{B}_{D}(\vec{x},\omega)|\bigr)\;\sinh\bigl(4|\mscr{B}_{D}(\vec{x},\omega)|\bigr)\;d\beta_{D}(\vec{x},\omega)\bigg)\times \\ \no &\times&
\bigg(\prod_{\{\vec{x},t\}}d\bigl(|\ovv{c}_{+}(\vec{x},\omega)|\bigr)\;\Bigl(\sin\bigl(2|\ovv{c}_{+}(\vec{x},\omega)|\bigr)\Bigr)^{2}\;d\varphi_{+}(\vec{x},\omega)\;
\times\;d\bigl(|\ovv{c}_{-}(\vec{x},\omega)|\bigr)\;\Bigl(\sin\bigl(2|\ovv{c}_{-}(\vec{x},\omega)|\bigr)\Bigr)^{2}\;d\varphi_{-}(\vec{x},\omega)\;\times \\ \no &\times&
d\bigl(|\mscr{C}_{D}(\vec{x},\omega)|\bigr)\;\sinh\bigl(2|\mscr{C}_{D}(\vec{x},\omega)|\bigr)\;d\gamma_{D}(\vec{x},\omega) \bigg)=  
\mcal{N}_{\delta\wt{\Sigma}\,\wt{K}}
\bigg(\prod_{\{\vec{x},t\}}d\bigl(\delta\lambda_{+}(\vec{x},\omega)\bigr)\,
d\bigl(\delta\lambda_{-}(\vec{x},\omega)\bigr)\;
\mbox{det}^{\!\boldsymbol{2}}\hspace*{-0.06cm}\Bigg[\hspace*{-0.2cm}
\bea{cc}\delta\lambda_{+}(\vec{x},\omega) \hspace*{-0.1cm}& \delta\lambda_{-}(\vec{x},\omega)  \\
\delta\lambda_{+}^{3}(\vec{x},\omega)\hspace*{-0.1cm} & \delta\lambda_{-}^{3}(\vec{x},\omega) \eea\hspace*{-0.2cm}
\Bigg]\hspace*{-0.16cm}\times  \\ \no &\times&\tfrac{1}{4}\;
d\Bigl(\cosh\bigl(4|\mscr{B}_{D}(\vec{x},\omega)|\bigr)\Bigr)\;d\beta_{D}(\vec{x},\omega)\bigg) \times \bigg(\prod_{\{\vec{x},t\}}
\tfrac{1}{2}\;d\Bigl(\cosh\bigl(2|\mscr{C}_{D}(\vec{x},\omega)|\bigr)\Bigr)\;d\gamma_{D}(\vec{x},\omega) \;\times \\ \no &\times&
\tfrac{1}{2}\;d\Bigl(|\ovv{c}_{+}(\vec{x},\omega)|-\tfrac{1}{4}\sin\bigl(4|\ovv{c}_{+}(\vec{x},\omega)|\bigr)\Bigr)\;d\varphi_{+}(\vec{x},\omega)\; \times\;
\tfrac{1}{2}\;d\Bigl(|\ovv{c}_{-}(\vec{x},\omega)|-\tfrac{1}{4}\sin\bigl(4|\ovv{c}_{-}(\vec{x},\omega)|\bigr)\Bigr)\;d\varphi_{-}(\vec{x},\omega)\bigg)\;;
\eeq 
\beq\lb{s4_33}
\lefteqn{\ovv{Z_{I}[\hat{\mscr{J}}]}=\bigg\langle\exp\bigg\{-\frac{1}{2\;R_{I}^{2}\;\frac{\mscr{N}_{t}}{\mcal{N}_{x}}}\sum_{p=\pm}\sum_{\vec{x}}
\Big(1-\mu_{p}^{(I)}\Big)\;
\trab\Big[\wt{J}_{\psi\psi;\vec{x};p}^{ab}\;\wt{\kappa}\;
\wt{J}_{\psi\psi;\vec{x};p}^{ba}\;\wt{\kappa}\Big]\bigg\}  
\int d[\delta\wt{\Sigma}_{\vec{x};pq}^{ab}(\omega)\;\wt{K}]     }   \\ \no &\times&
\exp\bigg\{-\frac{1}{8}\frac{1}{R_{I}^{2}\;\frac{\mscr{N}_{t}}{\mcal{N}_{x}}}
\int_{\breve{0}}^{+\breve{\Omega}}\frac{d\omega}{(\frac{2\pi}{T_{max}})}
\sum_{p,q=\pm}\sum_{\vec{x}}\Big(1-\delta_{pq}\;\mu_{p}^{(I)}\Big)\;
\trab\Big[\delta\wt{\Sigma}_{\vec{x};pq}^{ab}(\omega)\;\wt{K}\;
\delta\wt{\Sigma}_{\vec{x};qp}^{ba}(\omega)\;\wt{K}\Big]\bigg\}    \\ \no &\times&
\exp\bigg\{\frac{1}{2R_{I}^{2}\;\frac{\mscr{N}_{t}}{\mcal{N}_{x}} }
\int_{\breve{0}}^{+\breve{\Omega}}\frac{d\omega}{(\frac{2\pi}{T_{max}})}\sum_{p=\pm}\sum_{\vec{x}}
\eta_{p}\;\Big(1-\mu_{p}^{(I)}\Big)
\trab\Big[\delta\wt{\Sigma}_{\vec{x};pp}^{ab}(\omega)\;\wt{\kappa}\;
\wt{J}_{\psi\psi;\vec{x};p}^{ba}\;\wt{\kappa}\Big]\bigg\}     \\ \no &\times&
\exp\bigg\{-\frac{1}{2}\int_{\breve{C}_{\omega}}\frac{d\omega_{p}}{(\frac{2\pi}{T_{max}})}\eta_{p}\sum_{\vec{x}}\mcal{N}_{x}
\trab\ln\bigg[\hat{1}+\delta\breve{\mscr{H}}(\hat{T}^{-1},\hat{T})\;\bigl(\breve{\mfrak{H}}+\hat{\sigma}_{V_{0}}^{(0)}\bigr)^{-1}+
\bigl(\hat{T}^{-1}\hat{I}\tfrac{1}{\mcal{N}_{x}}\hat{\mscr{J}}\hat{I}\wt{K}\hat{T}\bigr)\;
\bigl(\breve{\mfrak{H}}+\hat{\sigma}_{V_{0}}^{(0)}\bigr)^{-1}\bigg]_{\vec{x}=\vec{x}\ppr;p=q}^{a=b}\hspace*{-1.3cm}(\omega,\omega\ppr=\omega)\bigg\} 
\\ \no &\times&
\exp\bigg\{\frac{\im}{2}\frac{\mscr{N}_{t}^{2}}{\hbar}\int_{\breve{0}}^{+\breve{\Omega}}\hspace*{-0.28cm}
\frac{d\omega}{(\frac{2\pi}{T_{max}})}\,\frac{d\omega\pppr}{(\frac{2\pi}{T_{max}})}\hspace*{-0.2cm}
\sum_{p\ppr,q\pppr=\pm}^{a\ppr,b\pppr=1,2}
\sum_{\vec{x},\vec{x}\pppr}\mcal{N}_{x}\;T_{max}^{2}\;J_{\psi;\vec{x}\pppr}^{+b\pppr}(\omega_{q\pppr}\pppr)\;\hat{I}\;\wt{K}\; 
\bigg[\int_{\breve{0}}^{+\breve{\Omega}}\hspace*{-0.28cm}\frac{d\omega\ppr}{(\frac{2\pi}{T_{max}})}\sum_{p,q,q\ppr=\pm}^{a,b,b\ppr=1,2}
\sum_{\vec{x}\ppr}\mcal{N}_{x} \;\times    \\  \no &\times&
\bigg\{\hat{T}_{q\pppr q\ppr}^{b\pppr b\ppr}(\vec{x}\pppr,\omega\pppr) \;
\bigl(\breve{\mfrak{H}}+\hat{\ovv{\sigma}}_{V_{0}}^{(0)}\bigr)_{\vec{x}\pppr,\vec{x}\ppr}^{-1;b\ppr=b}(\omega_{q\ppr}\pppr,\omega_{q}\ppr)\;\times
\\  \no &\times&
\Big[\hat{1}+
\delta\breve{\mscr{H}}(\hat{T}^{-1},\hat{T})\;\bigl(\breve{\mfrak{H}}+\hat{\ovv{\sigma}}_{V_{0}}^{(0)}\bigr)^{-1} +  
\bigl(\hat{T}^{-1}\hat{I}\tfrac{1}{\mcal{N}_{x}}\hat{\mscr{J}}\hat{I}\wt{K}\hat{T}\bigr)\;
\bigl(\breve{\mfrak{H}}+\hat{\ovv{\sigma}}_{V_{0}}^{(0)}\bigr)^{-1}\Big]_{\vec{x}\ppr,\vec{x}}^{-1;ba}(\omega_{q}\ppr,\omega_{p})\;
\hat{T}_{pp\ppr}^{-1;aa\ppr}(\vec{x},\omega)\bigg\}_{\vec{x}\pppr,\vec{x}}^{b\pppr a\ppr}\hspace{-0.6cm}(\omega_{q\pppr}\pppr,\omega_{p\ppr}) +
\\  \no &-&
\bigl(\breve{\mfrak{H}}+\hat{\ovv{\sigma}}_{V_{0}}^{(0)}\bigr)_{\vec{x}\pppr,\vec{x}}^{-1;b\pppr a\ppr}(\omega_{q\pppr}\pppr,\omega_{p\ppr})\bigg]
\hat{I}\;J_{\psi;\vec{x}}^{a\ppr}(\omega_{p\ppr})\bigg\}\bigg\rangle_{\boldsymbol{\hat{\ovv{\sigma}}_{V_{0}}^{(0)}};}
\eeq
\beq\lb{s4_34}
\lefteqn{\bigg\langle\Big(\mbox{coset fields and }\hat{\ovv{\sigma}}_{V_{0}}^{(0)}\Big)
\bigg\rangle_{\boldsymbol{\hat{\ovv{\sigma}}_{V_{0}}^{(0)}}}=
\int d[\ovv{\sigma}_{V_{0};p}^{(0)}(\vec{x})]\;\;
\exp\bigg\{-\frac{1}{4}\frac{R_{I}^{2}\,(\mscr{N}_{t}^{3}/\mcal{N}_{x})}{(\hbar\;V_{0})^{2}}\sum_{\vec{x}}
\Big(\ovv{\sigma}_{V_{0};+}^{(0)}(\vec{x})-\ovv{\sigma}_{V_{0};p}^{(0)}(\vec{x})\Big)^{2}\bigg\}   }  \\ \no &\times&
\exp\bigg\{\frac{\im}{2\hbar}\frac{T_{max}}{V_{0}-\im\;\ve_{p}}\sum_{\vec{x};p=\pm}\eta_{p}\;\;
\ovv{\sigma}_{V_{0};p}^{(0)}(\vec{x})\;\;\ovv{\sigma}_{V_{0};p}^{(0)}(\vec{x})
-\int_{\breve{C}_{\omega}}\frac{d\omega_{p}}{(\frac{2\pi}{T_{max}})}\,\eta_{p}\sum_{\vec{x}}\mcal{N}_{x}
\trpq\ln\Big[\Big(\breve{\mfrak{H}}^{11}+\hat{\ovv{\sigma}}_{V_{0}}^{(0)}\Big)_{\vec{x}=\vec{x}\ppr;p=q}^{11}(\omega,\omega\ppr=t)\Big]\bigg\} 
\\   \no &\times& \exp\bigg\{\im\frac{\mscr{N}_{t}^{2}}{\hbar}
\int_{\breve{0}}^{+\breve{\Omega}}\frac{d\omega}{(\frac{2\pi}{T_{max}})}\;\frac{d\omega\ppr}{(\frac{2\pi}{T_{max}})}
\sum_{p,q=\pm}\sum_{\vec{x},\vec{x}\ppr}
\mcal{N}_{x}\;T_{max}^{2}\;j_{\psi;\vec{x}\ppr}\pdag(\omega_{q}\ppr)\Big[\bigl(\breve{\mfrak{H}}^{11}+\hat{\sigma}_{V_{0}}^{(0)}\bigr)
\hat{\eta}\Big]_{\vec{x}\ppr,\vec{x}}^{-1}(\omega_{q}\ppr,\omega_{p})\;j_{\psi;\vec{x}}(\omega_{p})\bigg\}  \times  \\  \no &\times&
\Big(\mbox{coset fields and }\hat{\ovv{\sigma}}_{V_{0}}^{(0)}\Big)_{\mbox{.}}
\eeq
The corresponding Green function term of (\ref{s3_97}) and gradient operator (\ref{s3_98}) has a similar form in frequency space under
the simplifying restriction to stationary states
\beq\lb{s4_35}
\Big(\breve{\mfrak{H}}+\hat{\ovv{\sigma}}_{V_{0}}^{(0)}\Big)_{\vec{x},\vec{x}\ppr}^{ab}(\omega_{p},\omega_{q}\ppr) &=&
\Big(\hat{H}_{p}^{a}(\omega_{p})+\ovv{\sigma}_{V_{0}}^{(0)}(\vec{x})\Big)\delta_{ab}\;\delta_{pq}\;\delta_{\vec{x},\vec{x}\ppr}\;
\delta(\omega_{p}-\omega_{q}\ppr)\;;  \\ \lb{s4_36}
\delta\breve{\mscr{H}}_{\vec{x},\vec{x}\ppr}^{ab}(\hat{T}^{-1},\hat{T};\omega_{p},\omega_{q}\ppr) &=&
\hat{T}_{pp\ppr}^{-1;aa\ppr}(\vec{x},\omega)\;\breve{\mfrak{H}}_{\vec{x},\vec{x}\ppr}^{a\ppr=b\ppr}(\omega_{p\ppr},\omega_{q\ppr}\ppr)\;
\hat{T}_{q\ppr q}^{b\ppr b}(\vec{x}\ppr,\omega\ppr)-\breve{\mfrak{H}}_{\vec{x},\vec{x}\ppr}^{a=b}(\omega_{p},\omega_{q}\ppr)\;.
\eeq
Although it is straightforward to integrate over the density-related variables \(d|\mscr{B}_{D}(\vec{x},\omega)|\), \(d\beta_{D}(\vec{x},\omega)\)
and \(d\big(\delta\lambda_{\pm}(\vec{x},\omega)\,\big)\) within the Gaussian factors of  \(\ovv{Z_{I}[\hat{\mscr{J}}]}\) (\ref{s4_33}) in a direct manner, 
there occurs a further simplifying fact which concerns the parameter \(\mu_{p}^{(I)}\) (\ref{s4_5}); as one takes 
the limit \(\mscr{N}_{t}=(T_{max}/\sdelta t)\rightarrow\infty\) for an infinite
number of discrete time steps, the parameter \(\mu_{p}^{(I)}\) approaches zero for an even simpler calculation 
of the density-related Gaussian factors within  \(\ovv{Z_{I}[\hat{\mscr{J}}]}\) (\ref{s4_33}). Considering this simplifying aspect,
we finally attain the reduced path integral (\ref{s4_37}) for the case of static disorder with the appropriate scale of the disorder
parameter and the parameter \(\mcal{N}_{x}\) (\ref{s1_3}) for the number of spatial points which allows for a saddle
point approximation in the limit \(\mcal{N}_{x}\rightarrow\infty\)
\beq\lb{s4_37}
\lefteqn{\ovv{Z_{I}[\hat{\mscr{J}}]}=\bigg\langle\exp\bigg\{-\frac{1}{2}\frac{1}{R_{I}^{2}}\;\frac{\mcal{N}_{x}}{\mscr{N}_{t}}\sum_{p=\pm}\sum_{\vec{x}}
\trab\Big[\wt{J}_{\psi\psi;\vec{x};p}^{ab}\;\wt{\kappa}\;
\wt{J}_{\psi\psi;\vec{x};p}^{ba}\;\wt{\kappa}\Big]\bigg\}  
\int d[\delta\wt{\Sigma}_{\vec{x};pq}^{ab}(\omega)\;\wt{K}]     }   \\ \no &\times&
\exp\bigg\{-\frac{1}{8}\frac{1}{R_{I}^{2}}\;\mcal{N}_{x}
\int_{\breve{0}}^{+\breve{\Omega}}\frac{d\omega}{2\pi\,\Omega_{0}}
\sum_{p,q=\pm}\sum_{\vec{x}}
\trab\Big[\delta\wt{\Sigma}_{\vec{x};pq}^{ab}(\omega)\;\wt{K}\;
\delta\wt{\Sigma}_{\vec{x};qp}^{ba}(\omega)\;\wt{K}\Big]\bigg\}    \\ \no &\times&
\exp\bigg\{\frac{1}{2}\frac{1}{R_{I}^{2}}\;\mcal{N}_{x}
\int_{\breve{0}}^{+\breve{\Omega}}\frac{d\omega}{2\pi\,\Omega_{0}}\sum_{p=\pm}\sum_{\vec{x}}
\eta_{p}\;
\trab\Big[\delta\wt{\Sigma}_{\vec{x};pp}^{ab}(\omega)\;\wt{\kappa}\;
\wt{J}_{\psi\psi;\vec{x};p}^{ba}\;\wt{\kappa}\Big]\bigg\}     \\ \no &\times&
\exp\bigg\{-\frac{1}{2}\int_{\breve{C}_{\omega}}\frac{d\omega_{p}}{(\frac{2\pi}{T_{max}})}\eta_{p}\sum_{\vec{x}}\mcal{N}_{x}
\trab\ln\bigg[\hat{1}+\delta\breve{\mscr{H}}(\hat{T}^{-1},\hat{T})\;\bigl(\breve{\mfrak{H}}+\hat{\sigma}_{V_{0}}^{(0)}\bigr)^{-1}+
\bigl(\hat{T}^{-1}\hat{I}\tfrac{1}{\mcal{N}_{x}}\hat{\mscr{J}}\hat{I}\wt{K}\hat{T}\bigr)\;
\bigl(\breve{\mfrak{H}}+\hat{\sigma}_{V_{0}}^{(0)}\bigr)^{-1}\bigg]_{\vec{x}=\vec{x}\ppr;p=q}^{a=b}\hspace*{-1.3cm}(\omega,\omega\ppr=\omega)\bigg\} 
\\ \no &\times&
\exp\bigg\{\frac{\im}{2}\frac{\mscr{N}_{t}^{2}}{\hbar}\int_{\breve{0}}^{+\breve{\Omega}}\hspace*{-0.28cm}
\frac{d\omega}{(\frac{2\pi}{T_{max}})}\,\frac{d\omega\pppr}{(\frac{2\pi}{T_{max}})}\hspace*{-0.2cm}
\sum_{p\ppr,q\pppr=\pm}^{a\ppr,b\pppr=1,2}
\sum_{\vec{x},\vec{x}\pppr}\mcal{N}_{x}\;T_{max}^{2}\;J_{\psi;\vec{x}\pppr}^{+b\pppr}(\omega_{q\pppr}\pppr)\;\hat{I}\;\wt{K}\; 
\bigg[\int_{\breve{0}}^{+\breve{\Omega}}\hspace*{-0.28cm}\frac{d\omega\ppr}{(\frac{2\pi}{T_{max}})}\sum_{p,q,q\ppr=\pm}^{a,b,b\ppr=1,2}
\sum_{\vec{x}\ppr}\mcal{N}_{x} \;\times    \\  \no &\times&
\bigg\{\hat{T}_{q\pppr q\ppr}^{b\pppr b\ppr}(\vec{x}\pppr,\omega\pppr) \;
\bigl(\breve{\mfrak{H}}+\hat{\ovv{\sigma}}_{V_{0}}^{(0)}\bigr)_{\vec{x}\pppr,\vec{x}\ppr}^{-1;b\ppr=b}(\omega_{q\ppr}\pppr,\omega_{q}\ppr)\;\times
\\  \no &\times&
\Big[\hat{1}+
\delta\breve{\mscr{H}}(\hat{T}^{-1},\hat{T})\;\bigl(\breve{\mfrak{H}}+\hat{\ovv{\sigma}}_{V_{0}}^{(0)}\bigr)^{-1} +  
\bigl(\hat{T}^{-1}\hat{I}\tfrac{1}{\mcal{N}_{x}}\hat{\mscr{J}}\hat{I}\wt{K}\hat{T}\bigr)\;
\bigl(\breve{\mfrak{H}}+\hat{\ovv{\sigma}}_{V_{0}}^{(0)}\bigr)^{-1}\Big]_{\vec{x}\ppr,\vec{x}}^{-1;ba}(\omega_{q}\ppr,\omega_{p})\;
\hat{T}_{pp\ppr}^{-1;aa\ppr}(\vec{x},\omega)\bigg\}_{\vec{x}\pppr,\vec{x}}^{b\pppr a\ppr}\hspace{-0.6cm}(\omega_{q\pppr}\pppr,\omega_{p\ppr}) +
\\  \no &-&
\bigl(\breve{\mfrak{H}}+\hat{\ovv{\sigma}}_{V_{0}}^{(0)}\bigr)_{\vec{x}\pppr,\vec{x}}^{-1;b\pppr a\ppr}(\omega_{q\pppr}\pppr,\omega_{p\ppr})\bigg]
\hat{I}\;J_{\psi;\vec{x}}^{a\ppr}(\omega_{p\ppr})\bigg\}\bigg\rangle_{\boldsymbol{\hat{\ovv{\sigma}}_{V_{0}}^{(0)}}.}
\eeq

% Set the ending of a LaTeX document
\end{document}